\documentclass[pre,amssymb,superscriptaddress,showpacs,notitlepage,nofootinbib]{revtex4-1}
\usepackage{amsmath}
\usepackage{amsfonts}
\usepackage{amssymb}
\usepackage{epsfig}
\usepackage{graphicx}
\usepackage[normalem]{ulem}
\usepackage{comment}

\renewcommand{\phi}{\varphi}

\newcommand{\be}{\begin{equation}}
\newcommand{\ee}{\end{equation}}
\newcommand{\bea}{\begin{equnaray}}
\newcommand{\eea}{\end{equnaray}}
\newcommand{\ba}{\begin{align}}
\newcommand{\ea}{\end{align}}

\usepackage{color}

\def\br{{\bf r}}
\def\brpk{{\bf r}^{\preceq k}}
\def\brsk{{\bf r}^{\succ k}}
\def\Nd{N_{\rm d}}
\def\Nm{N_{\rm m}}
\def\Od{\Omega_{\rm d}}
\def\Om{\Omega_{\rm m}}
\definecolor{green}{rgb}{0.0, 0.44, 0.0}
\definecolor{red}{rgb}{1.0, 0.13, 0.32}
\definecolor{blue}{rgb}{0.06, 0.2, 0.65}
\definecolor{magenta}{rgb}{1.0, 0.0, 1.00}
\definecolor{purple}{rgb}{0.7, 0.0, 0.7}
\definecolor{cyan}{rgb}{0.0, 1.0, 1.0}

\begin{document}

\title{Creating equilibrium glassy states via random particle bonding}

\author{Misaki Ozawa}
\affiliation{Univ. Grenoble Alpes, CNRS, LIPhy, 38000 Grenoble, France}

\author{Jean-Louis Barrat}
\affiliation{Univ. Grenoble Alpes, CNRS, LIPhy, 38000 Grenoble, France}

\author{Walter Kob}
\affiliation{Department of Physics, University of Montpellier and CNRS, F-34095 Montpellier, France}

\author{Francesco Zamponi}
\affiliation{Dipartimento di Fisica, Sapienza Universit\`a di Roma, Piazzale Aldo Moro 2, 00185 Rome, Italy }

\begin{abstract}
Creating amorphous solid states by randomly bonding an ensemble of dense liquid monomers is a common procedure which is applied to create a variety of materials such as epoxy resins, colloidal gels, and vitrimers. The properties of the resulting solid do, however, {\it a priori} strongly depend on the preparation history. This can lead to substantial aging of the material, i.e., properties such as mechanical moduli and transport coefficients depend on the time elapsed since solidification, which can lead to a slow degradation of the material in technological applications.
It is therefore important to understand under which conditions random monomer bonding can lead to stable solid states, i.e., long-lived metastable states whose properties do not change over time. 
In this work, we present a theoretical and computational analysis of this problem, and introduce a random bonding procedure that guarantees the proper equilibration of the resulting amorphous states. Our procedure also provides a new route to investigate the fundamental properties of glassy energy landscapes by producing translationally-invariant ultrastable glassy states of simple particle models.
\end{abstract}

\maketitle


\section{Introduction}

Since glasses are out-of-equilibrium materials, their properties depend on their history of formation~\cite{ediger1996supercooled}. A glass quenched from its liquid state with a slow cooling rate will have higher kinetic and mechanical stability with lower energy (or enthalpy) than if cooled quickly, which reflects the fact that the system resides in a lower region of its rugged energy landscape~\cite{debenedetti2001supercooled,rodney2011modeling}. Producing stable glasses and hence accessing the deep minima inside this landscape is a crucial challenge for experiments as well as computer simulations, as it allows one to devise better materials and/or gain insight into the nature of the glassy state. In experiments, vapor deposition techniques with controlled substrate temperature can produce glass samples with extraordinary kinetic stability, a major breakthrough in the last decades~\cite{swallen2007organic,queen2013excess,yu2013ultrastable,yoon2018testing,raegen2020ultrastable,ediger2017perspective,rodriguez2022ultrastable}. On the atomistic simulation side, various sampling techniques have been developed, such as replica exchange methods~\cite{marinari1992simulated,hukushima1996exchange,yamamoto2000replica}, Monte-Carlo simulations with smart updates~\cite{santen2000absence,grigera2001fast,gutierrez2015static,ninarello2017models}, random pinning methods~\cite{kim2003effects,cammarota2012ideal}, as well as machine learning-assisted sampling techniques~\cite{noe2019boltzmann,wu2019solving,mcnaughton2020boosting,wu2021unbiased,hibat2021variational,gabrie2022adaptive,ciarella2023machine}. 
In particular, the swap-Monte Carlo~\cite{ninarello2017models}, its generalizations~\cite{brito2018theory,hagh2022transient}, and the random pinning approach~\cite{cammarota2012ideal,kob2013probing,ozawa2015equilibrium} allow to generate equilibrium configurations deep inside the glassy landscape. The key idea behind these approaches is to vary in a systematic manner certain degrees of freedom while maintaining the thermal equilibrium properties of the system. In the swap MC algorithm~\cite{ninarello2017models}, the diameters of the particles are the new dynamical variables that are allowed to fluctuate, accelerating significantly the relaxation dynamics. 
While this method revolutionized computational glass physics, it has so far remained in the realm of computer simulations, and extending it to real experiments has turned out to be challenging.
In the random pinning approach~\cite{cammarota2012ideal}, the positions of a fraction of the particles are permanently frozen, and hence the system composed of the remaining mobile particles enters a very glassy state with strong confinements due to the presence of the pinned particles. Higher concentrations of pinned particles lead to ideal glasses~\cite{kob2013probing,ozawa2015equilibrium,ozawa2018ideal}, whose thermodynamic behavior has been found to be consistent with the random first-order transition theory~\cite{biroli2023rfot}. 
A big advantage of this method is that it can be applied to experimental systems, such as colloids~\cite{gokhale2014growing,williams2018experimental} and molecular liquids~\cite{kikumoto2020towards,das2023soft}. 
However, it has been demonstrated that the dynamics is significantly altered from the bulk state, possibly due to the violation of translational invariance, leading to a strong decoupling between self and collective behavior~\cite{charbonneau2013decorrelation,ozawa2015equilibrium,chakrabarty2016understanding}, a decrease of fragility as the concentration of pinned particles is increased~\cite{kim2011slow,chakrabarty2015dynamics}, and a suppression of dynamical heterogeneity when the glass transition is approached~\cite{kim2011slow,jack2013dynamical,kob2014nonlinear,li2015decoupling}, in stark contrast to the behavior of standard bulk materials that are approaching their glass transition.

Recently, we have proposed the random bonding method, in which one creates a bond between a pair of randomly chosen nearest neighbor particles~\cite{Ozawa2023}. This idea is inspired by random pinning~\cite{cammarota2012ideal}, but random bonding has the big advantage that it preserves the translational invariance of the system and that it can also be realized without much difficulty in real experiments. 
In fact, random bonding is routinely used to prepare amorphous solids such as epoxy resins (via curing agents~\cite{johari1994dynamics} or stereolitography~\cite{corcione2006temperature}) and vitrimers~\cite{kloxin2013covalent,denissen2016vitrimers}. It has also been used to prepare colloidal or emulsion clusters of various shapes via programmable bond activation~\cite{duguet2011design,peng2013colloidal} using temperature control~\cite{mcmullen2022self}, salt addition~\cite{mcmullen2018freely}, or UV light~\cite{yuan2016synthesis}.
The problem is that, to the best of our knowledge, it is not clear whether and under which conditions these techniques can produce stable glasses, see e.g.~\cite{goldbart1996randomly,corcione2006temperature,carbas2014effect,corezzi2002bond,mereu2015interplay}.
In this work, using theoretical analysis and computer simulations to probe the kinetic and mechanical properties of the system, we demonstrate that the random bonding method does indeed create ultrastable glasses in the bulk~\cite{Ozawa2023}. We confirm and theoretically support the preliminary molecular dynamics simulations of Ref.~\cite{Ozawa2023}, which suggested that the relaxation dynamics does not show aging within numerical accuracy. This implies that right after bonding, the resulting configuration is indeed close to  equilibrium~\cite{Ozawa2023}, akin to random pinning~\cite{krakoviack2010statistical}.

More specifically, in this work
we demonstrate that while the system is in equilibrium if the bonded particles are chosen completely randomly (in this case, the bond lengths are thus arbitrary), the bonding from pairs from nearest neighbor particles (which corresponds to a more realistic situation) does not ensure strict equilibration. However, it turn out that in practice the deviation from equilibrium is very small, and it can be negligible for most practical purposes, which will be demonstrated by detailed molecular simulations. 

Subsequently, we present results on the (almost) equilibrium dynamics of randomly bonded glass-forming liquids and contrast our findings with the ones from the dynamics of randomly pinned systems. 
We find that self and collective correlation functions for the translational degrees of freedom, as well as the rotational correlation function, are strongly coupled. Besides, we observe that the kinetic fragility does not change by increasing the concentration of bonds. Finally, we find that dynamical heterogeneity keeps growing with approaching the glass transition. These trends are thus opposite to the behavior found in the dynamics of randomly pinned systems, highlighting the importance of the nature of the quenched disorder.

\section{Statistical mechanics of bonded systems}
\label{sec:statmech}

In this section,
we discuss the statistical mechanics of randomly-bonded glass formers composed of monomers and dimers.
We randomly choose pairs of neighboring particles from an equilibrium configuration and bond then together by introducing a rigid-body constraint~\cite{Ozawa2023}.
The main goal of this section is to demonstrate that the creation of these bonds does not perturb significantly the thermal equilibrium of the system, and we do this by using similar ideas as in the random pinning protocol that fixes the positions of particles~\cite{scheidler2004relaxation,krakoviack2010statistical}. However, since the case of random bonding is quite subtle, we begin with a review of the random pinning approach and we then discuss how to extend the idea to random bonding.

\subsection{Quiet freezing of variables}
\label{sec:frozen}

Consider a system whose degrees of freedom are arbitrarily split into two distinct vectors ${\bf x}$ and ${\bf y}$ with Hamiltonian $H({\bf x},{\bf y})$. The equilibrium probability distribution of the total system is $\rho({\bf x},{\bf y})=\exp[-\beta H({\bf x},{\bf y})]/Z$, where $Z$ is the partition function and $\beta$ is the inverse temperature.
Let us assume that an equilibrium configuration of the system can be generated by sampling from $\rho({\bf x},{\bf y})$. 
Consider now the degrees of freedom ${\bf y}$. Their statistics is described by
the marginal probability distribution (we add a subscript ``$f$'' because these are the degrees of freedom that will be frozen in the following)
\begin{equation}
    \rho_f({\bf y}) = \int \mathrm{d}{\bf x} 
    \rho({\bf x},{\bf y})
    = \frac{Z_f({\bf y})}{Z}
    \ \ \ \mbox{with} 
    \quad 
    Z_f({\bf y}) =\int \mathrm{d}{\bf x} \,
    e^{-\beta H({\bf x},{\bf y})} \ .
    \label{eq:def_dist_frozen}
\end{equation}
Next, consider a setting in which one first generates a configuration of the ${\bf y}$ degrees of freedom from their marginal probability distribution $\rho_f({\bf y})$, and then considers a system in which ${\bf y}$ are ``frozen'' and whose dynamical variables are the ${\bf x}$ degrees of freedom, with equilibrium probability
\begin{equation}\label{eq:condp}
    \rho({\bf x}|{\bf y}) = 
    \frac{e^{-\beta H({\bf x},{\bf y})}}{Z_f({\bf y})} \ .
\end{equation}
Here, the ${\bf y}$ variables play the role of a {\it frozen} or {\it pinned} quenched disorder, and one is interested in the thermal properties of the system described by ${\bf x}$.

Using the chain rule of probabilities, we have
\begin{equation}\label{eq:chain}
    \rho({\bf x},{\bf y}) = 
    \rho({\bf x}|{\bf y}) \rho_f({\bf y}) \ .
\end{equation}
Hence, a pair $\{{\bf x}$, ${\bf y}\}$ generated from the joint distribution $\rho({\bf x},{\bf y})$ can be considered either as an equilibrium configuration of the full system, or as a realization ${\bf y}$ of the quenched disorder of the frozen system obtained from $\rho_f({\bf y})$ together with a typical equilibrium realization ${\bf x}$ of that system obtained from $\rho({\bf x}|{\bf y})$. We conclude that one can generate a pair ${\{{\bf x}, {\bf y}\}}$ from $\rho({\bf x},{\bf y})$, then {\it freeze} (or {\it pin}) ${\bf y}$, and automatically obtain an equilibrium configuration ${\bf x}$ (but only a single one) of the distribution $\rho({\bf x}|{\bf y})$ that describes the frozen system with quenched disorder ${\bf y}$. 

We now define an ``annealed'' average $\langle \cdots \rangle$, a ``thermal'' average $\langle \cdots \rangle_{\bf y}$,
and a ``disorder" average $\overline{\cdots}$ by
\begin{equation}
\langle \cdots \rangle = \int \mathrm{d}{\bf x}
\mathrm{d}{\bf y} \rho({\bf x},{\bf y}) (\cdots)  \ ,
\qquad
\langle \cdots \rangle_{\bf y} = \int \mathrm{d}{\bf x}
 \rho({\bf x}|{\bf y}) (\cdots)  \ ,
 \qquad
 \overline{\cdots} = \int \mathrm{d}{\bf y}
 \rho_f({\bf y}) (\cdots) \ ,
 \label{eq:def_averages}
\end{equation}
respectively.
Furthermore we define the ``quenched'' average as $\overline{\langle \cdots \rangle_{\bf y}}$ and
because of the chain rule in Eq.~\eqref{eq:chain}, the annealed and quenched averages coincide:
\begin{equation}
    \langle \cdots \rangle =
    \overline{\langle \cdots \rangle_{\bf y}} \ .
    \label{eq:identity}
\end{equation}
Conversely, the validity of Eq.~\eqref{eq:identity} for an arbitrary observable implies Eq.~\eqref{eq:chain}. Equations~\eqref{eq:chain} and \eqref{eq:identity} provide the core idea behind the random pinning and swap Monte-Carlo simulations, as we will discuss in detail below.

Since the freezing of degrees of freedom eliminates possible relaxation channels~\cite{ninarello2017models,ikeda2017mean,szamel2019theory,kapteijns2019fast,hagh2022transient}, it can be expected that sampling ${\bf x}$ from $\rho({\bf x}|{\bf y})$ at fixed ${\bf y}$ is harder than sampling
${\bf x}$ and ${\bf y}$ together from $\rho({\bf x},{\bf y})$. 
(In some cases, however, the conditional sampling gives rise to a faster sampling, see, e.g., Ref.~\cite{marchand2022wavelet}.) In other words, a local dynamics for ${\bf x}$ that satisfies detailed balance with respect to $\rho({\bf x}|{\bf y})$ will in general have a much larger decorrelation time than a similar local dynamics that acts on
${\bf x}$ and ${\bf y}$ and satisfies detailed balance with respect to $\rho({\bf x},{\bf y})$. Generating additional independent configurations from $\rho({\bf x}|{\bf y})$ for the same ${\bf y}$ might thus be a hard task.
Still, having a single equilibrium configuration ${\bf x}$ of the pinned system allows one to run local dynamics starting from that configuration and thus obtain equilibrium dynamical properties without having to worry about the process of equilibration itself.

A few remarks on this construction are in order at this point:
\begin{itemize}

\item The distribution $\rho_f({\bf y})$ depends on temperature and hence a different ensemble of pinned systems is obtained at each preparation temperature. The configuration ${\bf x}$ is in equilibrium at the same temperature. Once ${\bf y}$ is frozen, one is still allowed to change the temperature of ${\bf x}$, but in that case equilibrium is not guaranteed anymore. 

\item In systems with quenched disorder, the thermal degrees of freedom ${\bf x}$ 
are usually described by $\rho({\bf x}|{\bf y})$ as
in Eq.~\eqref{eq:condp}, but the distribution of the quenched disorder ${\bf y}$, i.e., $\rho_q({\bf y})$, is chosen independently, and in general $\rho_q({\bf y})\neq Z_f({\bf y})/Z$, where $Z_f({\bf y})$ is defined in Eq.~\eqref{eq:def_dist_frozen}. 
Physically, this corresponds to the fact that the quenched disorder ${\bf y}$ represents impurities in the Hamiltonian $H({\bf x},{\bf y})$ (e.g., the location of the magnetic atoms in a magnetic alloy that forms a spin glass) whose dynamics is extremely slow, thus preventing them to equilibrate with the other degrees of freedom ${\bf x}$ and
as a consequence the annealed and quenched averages do not coincide. Note that usually one keeps $\rho_q({\bf y})$ fixed 
while changing the temperature associated to ${\bf x}$.
The choice $\rho_q({\bf y})=\rho_f({\bf y})= Z_f({\bf y})/Z$,
in which the quenched disorder depends on temperature,
guarantees the equality of annealed and quenched averages in Eq.~\eqref{eq:identity}. It is a very special choice, and is called {\it Nishimori condition} in the physics literature, {\it quiet planting} in optimization, and {\it Bayes optimal condition} in statistical inference. See Ref.~\cite{zdeborova2016statistical} for a pedagogical discussion. 
\item  The thermal average $\langle O({\bf x}) \rangle_{\bf y}$ of an extensive observable that is the sum of local terms of the form
\begin{equation}
O({\bf x}) = \sum_{i} o_1(x_i) 
+
\sum_{i,j} o_2(x_i,x_j) + \cdots
\end{equation}
is a random variable that depends on the realization of the disorder ${\bf y}$ (the ``sample''). However, in the thermodynamic limit, disordered systems usually display the so-called {\it self-averaging property}: The disorder-induced fluctuations of thermal averages vanish and as a consequence the thermal average for a single typical sample coincides with the average over samples, namely, $\langle \cdots \rangle_{\bf y}=\overline{\langle \cdots \rangle_{\bf y}}$. Furthermore, note that the observable $O({\bf x},{\bf y})$ can also depend on the frozen degrees of freedom ${\bf y}$ (e.g. if $O=H$), in which case its thermal average also depends explicitly on the ${\bf y}$ through $O$. This does not affect the proof of equivalence of the annealed and quenched averages in Eq.~(\ref{eq:identity}), and it also does not affect the self-averaging property.
Note that self-averaging only holds for extensive variables, i.e., averages of local terms over the whole system or at least a finite fraction of it, in the thermodynamic limit. As an  example, in a system that has been subjected to the random pinning procedure, the pair correlation function measured in a large configuration will be independent of the choice of the pinned particles, and is self averaging. On the other hand, the density in a specific point of space will depend on this choice, and is not self averaging.

\item A local dynamics on ${\bf x}$ that samples from $\rho({\bf x}|{\bf y})$ could be so slow that it becomes non-ergodic at an ideal glass transition. This has been shown to happen in randomly pinned mean field spin glasses~\cite{cammarota2012ideal,cammarota2013random} and finite-dimensional glasses~\cite{kob2013probing,ozawa2015equilibrium,ozawa2018ideal}. In this case, the system remains stuck forever around the initial equilibrium configuration ${\bf x}$, and generating independent additional configurations is impossible using standard simulation approaches such as simple Monte Carlo simulations or molecular dynamics.
\end{itemize}

It is crucial to stress that the above construction, which we will call {\it quiet freezing} of variables, supposes that the division of the degrees of freedom into ${\bf x}$ and ${\bf y}$ is done {\it before} the thermalized configuration is constructed. In other words, the proof described above does not hold if we first thermalize a system and then choose which degrees of freedom to freeze by using properties of the thermalized configuration, because this would introduce a bias that is not described by Eq.~\eqref{eq:def_dist_frozen}.
Keeping this in mind, we now discuss how this construction can be applied to randomly pinned particle systems, and how it can be generalized when pinning is not random.

\subsection{Random pinning and wall pinning}
\label{subsec:Random pinning and wall pinning}

The construction of Sec.~\ref{sec:frozen} can be applied in a straightforward manner to the {\it random pinning} procedure~\cite{kim2003effects,scheidler2004relaxation,krakoviack2010statistical}. 
We will consider a system of $N$ point particles in $d$ dimensions, such as the Kob-Andersen model~\cite{kob1995testing}, and we will make use of the particle permutation symmetry as a crucial ingredient.
Note that in most cases the systems of interest are polydisperse, including binary or ternary mixtures, and the particle permutation symmetry does not exist because particles have distinct sizes. Yet, one can always recover it if one considers the permutation of the particle species as additional degrees of freedom and sum up all the possible permutations in the partition function~\cite{hiroike1960new,morita1961new,ozawa2018configurational,brito2018theory}. This treatment does not change the thermodynamic properties that we discuss in this paper. A formally equivalent way of reintroducing the particle permutation symmetry is to assign to each particle an additional degree of freedom describing its size.
Thus, for simplicity, we focus only on positions (and momenta) as relevant degrees of freedom.

\subsubsection{Random pinning: Choosing at random the particles that will be pinned}

The Hamiltonian $H$ and the partition function $Z$ are given by
\begin{eqnarray}
H({\bf r}^N, {\bf p}^N) = \sum_{i=1}^N \frac{{\bf p}_i^2}{2m_i} + U({\bf r}^N) \ , \qquad
Z = \int(\prod_{i=1}^N \mathrm{d}{\bf r}_i \mathrm{d}{\bf p}_i) \exp[-\beta H({\bf r}^N, {\bf p}^N)] \ ,
\label{eq:original_H}
\end{eqnarray}
where ${\bf r}_i$, ${\bf p}_i$, and $m_i$ are the position, momentum, and mass of the $i$-th particle, respectively. $U$ is the potential energy, and $\beta=1/T$ is the inverse temperature.
Particles are supposed to be confined in a volume $V$ with some boundary conditions that do not need to be specified at this stage.
In this paper, we use a shorthand notation for a vector of $N$ variables, e.g., ${\bf r}^N=( {\bf r}_1, {\bf r}_2, ..., {\bf r}_N)$.
Note that we omit the combinatorial factors such as $N!$ and the Planck's constant $h$ in the definition of the partition function in Eq.~(\ref{eq:original_H}), because we are not concerned with the absolute value of the free energy or entropy.

One can then split the $N$ particles into a set of $N_f$ pinned particles, with coordinates ${\bf y} = \{ {\bf r}^{N_f}, {\bf p}^{N_f} \}$, and a set of $N-N_f$ unpinned (or mobile) particles with the remaining coordinates, ${\bf x}$. Because the labeling of particles is arbitrary, we can consider that the first $N_f$ particles are the pinned ones. Note that the splitting is here performed before any thermalization, as we emphasized in Sec.~\ref{sec:frozen}. One can then thermalize the whole system of $N$ particles, hence the joint set ${\{{\bf x}, {\bf y}\}}$, in the liquid phase. Because in such phase particles can freely diffuse, we expect the pinned particles to be uniformly distributed in the volume $V$. At this point, the positions of the $N_f$ particles, ${\bf r}^{N_f}$, are frozen and their velocities are set to zero. Because the distribution of momenta is a product of independent distributions for each particles, setting the momenta of pinned particles to zero does not affect the distribution of ${\bf x}$ (in other words, only the configurational part of the Hamiltonian matters). Hence, the particles ${\bf x}$ can be considered to be in equilibrium with the pinned particles ${\bf y}$, and the procedure allows one to generate an equilibrium configuration of the pinned system even in the case in which the density of pinned particles is so high that the pinned system is glassy. Note that in usual implementations of the random pinning procedure, one first generates a configuration of the full system of $N$ particles and then chooses at random the $N_f$ particles to be pinned. But, since from the point of view of the full system this is just a labeling that does not depend on the equilibrated configuration, the two operations (randomly choosing the particles that are going to be pinned and equilibrating the full system) can be safely exchanged. We thus conclude that random pinning belongs to the class of quiet freezing procedures, as it is well known and numerically validated.

\subsubsection{Wall pinning: Choosing the particles to be pinned according to a geometrical criterion}
\label{sec:wall}

\begin{figure}[t]
\includegraphics[width=0.3\columnwidth]{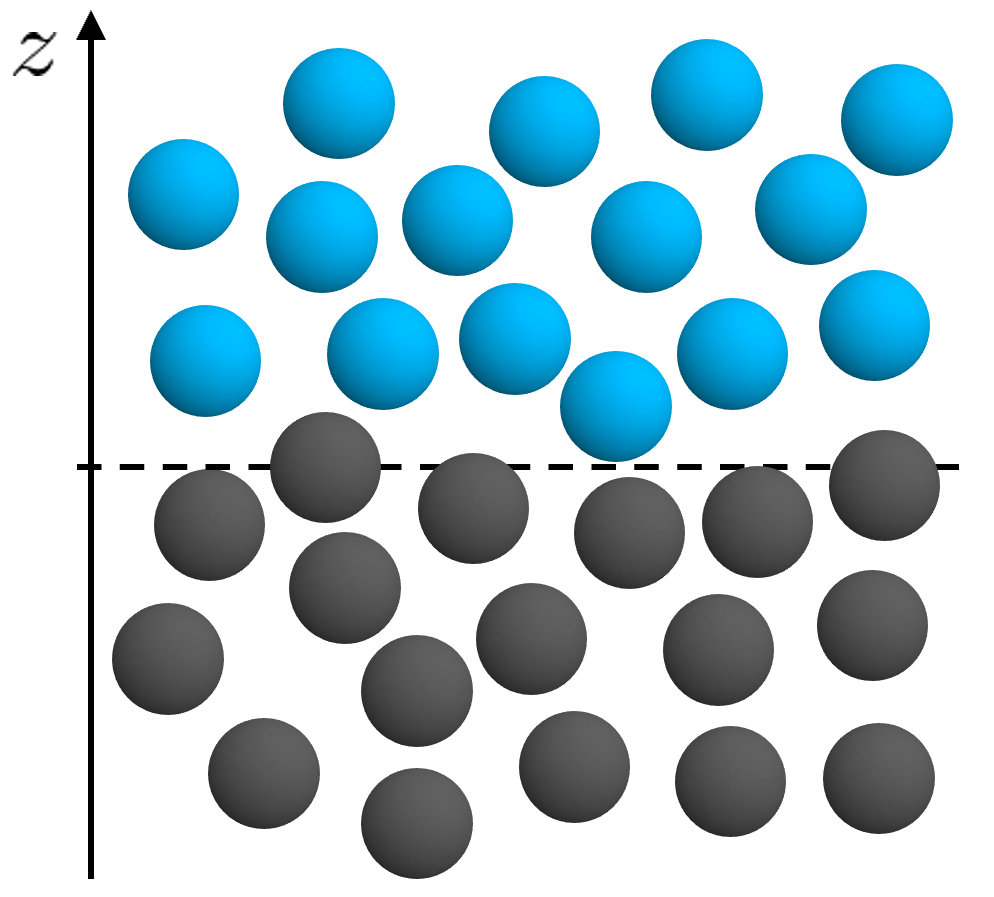}
\caption{
{\bf Illustration of the wall pinning procedure}, of which we give in the main text two different proofs. The first, as in Ref.~\cite{scheidler2004relaxation}, is based on the idea of separating the volume in the wall (bottom part) and fluid (top part) regions,  and freezing particles in the former region; the corresponding averages are given in Eq.~\eqref{eq:chaingen}.
The second corresponds to ordering the particles according to their $z$ component, and freezing the $k$ particles with the smallest $z$, as in Eq.~\eqref{eq:chaingen2}. The number $k$ has to be chosen such that the wall region has the desired size. 
}
\label{fig:wall}
\end{figure}

In Ref.~\cite{scheidler2004relaxation}, the random pinning construction has been extended to introduce the possibility of choosing which particles have to be frozen {\it after} the equilibrated configuration of the unpinned system is prepared. This ``wall pinning'' construction has been widely applied in the context of glass physics in order to probe the equilibrium properties of liquids~\cite{biroli2008thermodynamic,hocky2012growing,hocky2014crossovers,yaida2016point}. We briefly describe the proof for completeness.
Suppose that the total volume $V$ is split into a ``wall'' region $W$, in which particles will eventually be frozen, and a ``fluid'' region $F$, such that $V=W+F$, as schematically shown in Fig.~\ref{fig:wall}. The shape of these regions is completely arbitrary. 
We can then generate an equilibrium configuration of the full $N$-particle system, and freeze the particles that fall into the $W$ region. Because the choice of degrees of freedom that are going to be frozen depends on the equilibrated configuration, the proof of Sec.~\ref{sec:frozen} does not apply directly and must be generalized. We note that the momenta are always irrelevant (their distribution is a product over particles), and thus can be ignored. 
Given the total configurational partition function,
\begin{equation}
Z_c = \int(\prod_{i=1}^N \mathrm{d}{\bf r}_i) e^{-\beta U({\bf r}^N)} \ ,
\end{equation}
the average of an arbitrary observable
$O({\bf r}^N)$ that is invariant under permutations of the particle labels can be written as
\begin{equation}\begin{split}
\langle O \rangle &=
\frac1{Z_c} \int_{W+F} \mathrm{d}{\bf r}_1 \cdots \int_{W+F} \mathrm{d}{\bf r}_N \, O({\bf r}^N)
e^{-\beta U({\bf r}^N)} \\
&=
\frac1{Z_c} \left[ \int_{W} \mathrm{d}{\bf r}_1+\int_F \mathrm{d}{\bf r}_1\right] \cdots \left[\int_{W}\mathrm{d}{\bf r}_N+\int_F \mathrm{d}{\bf r}_N\right]  \, O({\bf r}^N)
e^{-\beta U({\bf r}^N)} \\
&=
\frac1{Z_c} \sum_{k=0}^N \binom{N}{k} \int_{W} \mathrm{d}{\bf r}_1 \cdots \int_{W} \mathrm{d}{\bf r}_k 
\int_F \mathrm{d}{\bf r}_{k+1}
\cdots \int_F \mathrm{d}{\bf r}_N
\, O({\bf r}^N)
e^{-\beta U({\bf r}^N)} \ ,
\end{split}\end{equation}
where $\binom{N}{k}$ is the binomial coefficient. In the last step we defined $k$ as the number of particles in the $W$ region, and we used the permutation symmetry of both $O({\bf r}^N)$ and $U({\bf r}^N)$, hence of the whole integrand, to relabel the first $k$ particles as being those in the $W$ region and the remaining $N-k$ as being those in the $F$ region.

Using the shorthand notation
${\bf r}^W = ({\bf r}_{1},\cdots,{\bf r}_k ) \in W^k = \Omega_W$
and
${\bf r}^F = ({\bf r}_{k+1},\cdots,{\bf r}_N ) \in F^{N-k} = \Omega_F$,
we can write
\begin{equation}\label{eq:chaingen}
\begin{split}
\langle O \rangle 
&=
\frac1{Z_c} \sum_{k=0}^N  \binom{N}{k}
\int_{\Omega_W} \mathrm{d}{\bf r}^W
\int_{\Omega_F} \mathrm{d}{\bf r}^F \,
O(\br^W,{\bf r}^F)
e^{-\beta U({\bf r}^W,{\bf r}^F)} \\
&=
 \sum_{k=0}^N \binom{N}{k}
\int_{\Omega_W} \mathrm{d}{\bf r}^W
\frac{Z_W({\bf r}^W)}{Z_c} \,
\int_{\Omega_F} \mathrm{d}{\bf r}^F O(\br^W,{\bf r}^F) \frac{
e^{-\beta U({\bf r}^W,{\bf r}^F)} 
}
{
Z_W({\bf r}^W) 
}
\qquad \text{with}
\qquad
Z_W({\bf r}^W) =
\int_{\Omega_F} \mathrm{d}{\bf r}^F \,
e^{-\beta U({\bf r}^W,{\bf r}^F)} 
\ .
\end{split}
\end{equation}

We conclude, as in Sec.~\ref{sec:frozen}, that we
can define an ``annealed'' average $\langle \cdots \rangle$, a ``thermal'' average $\langle \cdots \rangle_W$ conditioned to the wall,
and a ``disorder'' average $\overline{ \cdots }$ over the realizations of the wall, from
\begin{equation}\label{eq:ave2}
\langle \cdots \rangle = \int \mathrm{d}{\bf r}^N \
\frac{e^{-\beta U({\bf r}^N)}}{Z_c} (\cdots)  \ ,
\qquad
\langle \cdots \rangle_W = 
\int_{\Omega_F} \mathrm{d}{\bf r}^F \ \frac{ 
e^{-\beta U({\bf r}^W,{\bf r}^F)} 
}
{
Z_W({\bf r}^W) 
} (\cdots)
 \ ,
 \qquad
 \overline{\cdots} = 
  \sum_{k=0}^N \binom{N}{k}
\int_{\Omega_W} \mathrm{d}{\bf r}^W
\frac{Z_W({\bf r}^W)}{Z_c}
 (\cdots) \ ,
\end{equation}
and because of the chain rule in Eq.~\eqref{eq:chaingen}, the annealed and quenched averages coincide, i.e.,
$\langle \cdots \rangle =
    \overline{\langle \cdots \rangle_W}$.
Clearly, this proof can be generalized to whatever situation in which (i) there is permutation symmetry over $N$ degrees of freedom (recall that this also holds for polydisperse systems if one considers the particle species as additional degrees of freedom) and (ii) the integration space of each individual degree of freedom can be split {\it a priori} (i.e., independently of the configuration of the system) into two distinct regions $W$ and $F$.
One can then generate a full equilibrium configuration of the $N$ degrees of freedom, and freeze those falling into region $W$, which produces an equilibrium configuration of the remaining degrees of freedom conditioned to the frozen ones.
We note that the unfrozen (``fluid'') degrees of freedom should be constrained inside the $\Omega_F$ region during the thermal average. This constraint is nearly satisfied in dense particle systems where an excluded volume effect prevents the fluid particles from entering the $\Omega_W$ region. 
Yet this is not the case for dilute systems, and hence one has to impose an additional constraint on the dynamics of fluid particles such as a hard wall condition.

Note that the only requirement on the observable $O(\br^N)$ is that it is invariant under permutations of particles.
We also mention that one can construct an observable $O({\bf r}^N)$ that only depends on the particles that are in the $F$ region,
while keeping the global permutation symmetry of the observable.
This can be done, for example, by writing
\begin{equation}\label{eq:Owall}
O({\bf r}^N) = \sum_{i=1}^N o_1({\bf r}_i) \,
\mathbb{I}[{\bf r}_i\in F]
+
\sum_{i=1}^N\sum_{j=1}^N o_2({\bf r}_i,{\bf r}_j) \,
\mathbb{I}[{\bf r}_i\in F]
\mathbb{I}[{\bf r}_j\in F] + \cdots
\ ,
\end{equation}
where $\mathbb{I}[{\cal E}]$ is the indicator function of event ${\cal E}$, which is one if ${\cal E}$ is realized and zero otherwise.
Then, any observable of the form in Eq.~\eqref{eq:Owall} can be used to characterize the fluid system only, while keeping the quenched and annealed averages coincident.

\subsubsection{Sorted pinning: Choosing the particles to be pinned according to an ordering}    
\label{sec:sortedpinning}

We now consider an alternative proof for quiet freezing in a similar spirit by introducing ``sorted pinning'' described as follows. Suppose that an ordering relation between the $N$ particle positions can be defined. One example is to use the binary ordering relation ${\bf r}_i \prec {\bf r}_j$ if 
${\bf r}_i \cdot {\bf e} < {\bf r}_j\cdot {\bf e}$, where ${\bf e}$ is an arbitrary unit vector. Similarly, ${\bf r}_i \preceq {\bf r}_j$ if 
${\bf r}_i \cdot {\bf e} \leq {\bf r}_j\cdot {\bf e}$.
We will see that the proof given below can be generalized to any ordering operation that is able to uniquely sort a set of positions, $\{ {\bf r}_1, \cdots, {\bf r}_N \}$.
We now define the vector for the $k$ first ordered particles and the vector for the rest of the ordered particles as $\brpk=({\bf r}_1,\cdots,{\bf r}_k)$ and $\brsk=({\bf r}_{k+1},\cdots,{\bf r}_N)$, respectively.
Considering again only permutation-symmetric observables
$O({\bf r}^N)$,
and using the permutation symmetry, we can order the $N$ particles and write
\begin{equation}
\label{eq:chaingen2}
\begin{split}
\langle O \rangle &=
\frac{N!}{Z_c} \int_V \mathrm{d}{\bf r}_1
\int_{{\bf r}_2 \succeq {\bf r}_1} \mathrm{d}{\bf r}_2
\cdots 
\int_{{\bf r}_N \succeq {\bf r}_{N-1}} \mathrm{d}{\bf r}_N
 \, O({\bf r}^N)
e^{-\beta U({\bf r}^N)} \\
&=
\int_{\Omega_{\preceq k}} \mathrm{d}\brpk
\frac{N! \, Z_k(\brpk)}{Z_c} \,
\int_{\Omega_{\succ k}} \mathrm{d}\brsk \, \frac{
e^{-\beta U(\brpk,\brsk)} 
}
{
Z_k(\brpk) 
} O(\brpk,\brsk)
\quad \text{with}
\quad
Z_k(\brpk) =
\int_{\Omega_{\succ k}} \mathrm{d}\brsk \,
e^{-\beta U(\brpk,\brsk)} 
\ ,
\end{split}
\end{equation}
where 
\begin{eqnarray}
    \int_{\Omega_{\preceq k}} \mathrm{d}\brpk &=& \int_V \mathrm{d}{\bf r}_1
\int_{{\bf r}_2 \succeq {\bf r}_1} \mathrm{d}{\bf r}_2
\cdots 
\int_{{\bf r}_k \succeq {\bf r}_{k-1}} \mathrm{d}{\bf r}_k, \\
    \int_{\Omega_{\succ k}} \mathrm{d}\brsk &=& \int_{{\bf r}_{k+1} \succeq {\bf r}_{k}} \mathrm{d}{\bf r}_{k+1} \int_{{\bf r}_{k+2} \succeq {\bf r}_{k+1}} \mathrm{d}{\bf r}_{k+2} \cdots \int_{{\bf r}_{N} \succeq {\bf r}_{N-1}} \mathrm{d}{\bf r}_{N}.
\end{eqnarray}
We thus obtain a relation similar to Eq.~(\ref{eq:chaingen}). 

The core idea both in the ``wall pinning'' proof (Sec.~\ref{sec:wall}) and in the ``sorted pinning'' proof (this section) is that the integral regions for the pinned and unpinned (fluid) particles have a separation, either by a wall that we specify or by ranking via some ordering operation. Thus, as long as a clear-cut separation is enforced, the internal ordering constraint on the two vectors $\brpk$ and $\brsk$ in Eq.~(\ref{eq:chaingen2}) can be released.
This results in a similar expression to the wall pinning case, but using a separation between the $k$-th and $(k+1)$-th ordered particles:
\begin{equation}
\begin{split}
\langle O \rangle &=\binom{N}{k}
\int_{V^k} \mathrm{d}\brpk
\frac{ Z_k(\brpk)}{ Z_c} \,
\int_{ \substack{{\bf r}_i \succeq {\bf r}_{\rm max} \\ (\forall i>k)}  } \mathrm{d}\brsk \frac{
e^{-\beta U(\brpk,\brsk)} 
}
{
Z_k(\brpk) 
} O(\brpk,\brsk) \\
& \qquad \text{with} \qquad {\bf r}_{\rm max} = \max\{{\bf r}_1,\cdots,{\bf r}_k\} \qquad \text{and} 
\qquad
Z_k(\brpk) = 
\int_{ \substack{{\bf r}_i \succeq {\bf r}_{\rm max} \\ (\forall i>k)}  } \mathrm{d}\brsk \,
e^{-\beta U(\brpk,\brsk)} 
\ .
\end{split}\end{equation}
This relation is identical to Eq.~(\ref{eq:chaingen}) with the only difference that $k$ is now fixed and the boundary between the wall and fluid regions is fluctuating and determined
by the largest of the first $k$ vectors. Hence, this second proof corresponds to a ``fixed-$k$'' ensemble while the first proof corresponds to a
``fixed-boundary'' ensemble. As usual, the two ensembles become equivalent in the thermodynamic limit if the wall and fluid regions are both macroscopic, i.e., if $k\sim N$.
An illustration is given in Fig.~\ref{fig:wall}.

While the derivations presented here and in Sec.~\ref{sec:wall} concern the thermodynamics of the system, it is useful to also discuss their dynamical meaning.
Suppose that we are performing some dynamics of the $N$-particle system that results, at a given time $t=0$, in an equilibrium configuration $\br^N$. At that instant, we can perform the sorting
of the particles (e.g., according to their $z$ component, setting ${\bf e}={\bf e}_z$) and identify the first $k$ ``wall'' particles and the last $N-k$ ``fluid'' ones. Now, if we let the system evolve in an unconstrained
 way, the $z$ coordinate of the $k$-th particle might cross the $(k+1)$-th one. But if this happens, we would just relabel the particles by exchanging $k \leftrightarrow k+1$,
such that at any time, the first $k$ particles would have the smallest $z$ values. This dynamics would result in the ``annealed'' thermodynamic average.
Alternatively, at time $t=0$ we can freeze the positions of the first $k$ particles, and only let the remaining $N-k$ evolve. We know that these $N-k$ particles start in equilibrium with the
wall, but then, we have the additional
hard constraint that the $z$ coordinates of the evolving particles must be at any time $t>0$ larger than $z_{\rm max} = \max\{z_1,\cdots, z_k\}$. This hard constraint can be implemented in
a Monte Carlo simulation by rejecting moves that would bring a particle at $z<z_{\rm max}$, or in Molecular Dynamics by adding a reflecting wall at $z=z_{\rm max}$. In both cases,
the resulting dynamics will lead to the ``thermal'' average $\langle \cdots \rangle_W$ over the fluid particles. One should then either perform the disorder average over the wall particles, 
i.e., $\overline{\cdots}$, by repeating the freezing procedure many times, or invoke the self-averaging properties for large $N$ (assuming $k$ to be of order $N$) to claim that one single
run is representative of the average.

Note that, as in the wall pinning case, we can construct an observable
$O(\br^N) = O(\brsk)$ that depends only on the last $N-k$ ordered vector $\brsk = ({\bf r}_{k+1},\cdots, {\bf r}_N)$. Such an observable is still invariant under permutations of the $N$ particles, and it can describe the fluid region without an explicit dependence on the frozen particles.

\subsection{Random bonding procedures}

\subsubsection{Variable transformation to create virtual dimers}
\label{sec:vartrans}

We now show that the ideas presented in Sec.~\ref{subsec:Random pinning and wall pinning} can be implemented within the random bonding approach.
Suppose that we want to create $N_{\rm d}$ dimers and $N_{\rm m}$ monomers from our system with $N=N_{\rm m}+2N_{\rm d}$ particles.
To this aim, given a configuration ${\bf r}^N$ of
the $N$ monomers,
what we need is a procedure that creates $\Nd$ dimers in a permutation-invariant way. 
More precisely, the criteria that are used to decide which particles are going to be in the dimers should not depend on the particle labeling itself.
 The remaining $N_{\rm m}$ particles are left unbonded.
We can then exploit the permutation symmetry to indicate the indices of dimers and monomers as belonging to sets $i \in \mathcal{D}=\{1,3,5,\cdots,2\Nd -1 \}$ and $i \in \mathcal{M} = \{2\Nd+1,\cdots, N\}$, respectively, with dimer $i$ being composed by particles $i$ and $i+1$. 
We then denote $\br^{2\Nd} = (\br_1,\br_2,\cdots,\br_{2\Nd})$ and 
${\bf r}^{N_{\rm m}}=(\br_{2\Nd+1},\cdots,\br_N)$ the dimer and monomer coordinates, respectively. 
We stress that this is just a sorting operation of the particle labels, and no physical constraint is imposed on the system at this stage.

The splitting of all particles into dimers and monomers defines an integration space for both sets of variables, which we denote as $\Od$ for the dimers
and $\Om$ for the monomers.
Because of permutation invariance, there are 
\begin{equation}
    \Pi_{\rm m,d} = \frac{N!}{2^{N_{\rm d}} N_{\rm d}! N_{\rm m}!}
    \label{eq:combination_dimers}
\end{equation}
distinct equivalent ways of constructing the dimers\footnote{There are $\binom{N}{2N_{\rm d}}=\frac{N!}{(2N_{\rm d})!N_{\rm m}!}$ ways to choose particles associated with dimers. Then there are $(2N_{\rm d}-1)!!=\frac{(2N_{\rm d})!}{2^{N_{\rm d}}N_{\rm d}!}$ ways to construct pairs among $2N_{\rm d}$ particles.
Thus $\Pi_{\rm m, d}=\binom{N}{2N_{\rm d}} (2N_{\rm d}-1)!!$, giving Eq.~(\ref{eq:combination_dimers}).}, and the sum over all these equivalent possibilities reconstructs the whole integration volume of the original monomer system\footnote{
To fix ideas by an example, consider the `wall pinning' setting of Sec.~\ref{sec:wall}. The criterion to decide whether a particle belongs to the wall or to the fluid regions is $\br_i \in W$ and $\br_i\in F$, respectively. So the integration space for wall particles is $\br^W \in \Omega_W = W^k$ 
and that for fluid particles is $\br^F \in \Omega_F = F^{N-k}$.
Of course, the union of all the spaces obtained by permuting the particle identities reconstructs the original space $\br^N \in V^N$, because of the trivial identity
\begin{equation}
    V^N = (W+F)^N = \sum_{k=0}^N \binom{N}{k} W^k \times F^{N-k} \ .
\end{equation}
}.
Taking advantage of the permutation symmetry of the problem, we can then write
\begin{equation}
\langle O \rangle =
\frac{\Pi_{\rm m,d}}{Z} \int (\prod_{i=1}^N \mathrm{d}{\bf p}_i) \int_{\Od} \mathrm{d}{\bf r}^{2\Nd}
\int_{\Omega_{\rm m}} \mathrm{d}{\bf r}^{N_{\rm m}}
 \, O({\bf r}^{N})
e^{-\beta H({\bf r}^N, {\bf p}^N)}  \ .
\label{eq:temp}
\end{equation}

\begin{figure}[t]
\includegraphics[width=0.5\columnwidth]{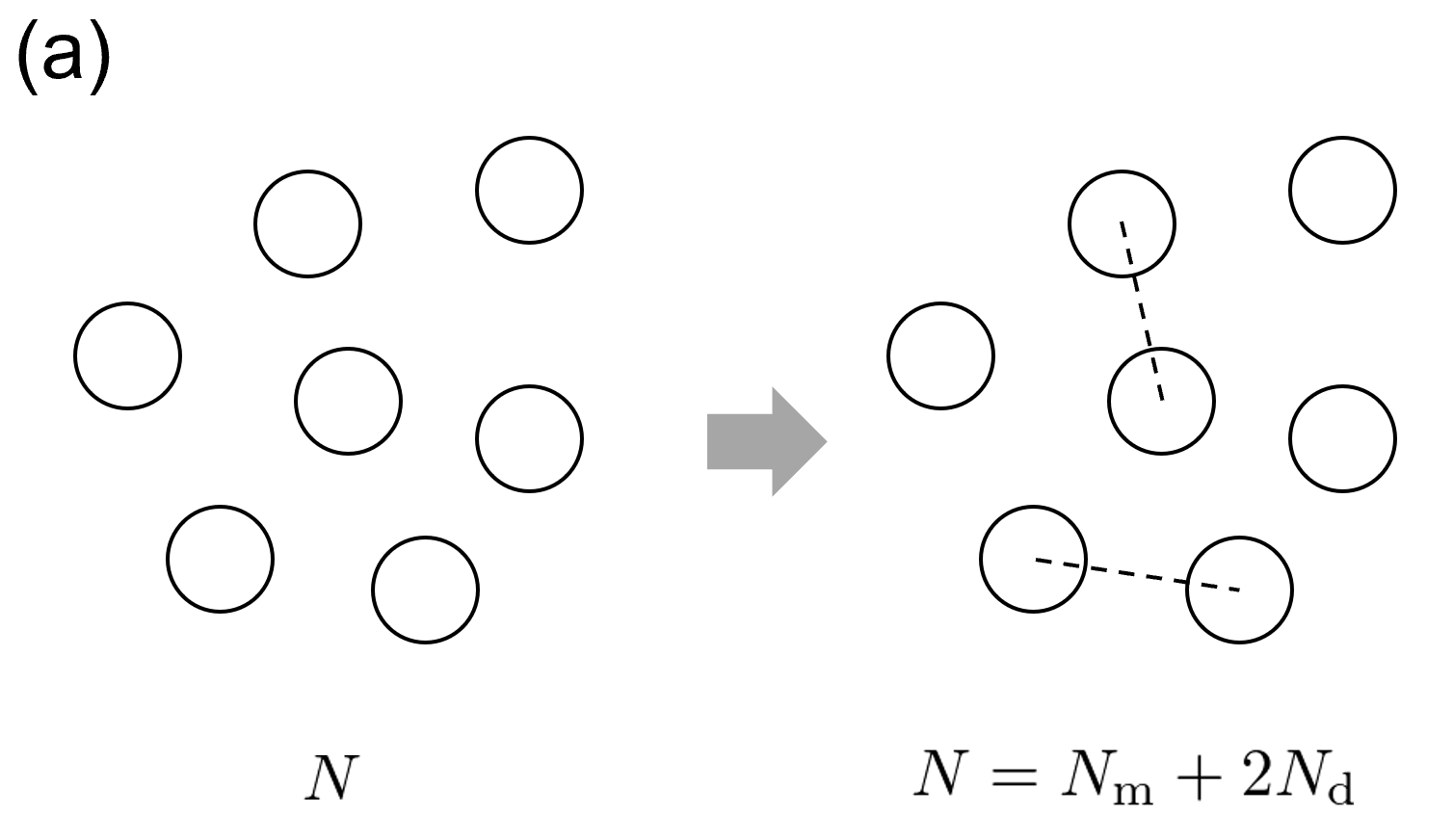}
\qquad \qquad
\includegraphics[width=0.3\columnwidth]{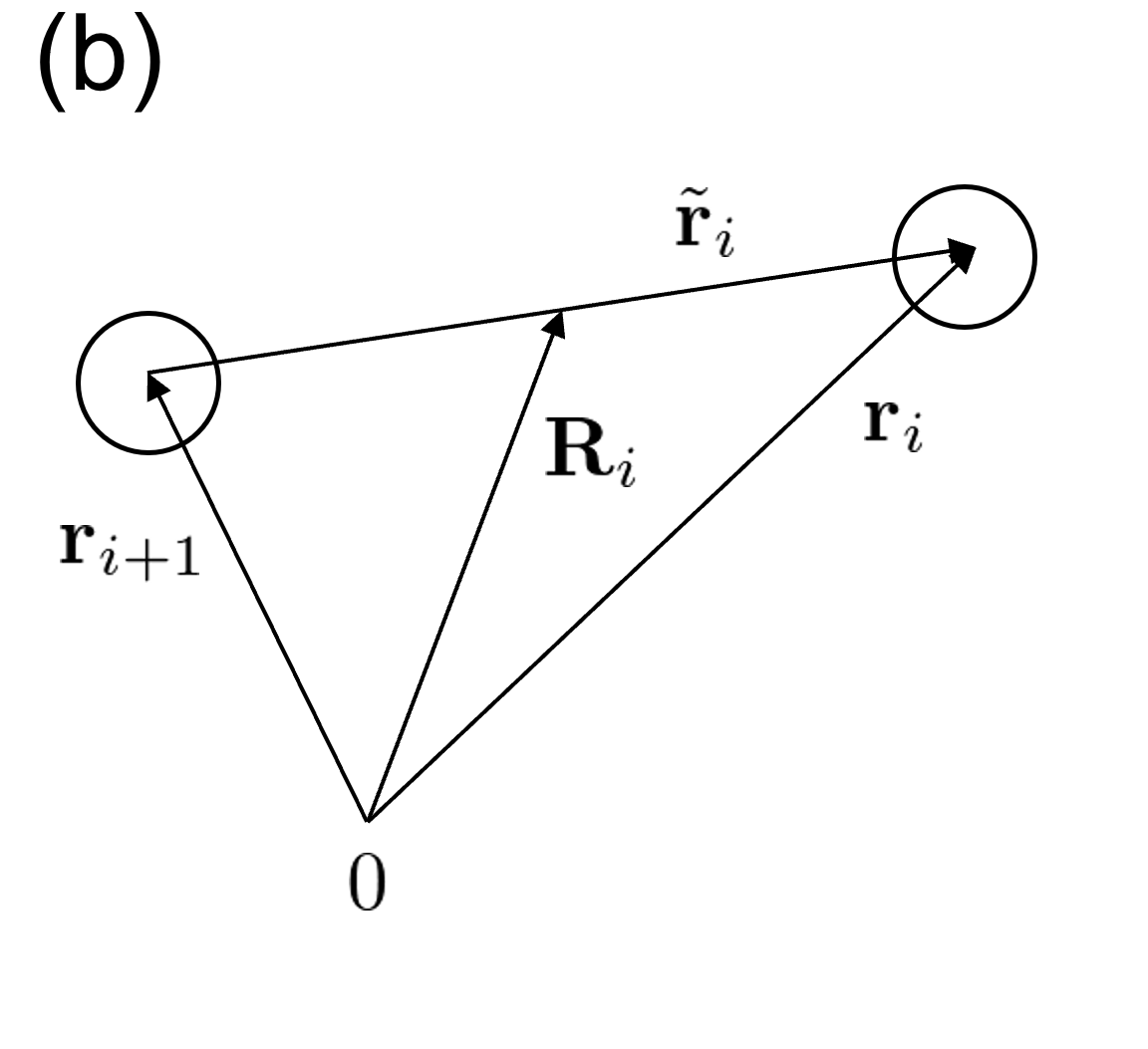}
\caption{{\bf Illustration of the creation of virtual bonds and the associated change of coordinates.}
(a) Schematic plot for making virtual bonds. (b) The center of mass and relative position describe a dimer connecting particles $i$ and $i+1$.}
\label{fig:virtual}
\end{figure}

We next consider a variable transformation from Cartesian coordinates to Jacobi coordinates for the dimers. 
We assign virtual bonds between particles in each dimer,
 as schematically shown in Fig.~\ref{fig:virtual}(a).
Once again, we emphasize that this is a virtual operation, and the actual system is not altered at all, i.e., this is merely a variable transformation.
For the monomers, we continue to use the Cartesian coordinates.
For the dimers, instead, we use the Jacobi coordinates for the two-body problem (Fig.~\ref{fig:virtual}(b)), using the center of mass and relative position 
for the dimer $i \in \mathcal{D}=\{1,3,5,\cdots,2\Nd -1 \}$ composed of the $i$-th and $(i+1)$-th particles. These are given by ${\bf R}_{i}=\frac{m_i {\bf r}_i + m_{i+1} {\bf r}_{i+1}}{m_i+m_{i+1}}$ and $\tilde{{\bf r}}_{i}={\bf r}_i-{\bf r}_{i+1}$, respectively. 
The corresponding momenta are denoted by ${\bf P}_{i}$ and $\tilde{{\bf p}}_{i}$, respectively.
Also, the dimer total mass and the reduced mass are given by $M_{i}=m_i+m_{i+1}$ and $\mu_{i}=\frac{m_i m_{i+1}}{m_i+m_{i+1}}$, respectively.
This variable transformation is formally written as $({\bf r}^N, {\bf p}^N) \to ( {\bf R}^{N_{\rm d}}, {\bf P}^{N_{\rm d}}, \tilde{{\bf r}}^{N_{\rm d}}, \tilde{{\bf p}}^{N_{\rm d}}, {\bf r}^{N_{\rm m}}, {\bf p}^{N_{\rm m}} )$.

Making this variable transformation, we can rewrite the Hamiltonian $H$ and the partition function $Z$ as
\begin{equation}
H( {\bf R}^{N_{\rm d}}, {\bf P}^{N_{\rm d}}, \tilde{{\bf r}}^{N_{\rm d}}, \tilde{{\bf p}}^{N_{\rm d}}, {\bf r}^{N_{\rm m}}, {\bf p}^{N_{\rm m}} ) = \sum_{i \in \mathcal{D}} \left( \frac{{\bf P}_{i}^2}{2 M_{i}} + \frac{\tilde{{\bf p}}_{i}^2}{2 \mu_{i}} \right) + \sum_{i \in \mathcal{M}} \frac{{\bf p}_i^2}{2 m_i} + U( {\bf R}^{N_{\rm d}}, \tilde{{\bf r}}^{N_{\rm d}}, {\bf r}^{N_{\rm m}} ),
\label{eq:H_Jacobi}
\end{equation}
\begin{equation}
Z =\Pi_{\rm m,d}
\int (\prod_{i \in \mathcal{D} } \mathrm{d} {\bf R}_{i} \mathrm{d} {\bf P}_{i}) 
\int_{\Od} (\prod_{i \in \mathcal{D} } \mathrm{d} \tilde{{\bf r}}_{i}\mathrm{d} \tilde{{\bf p}}_{i}) 
\int_{\Om}(\prod_{i \in \mathcal{M}} \mathrm{d}{\bf r}_i \mathrm{d}{\bf p}_i)
\exp[-\beta H( {\bf R}^{N_{\rm d}}, {\bf P}^{N_{\rm d}}, \tilde{{\bf r}}^{N_{\rm d}}, \tilde{{\bf p}}^{N_{\rm d}}, {\bf r}^{N_{\rm m}}, {\bf p}^{N_{\rm m}} )].
\label{eq:temp2}
\end{equation}
Note that the integration spaces $\Omega_{\rm d}$ and $\Omega_{\rm m}$ in Eq.~\eqref{eq:temp2} are not exactly the same as the ones in Eq.~(\ref{eq:temp}), they are the images of those spaces under the change of variable. Yet, with a little abuse of notation, we keep the same notation for both. 

We further proceed with the variable transformation for the relative movements of dimers by using spherical coordinates in $d=3$, which is formally written as
$(\tilde{\bf r}^{N_{\rm d}}, \tilde{\bf p}^{N_{\rm d}}) \to (\tilde{r}^{N_{\rm d}}, \theta^{N_{\rm d}}, \varphi^{N_{\rm d}}, p_{\tilde{r}}^{N_{\rm d}}, p_{\theta}^{N_{\rm d}}, p_{\varphi}^{N_{\rm d}})$.
We note that because of Liouville's theorem the Jacobian is unity for this transformation, namely, $\mathrm{d} \tilde{{\bf r}}_{i}\mathrm{d} \tilde{{\bf p}}_{i}=\mathrm{d} \tilde{r}_{i} \mathrm{d}\theta_{i}\mathrm{d}\varphi_{i} \mathrm{d}p_{\tilde{r}_{i}} \mathrm{d}p_{\theta_{i}} \mathrm{d}p_{\varphi_{i}}$. 
Also, the kinetic part of the Hamiltonian can be written as
\begin{equation}
\frac{\tilde{\bf p}_{i}^2}{2 \mu_{i}} = \frac{p_{\tilde{r}_{i}}^2}{2 \mu_{i}} + \frac{p_{\theta_{i}}^2}{2I_{i}} + \frac{p_{\varphi_{i}}^2}{2 I_{i} \sin^2\theta_{i}},
\label{eq:kinetic_spherical}
\end{equation}
where $I_{i}=\mu_{i}\tilde{r}_{i}^2$ is the moment of inertia, $p_{\tilde{r}_{i}} = \mu_{i} \dot{ \tilde{r}}_{i}$, $p_{\theta_{i}} = I_{i} \dot{\theta}_{i}$, and $p_{\varphi_{i}} = I_{i} (\sin^2\theta_{i}) \dot{\varphi}_{i}
$ are the new momenta and 
the dot denotes a time derivative. 
We note that as one can see in Eq.~(\ref{eq:kinetic_spherical}), the kinetic term also depends on the coordinates.
Thus we cannot treat the momenta and positions separately, unlike in the random pinning setting.

We finally arrive at the expressions for the Hamiltonian $H$ and the partition function $Z$ that are suitable for our study:
\begin{eqnarray}
H( {\bf R}^{N_{\rm d}}, {\bf P}^{N_{\rm d}}, \tilde{r}^{N_{\rm d}}, \theta^{N_{\rm d}}, \varphi^{N_{\rm d}}, p_{\tilde{r}}^{N_{\rm d}}, p_{\theta}^{N_{\rm d}}, p_{\varphi}^{N_{\rm d}}, {\bf r}^{N_{\rm m}}, {\bf p}^{N_{\rm m}} ) = \qquad \qquad \qquad \qquad \qquad \qquad \qquad \nonumber \\  \sum_{i \in \mathcal{D}} \left( \frac{{\bf P}_{i}^2}{2 M_{i}} + \frac{p_{\tilde{r}_{i}}^2}{2 \mu_{i}} + \frac{p_{\theta_{i}}^2}{2I_{i}} + \frac{p_{\varphi_{i}}^2}{2 I_{i}\sin^2\theta_{i}} \right) + \sum_{i \in \mathcal{M}} \frac{{\bf p}_i^2}{2 m_i} + U( {\bf R}^{N_{\rm d}}, \tilde{r}^{N_{\rm d}}, \theta^{N_{\rm d}}, \varphi^{N_{\rm d}}, {\bf r}^{N_{\rm m}} ),
\end{eqnarray}
\begin{eqnarray}
Z =\Pi_{\rm m,d}  
 \int (\prod_{i \in \mathcal{D}} \mathrm{d} {\bf R}_{i}\mathrm{d} {\bf P}_{i}) 
 \int_{\Od} (\prod_{i \in \mathcal{D}} \mathrm{d} \tilde{r}_{i} \mathrm{d}\theta_{i}\mathrm{d}\varphi_{i} \mathrm{d}p_{\tilde{r}_{i}} \mathrm{d}p_{\theta_{i}} \mathrm{d}p_{\varphi_{i}}) 
 \int_{\Om}(\prod_{i \in \mathcal{M}} \mathrm{d}{\bf r}_i \mathrm{d}{\bf p}_i) 
  \nonumber \\ \times \exp[-\beta H( {\bf R}^{N_{\rm d}}, {\bf P}^{N_{\rm d}}, \tilde{r}^{N_{\rm d}}, \theta^{N_{\rm d}}, \varphi^{N_{\rm d}}, p_{\tilde{r}}^{N_{\rm d}}, p_{\theta}^{N_{\rm d}}, p_{\varphi}^{N_{\rm d}}, {\bf r}^{N_{\rm m}}, {\bf p}^{N_{\rm m}} )].
\end{eqnarray}
We can then define the total (or annealed) average as
\begin{eqnarray}
\langle O \rangle =
\frac{\Pi_{\rm m,d}}{Z} 
 \int (\prod_{i \in \mathcal{D}} \mathrm{d} {\bf R}_{i}\mathrm{d} {\bf P}_{i}) 
 \int_{\Od} (\prod_{i \in \mathcal{D}} \mathrm{d} \tilde{r}_{i} \mathrm{d}\theta_{i}\mathrm{d}\varphi_{i} \mathrm{d}p_{\tilde{r}_{i}} \mathrm{d}p_{\theta_{i}} \mathrm{d}p_{\varphi_{i}}) 
 \int_{\Om}(\prod_{i \in \mathcal{M}} \mathrm{d}{\bf r}_i \mathrm{d}{\bf p}_i) 
  \nonumber \\ \times  O({\bf r}^{N})\, \exp[-\beta H( {\bf R}^{N_{\rm d}}, {\bf P}^{N_{\rm d}}, \tilde{r}^{N_{\rm d}}, \theta^{N_{\rm d}}, \varphi^{N_{\rm d}}, p_{\tilde{r}}^{N_{\rm d}}, p_{\theta}^{N_{\rm d}}, p_{\varphi}^{N_{\rm d}}, {\bf r}^{N_{\rm m}}, {\bf p}^{N_{\rm m}} )].
  \label{eq:totavedim}
\end{eqnarray}
We note again that up to this point, we have just discussed an exact variable transformation in the statistical mechanics expectation values. We did not modify the system itself at all.

From a dynamical point of view, similarly to what has been discussed at the end of Sec.~\ref{sec:wall}, one should imagine the following procedure: First, an equilibrium configuration
of $\br^N$ is generated at time $t=0$. Second, the $\Nd$ dimers are defined using the  permutationally-invariant procedure described in the first paragraph of this section.
Third, some dynamics is run and the configuration of the system evolves in time.
As long as the dimer and monomer variables remain into their respective domains\footnote{The reader should keep in mind that while the dimer constraints are always the same, they are formally represented in different ways depending on the choice of coordinates, hence leading to formally different spaces $\Od$ and $\Om$.}, $\br^{2\Nd}\in\Od$ and $\br^{\Nm}\in\Om$, one can keep running the dynamics.
If at some point one variable goes out of its domain,
 then one should run again the procedure to relabel the monomers and dimers according to the new particle positions. Once again, this is just a relabeling procedure that does not alter the system in any way.
Doing this results in the annealed average, as in Sec.~\ref{sec:wall}.

\subsubsection{Random bonding}
\label{sec:bonding}

Using the notations introduced in Sec.~\ref{sec:vartrans}, it is now straightforward to perform random bonding, i.e., freeze the dimer bond lengths to define a quenched average.
We consider the
dimer lengths $\tilde{r}^{N_{\rm d}}$ as being frozen, i.e.,
$\tilde{r}^{N_{\rm d}}$ plays the same role as the positions of the pinned particles in the random pinning approach, and of the frozen degrees of freedom ${\bf y}$ in the general discussion of Sec.~\ref{sec:frozen}.
We can thus consider the statistical mechanics of the remaining degrees of freedom, namely a composite of monomers and dimers for a particular realization of $\tilde{r}^{N_{\rm d}}$,
that play the role of the thermal degrees of freedom ${\bf x}$ in Sec.~\ref{sec:frozen}.
Note that the kinetic term associated to the frozen bond lengths is decoupled from all the other degrees of freedom. 
Hence, like in the particle pinning case, we can set the momenta $p_{\tilde r_i}$ to zero at the instant at which the bonds are frozen. 
Nevertheless, we include  in
the partition function the kinetic energy associated with  the non-frozen variables that, for a given realization $\tilde{r}^{N_{\rm d}}$, is then given by
\begin{eqnarray}
Z_{\tilde{r}}(\tilde{r}^{N_{\rm d}}) = 
\int (\prod_{i \in \mathcal{D}} \mathrm{d} {\bf R}_{i}\mathrm{d} {\bf P}_{i})
 \int (\prod_{i \in \mathcal{D}} \mathrm{d}\theta_{i}  \mathrm{d}\varphi_{i}
 \mathrm{d}p_{\tilde r_i}
 \mathrm{d}p_{\theta_{i}} \mathrm{d}p_{\varphi_{i}}) 
\int_{\Om}(\prod_{i \in \mathcal{M}} \mathrm{d}{\bf r}_i \mathrm{d}{\bf p}_i) 
 \nonumber \\ 
\times \exp[-\beta H( {\bf R}^{N_{\rm d}}, {\bf P}^{N_{\rm d}}, \tilde{r}^{N_{\rm d}}, \theta^{N_{\rm d}}, \varphi^{N_{\rm d}}, p_{\tilde{r}}^{N_{\rm d}}, p_{\theta}^{N_{\rm d}}, p_{\varphi}^{N_{\rm d}}, {\bf r}^{N_{\rm m}}, {\bf p}^{N_{\rm m}} )] \ ,
\label{eq:Z'}
\end{eqnarray}
and the thermal average of an observable for a particular realization $\tilde{r}^{N_{\rm d}}$ is defined by
\begin{eqnarray}
\langle \cdots \rangle_{\tilde{r}^{N_{\rm d}}} &=&
 \frac1{Z_{\tilde{r}}(\tilde{r}^{N_{\rm d}})} 
 \int (\prod_{i \in \mathcal{D}} \mathrm{d} {\bf R}_{i}\mathrm{d} {\bf P}_{i}) 
 \int (\prod_{i \in \mathcal{D}} \mathrm{d}\theta_{i}\mathrm{d}\varphi_{i} \mathrm{d}p_{\tilde r_i} \mathrm{d}p_{\theta_{i}} \mathrm{d}p_{\varphi_{i}})
 \int_{\Om}(\prod_{i \in \mathcal{M}} \mathrm{d}{\bf r}_i \mathrm{d}{\bf p}_i) 
  \nonumber \\ &\quad& 
\times \exp[-\beta H( {\bf R}^{N_{\rm d}}, {\bf P}^{N_{\rm d}}, \tilde{r}^{N_{\rm d}}, \theta^{N_{\rm d}}, \varphi^{N_{\rm d}}, p_{\tilde{r}}^{N_{\rm d}}, p_{\theta}^{N_{\rm d}}, p_{\varphi}^{N_{\rm d}}, {\bf r}^{N_{\rm m}}, {\bf p}^{N_{\rm m}} )] 
(\cdots) \ .
\label{eq:thermal_average}
\end{eqnarray}
Comparing this with Eq.~\eqref{eq:totavedim},
we can write
\begin{equation}
\begin{split}
\langle O \rangle =&
\frac{\Pi_{\rm m,d}}{Z} 
\int_{\Od} (\prod_{i \in \mathcal{D}} \mathrm{d} \tilde{r}_{i} )
 \int (\prod_{i \in \mathcal{D}} \mathrm{d} {\bf R}_{i}\mathrm{d} {\bf P}_{i}) 
 \int (\prod_{i \in \mathcal{D}} \mathrm{d}\theta_{i}\mathrm{d}\varphi_{i}
 \mathrm{d}p_{\tilde{r}_{i}} \mathrm{d}p_{\theta_{i}} \mathrm{d}p_{\varphi_{i}}) 
 \int_{\Om}(\prod_{i \in \mathcal{M}} \mathrm{d}{\bf r}_i \mathrm{d}{\bf p}_i)  
 \\ 
 &\times  
  O({\bf r}^N) \, \exp[-\beta H( {\bf R}^{N_{\rm d}}, {\bf P}^{N_{\rm d}}, \tilde{r}^{N_{\rm d}}, \theta^{N_{\rm d}}, \varphi^{N_{\rm d}}, p_{\tilde{r}}^{N_{\rm d}}, p_{\theta}^{N_{\rm d}}, p_{\varphi}^{N_{\rm d}}, {\bf r}^{N_{\rm m}}, {\bf p}^{N_{\rm m}} )] \\
 =&
\frac{\Pi_{\rm m,d}}{Z} 
\int_{\Od}  \mathrm{d} \tilde{r}^{\Nd} 
Z_{\tilde{r}}(\tilde{r}^{N_{\rm d}}) 
\langle O \rangle_{\tilde{r}^{N_{\rm d}}}
   \ .
  \end{split}
\end{equation}
From this expression, we can deduce 
the probability distribution of the frozen variables $\tilde{r}^{N_{\rm d}}$, which is given by
\begin{equation}
\rho_f(\tilde{r}^{N_{\rm d}}) =\Pi_{\rm m,d} \frac{Z_{\tilde{r}}(\tilde{r}^{N_{\rm d}})}{Z} \ ,
\label{eq:prob_length}
\end{equation}
and define a ``disorder'' average over the realization of $\tilde r^{N_{\rm d}}$ as \begin{equation}
\overline{\cdots} = 
\int_{\Od} \mathrm{d} \tilde{r}^{\Nd} \rho_f(\tilde{r}^{N_{\rm d}}) (\cdots).
\label{eq:quenched_average}
\end{equation}
Finally, as it can be expected from the discussion of
Sec.~\ref{sec:frozen}, we can obtain the identity that is at the basis of the random pinning procedure~\cite{scheidler2004relaxation,krakoviack2010statistical}: The quenched average, i.e., the disorder average over the frozen degrees of freedom taken after the thermal average over all remaining degrees of freedom conditioned to the frozen ones, corresponds exactly to the annealed average of the bulk system without bonding.
By using Eqs.~(\ref{eq:Z'}) -
(\ref{eq:quenched_average}), we get
\begin{equation}
\langle O \rangle =\int_{\Od} \mathrm{d} \tilde{r}^{\Nd} \rho_f(\tilde{r}^{N_{\rm d}}) \langle O \rangle_{\tilde{r}^{N_{\rm d}}}
=\overline{\langle O \rangle_{\tilde{r}^{N_{\rm d}}}} \ .
\end{equation}

The random bonding procedure then goes as follows. First, one generates an equilibrium configuration of $N$ particles. Then, one performs the permutationally-invariant procedure described at the beginning of section \ref{sec:vartrans}
to identify the $\Nd$ dimers. The bond lengths are finally frozen, i.e., taken as a quenched disorder, and one considers the statistical mechanics of the remaining degrees of freedom as thermal. With this construction, all the hypotheses discussed in section \ref{sec:frozen} are fulfilled, and we can rigorously state that an equilibrium configuration of a system containing $N_{\rm d}$ dimers and $N_{\rm m}$ monomers has been created. However, it should also be noted that the resulting system is a rather unphysical one: Due to the unbiased manner in which the choice of the dimers has been made (based purely on the labels, and not on a property of the initial configuration), their length distribution will be proportional  to the pair correlation function of the initial system of monomers, and arbitrarily long dimers (up to the system size) will be present\footnote{The case in which such long range interactions are soft has been considered in Refs.~\cite{nandi2021connecting,nandi2022thermodynamics} and it was found that such a system does show a strong slowing down of the dynamics when the number of links is increased.}. It is clear that in order to produce an equilibrium, or nearly equilibrium, configuration of a more realistic system, a bias must be introduced in the choice of the frozen dimers. The Nishimori (quiet freezing) condition (see section \ref{sec:frozen}) will therefore not be strictly respected, and the consequences of this choice have to be assessed. In the following we discuss several possible choices, and study two of them numerically.

\subsubsection{`Sorted bonding' from the shortest to the largest bond}

We start by considering a seemingly ``natural'' procedure, which will however turn out to give unsatisfactory results. In this procedure, the dimers are created between those pairs of particles that have the smallest inter-particle distance.
Specifically, we find the two particles with the smallest distance
$|\tilde{{\bf r}}_{ij}|=|{\bf r}_i-{\bf r}_j|$ and we relabel them as ${\bf r}_1$ and ${\bf r}_2$ (which particle is 1 and which one is 2 is irrelevant). 
Then we look at the remaining particles (excluding  1 and 2) and find once again the pair
with the smallest inter-particle distance among particles, which we relabel as ${\bf r}_3$ and ${\bf r}_4$. 
We continue iterating this procedure until $N_{\rm d}$ pairs have been sorted. This is a permutation-invariant procedure.
We then make $N_{\rm d}$ dimers by freezing the bond lengths. 

We note that the configuration that has been created is \emph{not} an equilibrium configuration of the systems of dimers and monomers interacting with the original Hamiltonian. To see this, one simply has to imagine the evolution of the system starting from the initial configuration just after freezing. In this initial configuration, all monomer-monomer distances or monomer-dimer distances are, by construction, larger than the largest dimer size. The dynamics will, obviously, not preserve this constraint and the ``hole'' created by the freezing procedure in the (say) monomer-monomer correlation function, which is an intensive, self averaging quantity, will disappear progressively. The system will show aging of a static observable, a hallmark of nonequilibrium dynamics.

A short reflection shows that the above procedure for creating dimers can in fact be used for creating an equilibrium monomer-dimer mixture if one allows a modification of the interaction energies between the particles.  This energy function should preserve the following properties:
 \begin{enumerate}
\item 
$\Od$ is  
defined by the constraint (as above, the tilde notation indicates a distance between a pair of particles, rather than an absolute position):
\begin{equation}
 |\tilde{{\bf r}}_{12}| < |\tilde{{\bf r}}_{34}| < |\tilde{{\bf r}}_{56}| < \cdots 
< |\tilde{\br}_{2\Nd-1,2\Nd}| = R_{\rm max} \ ,
\end{equation}
i.e., the integration over the $N_{\rm d}$ dimers (first $2N_{\rm d}$ particles) is constrained in such a way that each dimer has a larger inter-particle distance than the previous one.
Here, $R_{\rm max}$ is the maximal bond length of the dimers.

\item Additional constraints have to be imposed in $\Od$ on the distances between particles belonging to distinct dimers. For example, 
\begin{equation}
    |\tilde{{\bf r}}_{12}| < |\tilde{{\bf r}}_{13}| \ ,
    \qquad    
    |\tilde{{\bf r}}_{12}| < |\tilde{{\bf r}}_{14}| \ ,
\end{equation}
in order to guarantee that $|\tilde{{\bf r}}_{12}|$ remains the shortest distance. Similar constraints are needed for particles with higher labels.

\item
The integration over the remaining $N_{\rm m}$ monomers is over the space $\Omega_{\rm m}$ such that all of the $N_{\rm m}(N_{\rm m}-1)/2$ monomer-monomer distances
and each of the $\Nm \Nd$ monomer-dimer distances
 are larger than $R_{\rm max}$.
This must be true because, otherwise, in the sorting
procedure we would have chosen one of the monomers to belong to a dimer. 
The positions of the monomers are then constrained to the space $\Om$, i.e., the smallest monomer-monomer and monomer-dimer distance must stay larger than the largest dimer bond length $R_{\rm max}$. 
\end{enumerate}

In order to perform the quenched average dynamically, the above constraints must be implemented either by rejecting moves that violate them in a Monte Carlo simulation, or
by adding a hard wall term that would reflect the relative velocity of two monomers or a monomer-dimer pair if they reach the minimal distance in a Molecular Dynamics simulation.
The time-average under this constraint would then result in a proper quenched average over the $\Om$ space.

The positive aspect of this construction is that we can construct permutationally-invariant observables $O(\br^N)$ that only depend on the unfrozen degrees of freedom. However,  if the number of bonds is large,  such that $R_{\rm max}$
reaches the first peak of the radial distribution of monomer distances $g(r)$, the resulting constraints on the monomers will be quite strong, and the corresponding dynamics becomes unphysical, since, e.g., the interactions between the particles are non-zero even for arbitrarily large distances.

\subsubsection{`Random bonding' within a cutoff}
\label{sec:random_bonding_cutoff}

We now consider a different procedure, close  to the rigorous ``random bonding'' described at the end of section \ref{sec:bonding}, but now imposing that the length  of the formed dimers does not exceed a maximum value $R_{\rm b}$ (see Fig.~\ref{fig:bonding_schematic}(a)). This is the method used in Ref.~\cite{Ozawa2023}, where it was assumed to produce an equilibrium configuration of the dimer-monomer mixture.

The procedure is as follows: First, we select at random the first particle in each of the $\Nd$ dimer.
Second, for each of these, we choose a particle at random among those at distance smaller than $R_{\rm b}$ from it to be its partner in the dimer.
If there is no viable partner, we discard that candidate and choose at random a new one. This procedure is also permutation invariant with the following properties:

\begin{enumerate}
    \item The space $\Od$ is defined by the only constraint that the dimer bond lengths $\tilde r_i \leq R_{\rm b}$, which is trivially satisfied if the bond lengths are frozen. For the rest, dimers are free to translate and rotate.
\item The monomer space $\Omega_{\rm m}$ is, seemingly, unconstrained because in the dimer construction procedure nothing is implied about the monomers.
    
\end{enumerate}

However, the choice of the $\Nd$ dimers is now not only based on the labels of the particles, but depends on a property of the equilibrated configuration of the system of monomers, so that the Nishimori condition is, again, not strictly respected. The distribution of the frozen variables (bond lengths) is not strictly the one in an equilibrium, unconstrained system, but will converge to it as $R_{\rm b}$ is increased. As a result, the initial configuration will also display subtle correlations between non-bonded particles, which will decay with time if the system is propagated with the original interaction potential. To see this, let us consider the total pair correlation function of the system just after the freezing of the bonds, $g(r)$. This correlation function can be written as
\begin{equation}
    g(r)= g_{\rm inter}(r) + g_{\rm intra}(r) \quad ,
\end{equation}

\noindent
where $g_{\rm inter}$ is the contribution of pairs of particles that are not bonded together by a frozen bond, and $g_{\rm intra}(r)$ is the contribution of the pairs of particles that are connected. 
By construction, $g_{\rm intra}(r)$ will have a discontinuity at $R_{\rm b}$, above which it will jump to zero. On the other hand, $g(r)$ is the pair correlation of an equilibrium system of monomers, and is continuous. We conclude that $g_{\rm inter}(r)$ will initially have a discontinuity at $R_{\rm b}$, which will not be preserved by the dynamics. 
The situation is similar to the one described above in the ``sorted bonding'' case, but more subtle. In the following numerical study, we will see that the nonequilibrium  character of the initial configuration is extremely small, and disappears very rapidly. The choice of this bond distribution, for $R_{\rm b}$ of the order of the interparticle distance, leads to a physically reasonable system which is nearly at equilibrium,   and can be considered to be close to optimal.

\subsubsection{Random bonding with directional alignment}

\label{sec:directional_alignment}

As a final example, we will consider a situation in which it is intuitively more obvious that the initial configuration is out of equilibrium, and will display aging over a measurable time scale. The bonding scheme is similar to the random bonding within a cutoff $R_{\rm b}$ discussed in the previous section, however an additional constraint is imposed to the orientation of the frozen dimers, which are restricted to an angular sector $\theta_b$ around the $z$ direction (see Fig.~\ref{fig:bonding_schematic}(b)). The initial configuration therefore displays nematic order of the dimers, which obviously will not be preserved by the dynamics. The numerical study will allow us to estimate the persistence time of this nonequilibrium feature, in comparison to the isotropic case.

\section{Simulation methods}

\subsection{Model}

We employ the Kob-Andersen binary mixture~\cite{kob1995testing}, in which particles interact through the Lennard-Jones pair potential,
\begin{equation}\label{eqn:lj}
u_{\alpha\beta}(r) = 4\epsilon_{\alpha\beta}\left[
  {\left( \frac{\sigma_{\alpha\beta}}{r} \right)}^{12} - {\left(
    \frac{\sigma_{\alpha\beta}}{r} \right)}^6 \right],
\end{equation}
where $\alpha, \beta \in \{A,B\}$ are species indexes.
Both species have the same mass, which is set to $m=1$.
The value of the parameters $\sigma_{\alpha\beta}$ and $\epsilon_{\alpha\beta}$ are given in Ref.~\cite{kob1995testing}.
The units of length and energy are set by the parameters $\sigma=\sigma_{\rm AA}=1$ and $\epsilon=\epsilon_{\rm AA}=1$, respectively, and we put the Boltzmann constant $k_{\rm B}=1$.
The potentials are cut and shifted at a distance $2.5\sigma_{\alpha\beta}$.
We simulate systems composed of $N$ particles in a cubic box of side $L$ with periodic boundary conditions at a number density $\rho=N/V=1.2$.
We use the system size $N=1200$.
We perform constrained molecular dynamics simulations  via the RATTLE algorithm~\cite{andersen1983rattle} with a simple Nos\'e-Hoover thermostat~\cite{Ozawa2023}.

\begin{figure}[t]
\includegraphics[width=0.7\columnwidth]{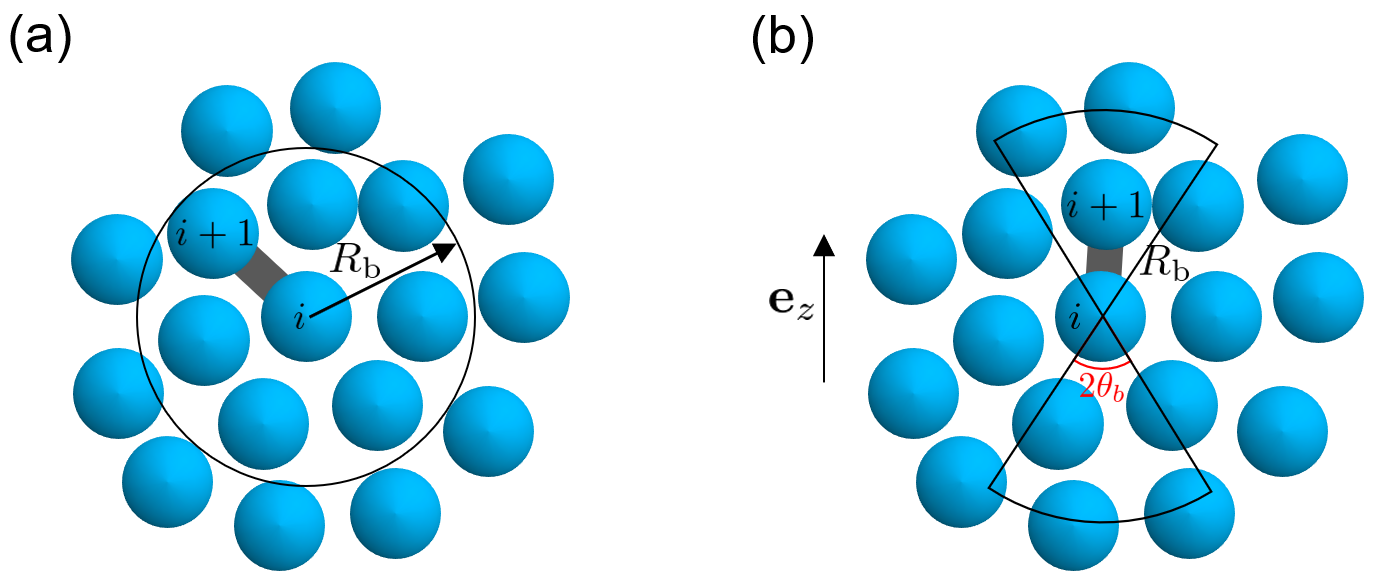}
\caption{
{\bf Schematic illustration of the construction of a bond connecting particles $i$ and $i+1$.} (a): Random bonding with spherical cut-off. The sphere at the center of the particle $i$ with the radius $R_{\rm b}$ defines its neighborhood.
(b): Random bonding with directional alignment. 
An additional constraint is imposed by a cut-off $P_{\rm b}=\cos \theta_{\rm b}$ associated with directional alignment.
}
\label{fig:bonding_schematic}
\end{figure}

\subsection{Making randomly bonded systems}

Starting from an equilibrium configuration of the original (bulk) KA model with $N$ particles described above, we generate a randomly bonded system composed of monomers and dimers.
We consider the following two protocols to do so.

1) {\it Random bonding with spherical cut-off:}
First, we choose a particle randomly, say particle $i$. We then randomly pick another particle $j$ among the neighboring particles of particle $i$, 
located inside a sphere with a cut-off radius $R_{\rm b}$ and which is not yet bonded.
We then relabel the particle $j$ as $i+1$. 
This process is schematically shown in Fig.~\ref{fig:bonding_schematic}(a). 
We set $R_{\rm b}=1.5$, which is near the first minimum of the radial distribution function, thus corresponding roughly to the boundary of the first coordination shell.
We then permanently freeze the distance between the two particles, $\tilde r_{i} = |{\bf r}_i-{\bf r}_{i+1}|$, which means that the particles $i$ and $i+1$ now form a dimer by a rigid body constraint. 
We repeat the above process for the remaining monomer particles until the number of dimers, $N_{\rm d}$, reaches the target value.
By construction, we have $N=N_{\rm m} + 2N_{\rm d}$, where $N_{\rm m}$ is the number of monomers. We introduce the control parameter $c=\frac{2N_{\rm d}}{N}=\frac{N-N_{\rm m}}{N}$, such that
$c=0$ corresponds to the system with only monomers (hence the original bulk model), whereas $c=1$ corresponds to a system having only dimers.
Using the algorithm explained above, it is difficult in practice to reach $c=1$, because at some point one runs out of neighboring pairs, leaving a few percent of monomer particles. Thus in the present work we use $c=0.95$ as the maximum value.

2) {\it Random bonding with directional alignment:}
In this protocol we aim to prepare an initial bonded configuration such that the orientation of the dumbbell molecules tend to align along the unit vector of the $z$-direction, ${\bf e}_z=(0,0,1)$.
We define a unit vector of the orientation of each dumbbell molecule composed of particles $i$ and $i+1$ by ${\bf n}_i=\tilde {\bf r}_i/|\tilde {\bf r}_i|$, where $\tilde {\bf r}_i=(x_i-x_{i+1}, y_i-y_{i+1}, z_i-z_{i+1})$.
In the protocol 1) we make a bond between the $i$-th and $(i+1)$-th particles randomly among the neighbor particles inside a spherical shell with the radius $R_{\rm b}$ centered at the position of the $i$-th particle. We now impose a further limitation on the neighbor region by taking into account the orientation of the molecule. 
To this end, we define a magnitude of alignment (or polarization) for each molecule, $P_i$, given by
\begin{equation}
    P_i = |{\bf e}_z \cdot {\bf n}_i| = |z_i-z_{i+1}|/|\tilde {\bf r}_i|=|\cos\theta_i|.
\end{equation}
We then introduce another cut-off threshold $P_{\rm b}$ such that particles having $P_i>P_{\rm b}=\cos \theta_b$ can be considered as candidate neighbors for bonding.
This new neighbor region is schematically shown in Fig.~\ref{fig:bonding_schematic}(b).
As in procedure 1), we repeat the above process for the remaining monomer particles until either the number of dimers reaches the target value, or we run out of candidate pairs for bonding. In practice, we set $c=0.75$ as the maximum value for this protocol with $P_{\rm b}=0.9$. 

\begin{figure}[t]
\includegraphics[width=0.35\columnwidth]{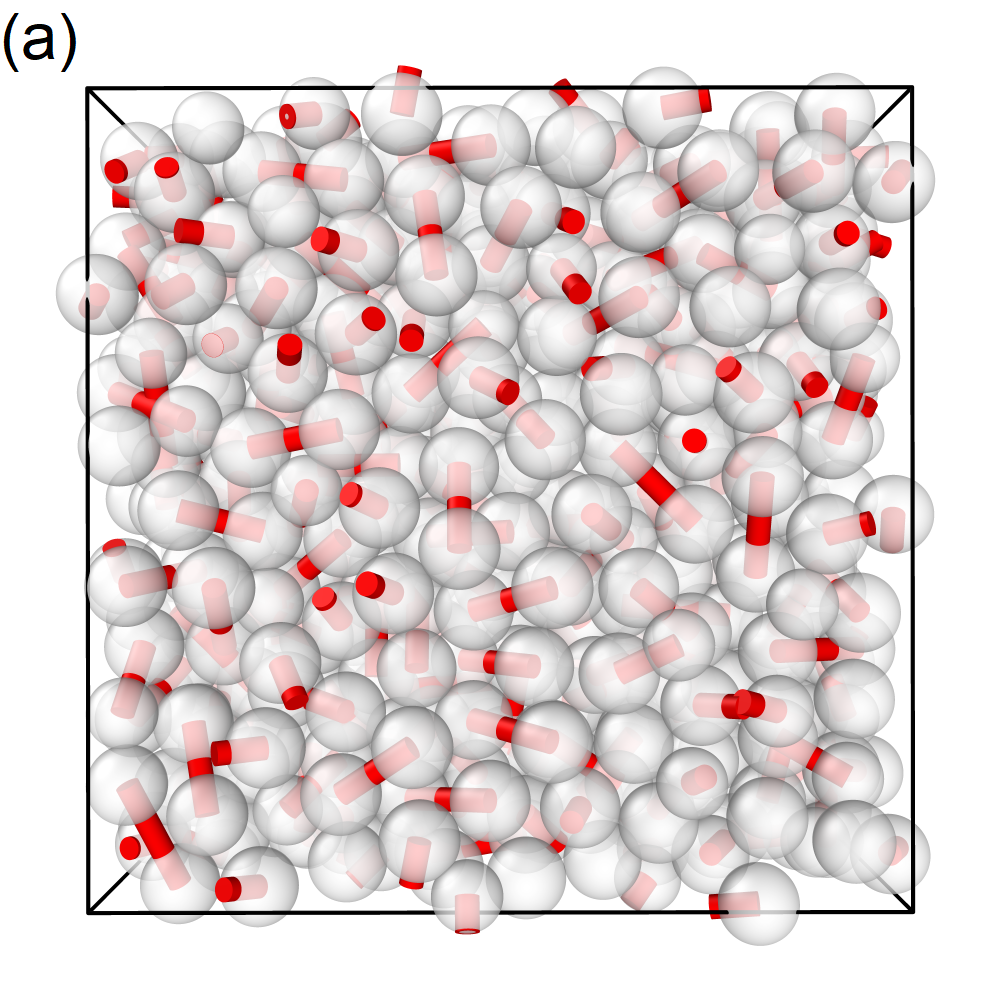}
\includegraphics[width=0.35\columnwidth]{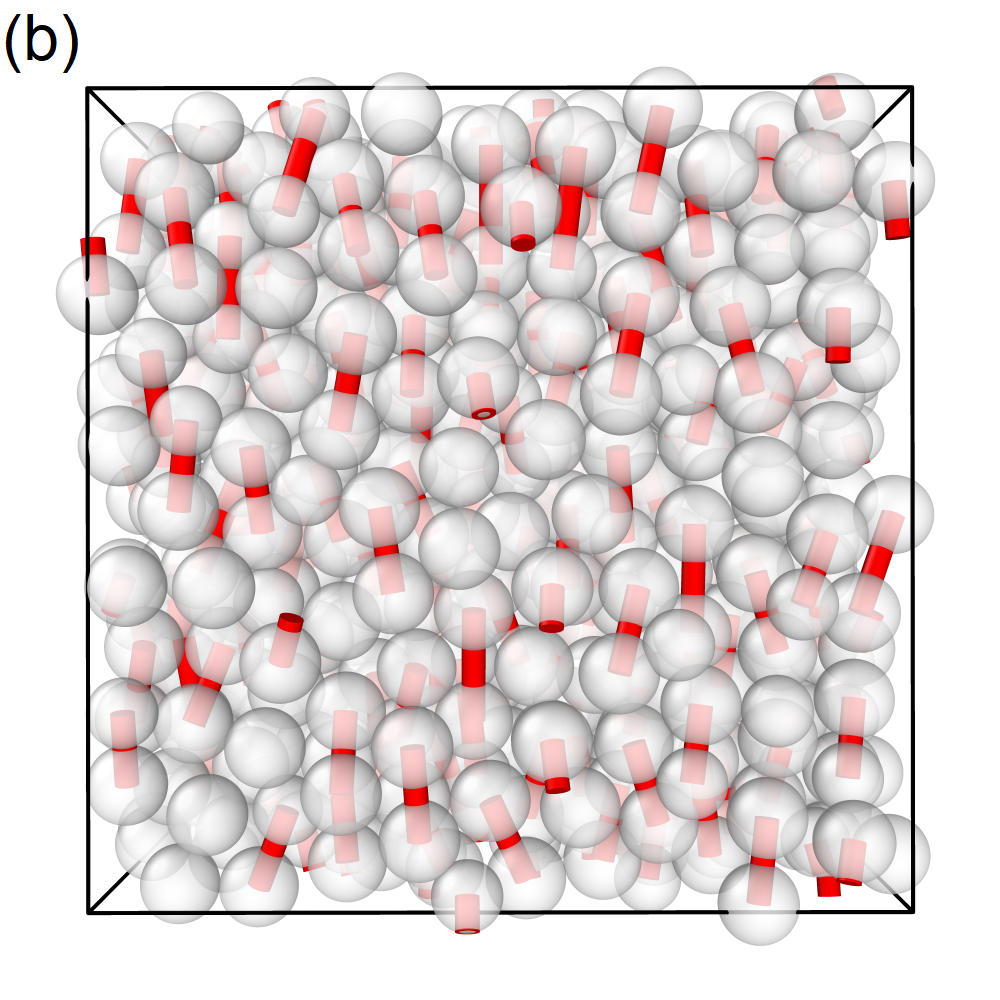}
\caption{{\bf Examples of initial bonded configurations},
here for $T=0.6$, $c=0.5$, and $R_{\rm b}=1.5$.
Left: $P_{\rm b}=0$ (random bonding with spherical cut-off). Right: $P_{\rm b}=0.9$ (random bonding with directional alignment).
}
\label{fig:initial}
\end{figure}

In Fig.~\ref{fig:initial}, we show initial bonded configurations for $T=0.6$, $c=0.5$, and $R_{\rm b}=1.5$.
The left panel is $P_{\rm b}=0.0$ (random bonding with spherical cut-off), and the right panel is $P_{\rm b}=0.9$ (random bonding with directional alignment).

\section{Structural properties}
\label{sec:static}

In this section, we study how the bonding process affects the equilibrium properties of the system in terms of the static structure. 

\subsection{Radial distribution function}

\begin{figure}[t]
\includegraphics[width=0.32\columnwidth]{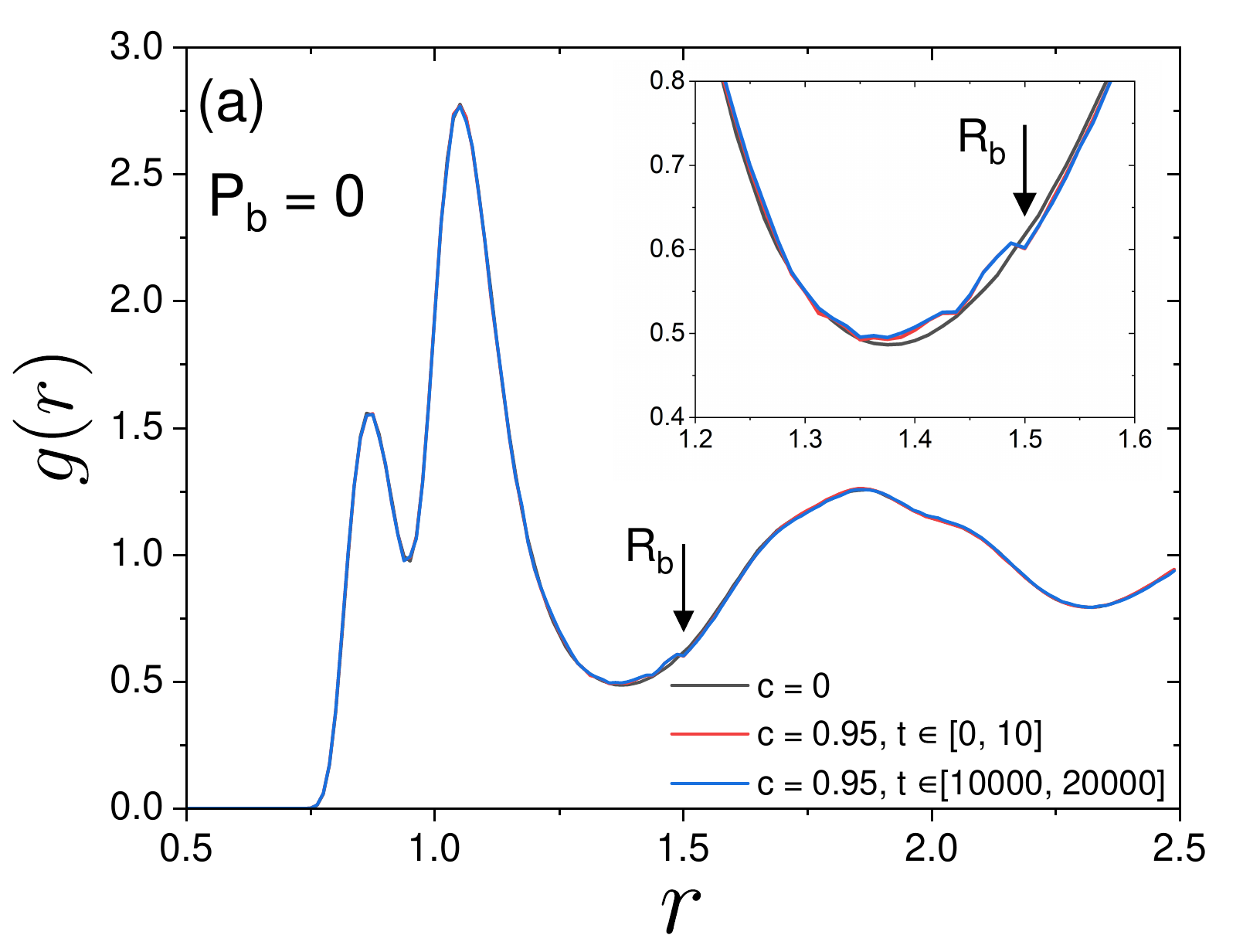}
\includegraphics[width=0.33\columnwidth]{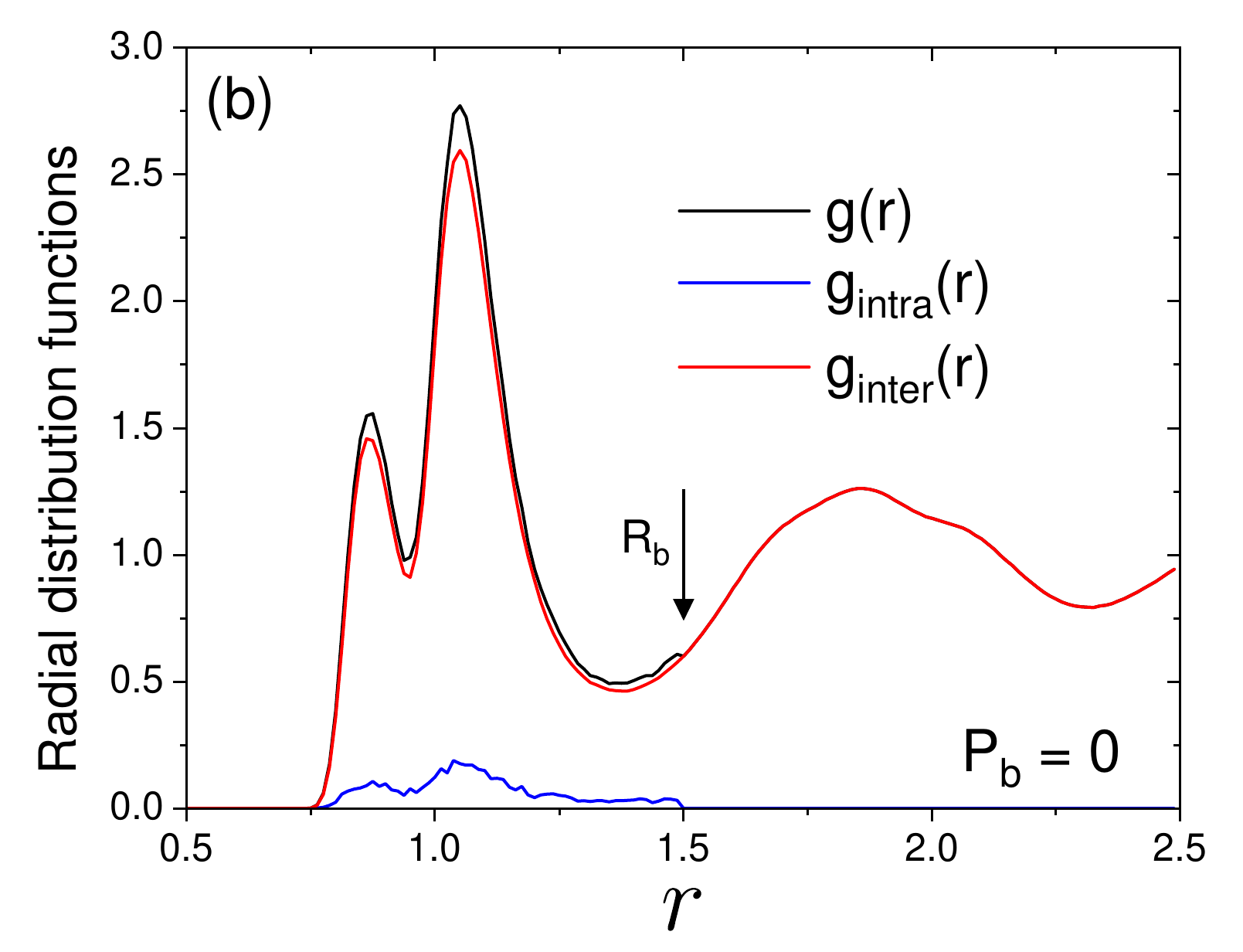}
\includegraphics[width=0.32\columnwidth]{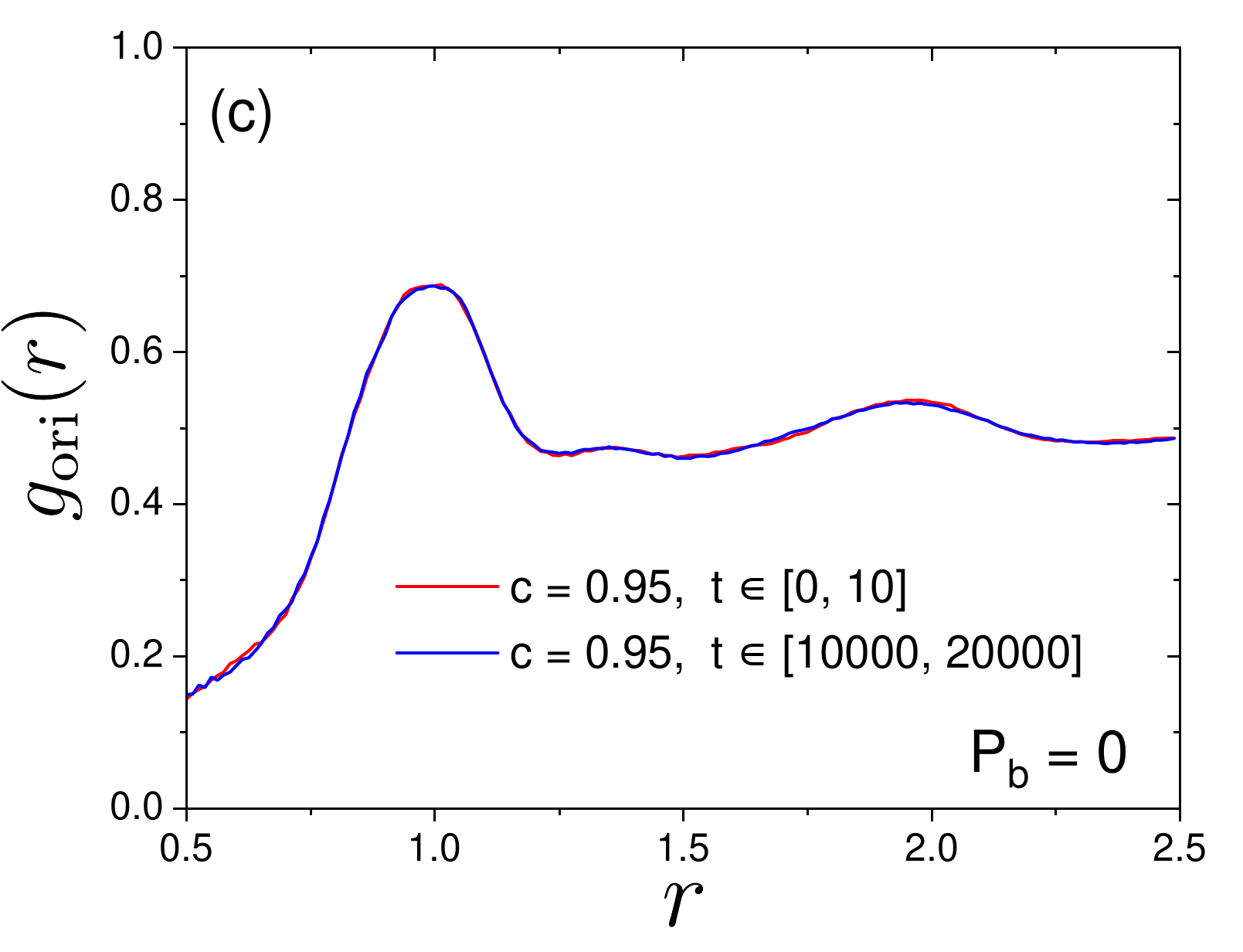}
\caption{{\bf Static properties in the random bonding with spherical cut-off.}
(a): Radial distribution function $g(r)$ for the original system with $c=0$ at $T=0.6$ (black curve), and bonded systems generated by the spherical cut-off protocol ($P_{\rm b}=0$) with $c=0.95$ at $T=0.6$ measured in time windows $t \in [0, 10]$ (red curve) and $t \in [10000, 20000]$ (blue curve). The arrow indicates the location of the cut-off, $R_{\rm b}=1.5$. The inset shows a zoomed plot near $R_{\rm b}$. 
(b) Decomposition of $g(r)$ into the intra molecule contribution $g_{\rm intra}(r)$ and inter molecule contribution $g_{\rm inter}(r)$ for $c=0.95$ at $T=0.6$ in the time window $t \in [0, 10]$.
(c): Orientational correlation function $g_{\rm ori}(r)$ for the bonded systems presented in (a), measured in two different time windows.
}
\label{fig:static_Pb0}
\label{fig:g_intra_and_inter}
\end{figure}

We first characterize the static structure by the radial distribution function $g(r)$ for all particles, which is defined by
\begin{equation}
g(r) = \frac{L^3}{4 \pi r^2 \Delta r N(N-1)} 
\overline{\left\langle
\sum_{ \substack{ i, j \\ (i \neq j) }  } \delta \left( r- | {\bf r}_i - {\bf r}_j |\right)
\right\rangle_{\tilde{r}^{N_{\rm d}}} } ,
\end{equation}
where $\Delta r$ is the bin size for computation. We use $\Delta r=0.0125$.
Figure~\ref{fig:static_Pb0}(a) shows $g(r)$ for the original system ($c=0$) in equilibrium and a bonded system ($c=0.95$) with $P_{\rm b}=0$ measured in a short time window ($t \in [0, 10]$) starting right after bonding ($t=0$) and a longer time window ($t \in [10000, 20000]$) starting after a waiting time $t_{\rm w}=10000$.
The former time scale corresponds to a vibrational one, whereas the latter timescale corresponds to the time scale for escaping from the cage (see Fig.~\ref{fig:absence_aging}(a)).
We note that the first and second peaks correspond to the nearest neighbor contacts for $A-B$ and $A-A$ pairs, respectively.
Overall, the three $g(r)$'s superimposed very well. However, around the cut-off distance $R_{\rm b}=1.5$ there are tiny but distinct differences between the original and bonded system, as emphasized in the inset. This means that the bonding process disturbs the initial equilibrium state, and the system enters an out-of-equilibrium state and thus will age. Yet this aging process is very fast, within the vibrational timescale, as can be recognized from the fact that in the inset the curves for $t \in [0, 10]$ and $t \in [10000, 20000]$ superimpose very well, i.e., the system ends up in a new stationary state very quickly. 

In order to understand better the differences between the bulk and bonded systems around $r=R_{\rm b}$, we decompose $g(r)$ into the contributions from the bonded pair (intra molecule contribution, $g_{\rm intra}(r)$) and the rest (inter molecule contribution, $g_{\rm inter}(r)$), following the discussion in Sec.~\ref{sec:random_bonding_cutoff}. By construction, $g(r)=g_{\rm intra}(r)+g_{\rm inter}(r)$.
In Fig.~\ref{fig:g_intra_and_inter}(b) we present $g(r)$, $g_{\rm intra}(r)$, and $g_{\rm inter}(r)$ for $c=0.95$ in the time window $t \in [0, 10]$.
This graph shows that $g_{\rm intra}(r)$ is only a small contribution to the total radial distribution and, as expected, vanishes beyond $R_{\rm b}$. 
Since the bonding we employ in this work is a rigid body constraint, $g_{\rm intra}(r)$ is completely frozen at time $t=0$, and it never changes during the simulation for any $t>0$.
We also find that $g_{\rm inter}(r)$, at strictly $t=0$, has a small dip at $r=R_{\rm b}$ (not shown), compensating the sharp edge in $g_{\rm intra}(r)$, such that the total $g(r)$ is smooth as in the original system.
At $t>0$, only $g_{\rm inter}(r)$ evolves with time, which removes the dip around $r=R_{\rm b}$ and ends up in a smooth curve as shown in Fig.~\ref{fig:g_intra_and_inter}(b). Consequently, the total contribution, $g(r)=g_{\rm intra}(r)+g_{\rm inter}(r)$, has a bump around $r=R_{\rm b}$ at $t>0$. Indeed, this out-of-equilibrium effect originates from the bonding process with a cut-off $R_{\rm b}$, but we show that it is a very minor perturbation of the whole system even for a very high value of the dimer concentration $c=0.95$.

\subsection{Orientational correlation function}

The radial distribution function $g(r)$ considers only the particle positions irrespective of molecular orientations. To characterize the configuration of molecules in more detail, we compute an orientational correlation function given by
\begin{equation}
g_{\rm ori}(r) = \frac{ 
\overline{\left\langle
\sum_{ \substack{ i, j \in \mathcal{D} \\ (i \neq j) }  } |{\bf n}_i \cdot {\bf n}_j| \delta \left( r- | {\bf R}_i - {\bf R}_j |\right)
\right\rangle_{\tilde{r}^{N_{\rm d}}} }}{ 
\overline{\left\langle
\sum_{ \substack{ i, j \in \mathcal{D} \\ (i \neq j) }  } \delta \left( r- | {\bf R}_i - {\bf R}_j |\right)
\right\rangle_{\tilde{r}^{N_{\rm d}}} }}  \quad ,
\end{equation}

\noindent
where ${\bf R}_i$ is the center of mass of molecule composed of the $i$-th and $(i+1)$-th particles.
In Fig.~\ref{fig:static_Pb0}(c), we present $g_{\rm ori}(r)$ for shorter and longer time windows, respectively.
We find that the two curves are superimposed very well. No aging is thus detected right after a quick relaxation within a vibrational timescale in terms of molecular orientational correlations.

\subsection{Random bonding with directional alignment}

\begin{figure}[t]
\includegraphics[width=0.32\columnwidth]{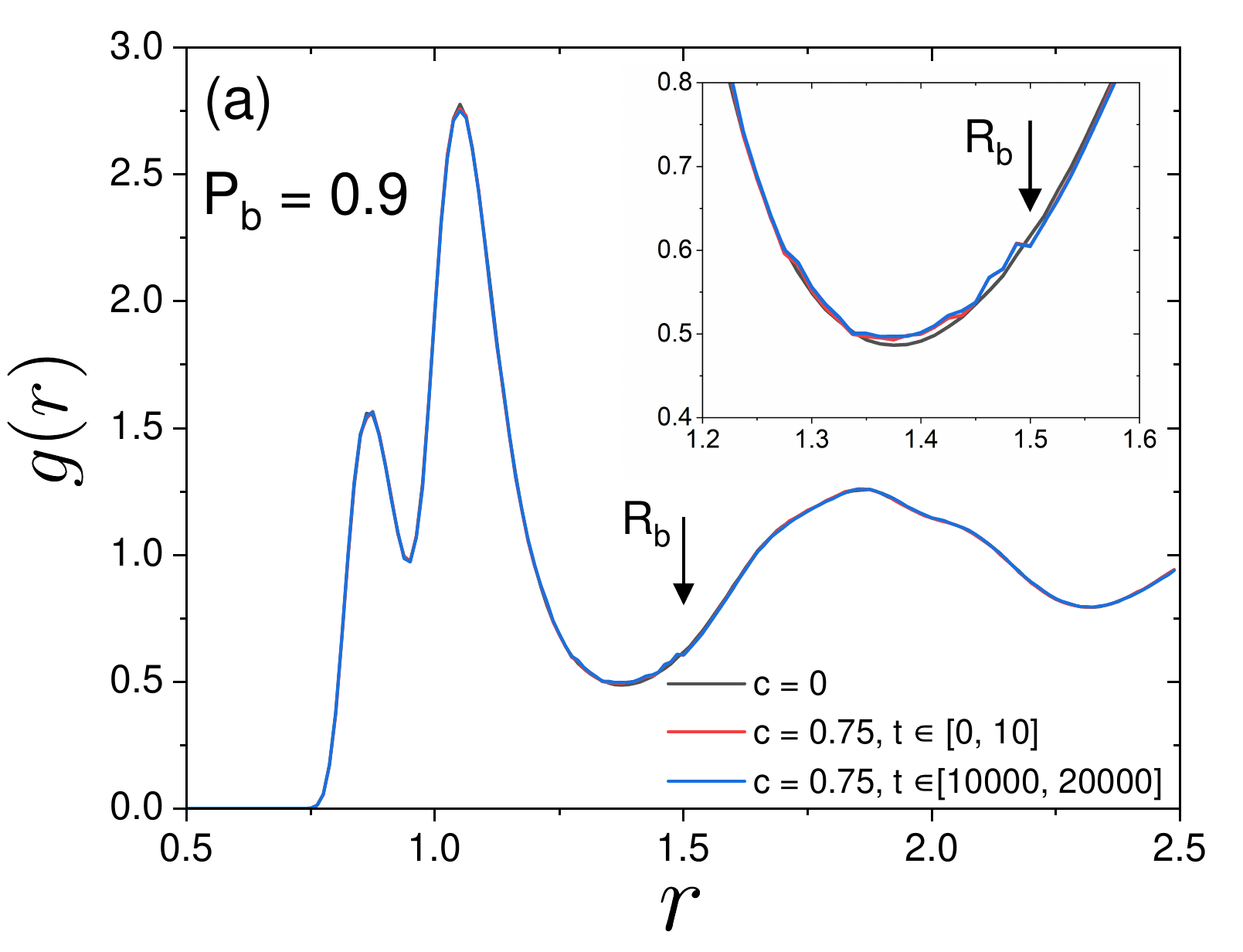}
\includegraphics[width=0.32\columnwidth]{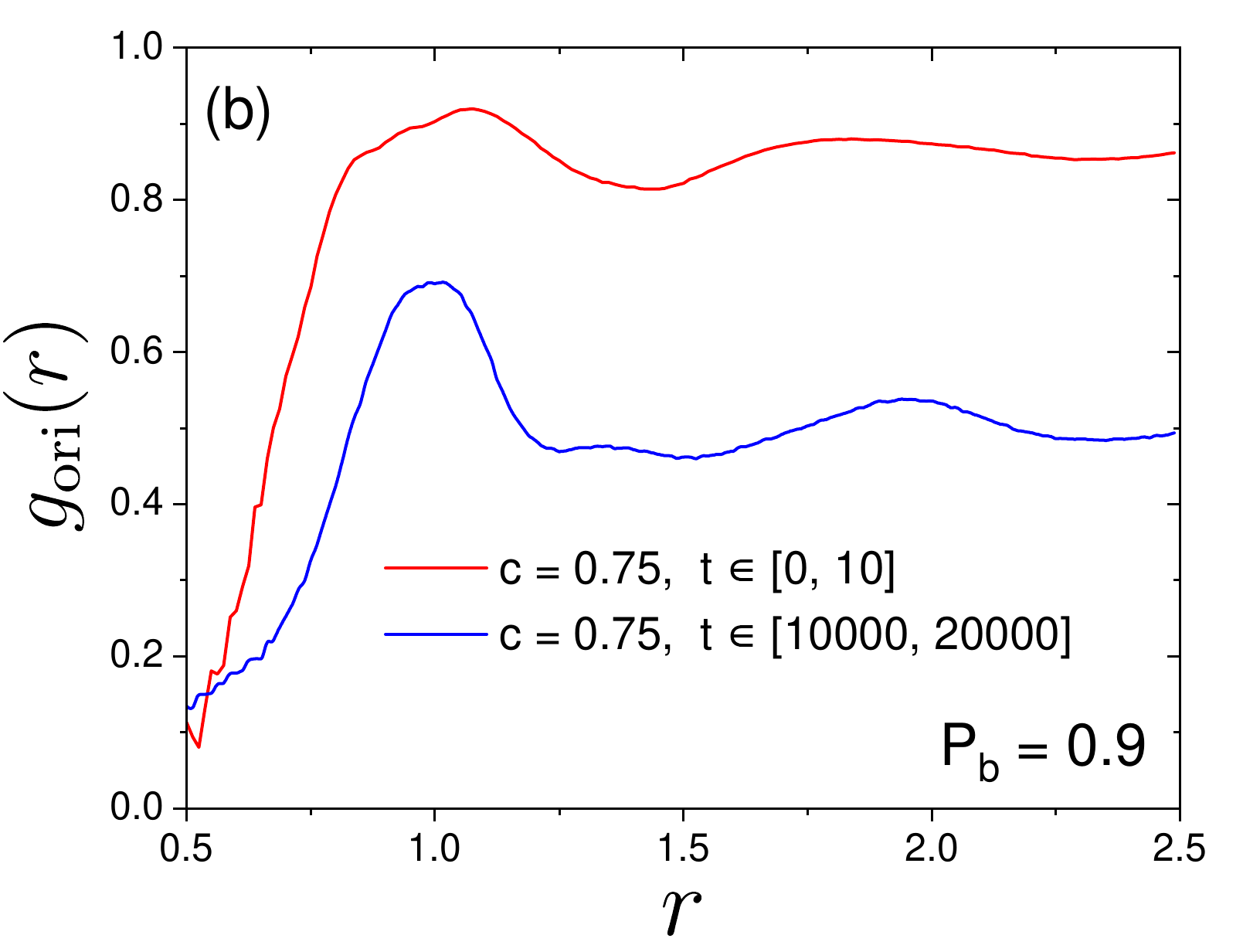}
\includegraphics[width=0.32\columnwidth]{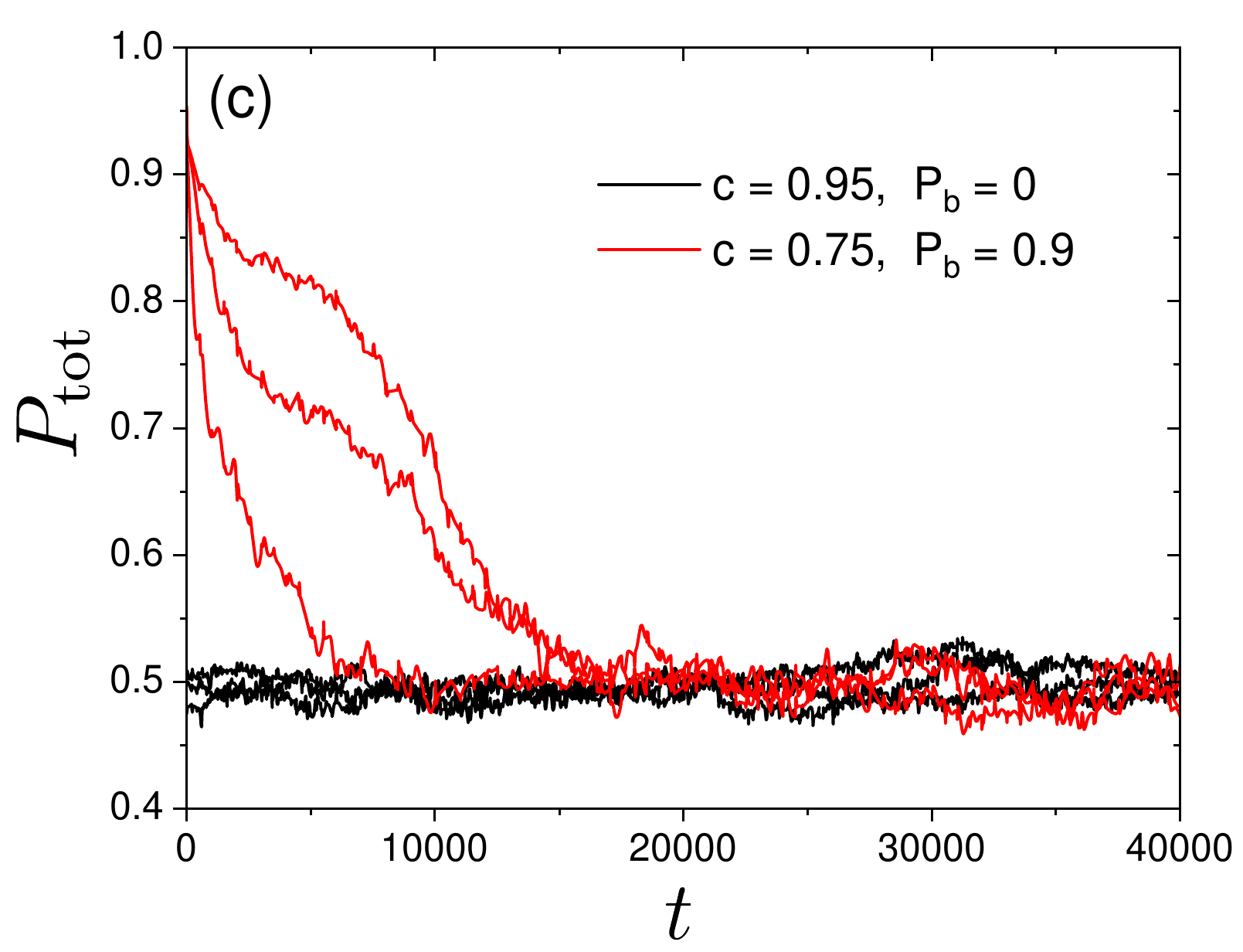}
\caption{{\bf Random bonding with directional alignment.}
(a): Radial distribution function $g(r)$ for the original system with $c=0$ at $T=0.6$ (black curve), bonded systems generated by the spherical and directional cut-off protocol ($P_{\rm b}=0.9$) with $c=0.75$ measured in time windows $t \in [0, 10]$ (red curve) and $t \in [10000, 20000]$ (blue curve). The arrow indicates the location of the spherical cut-off, $R_{\rm b}=1.5$. The inset shows a zoomed plot near $R_{\rm b}$. (b): Orientational correlation function $g_{\rm ori}(r)$ for the bonded systems presented in (a), measured in two different time windows. 
(c): Time evolution of the total polarization $P_{\rm tot}$ for the spherical cut-off ($c=0.95$ and $T=0.6$ with $P_{\rm b}=0$) and directional alignment ($c=0.75$ and $T=0.6$ with $P_{\rm b}=0.9$) protocols. Three representative trajectories are shown for each protocol.
}
\label{fig:static_Pb0.9}
\label{fig:P_tot}
\end{figure}

The above simulation results suggest that although it is not strictly in equilibrium, the bonding protocol with a spherical cut-off produces a nearly equilibrium state, and its aging process is very fast.
We now test whether this is also the case for an arbitrary bonding protocol. 
To this end, we study the bonding protocol with directional alignment illustrated in Fig.~\ref{fig:bonding_schematic}(b) (see also the discussion in Sec.~\ref{sec:directional_alignment}).
In Fig.~\ref{fig:static_Pb0.9}(a), we show $g(r)$ for the original system ($c=0$) in equilibrium and a bonded system ($c=0.75$) with $P_{\rm b}=0.9$ measured in a shorter time window ($t \in [0, 10]$) starting right after the bonding ($t=0$) and a longer time window ($t \in [10000, 20000]$) after a waiting time of $t_{\rm w}=10000$.
Similar to the spherical cut-off case, all $g(r)$'s superimpose well and the tiny discrepancy can be recognized only in the zoomed-in plot near~$R_{\rm b}$. 
However, a strong aging effect can be seen in $g_{\rm ori}(r)$ presented in  Fig.~\ref{fig:static_Pb0.9}(b).
At the shorter time scale, $g_{\rm ori}(r)$ is overall larger, due to the initially aligned molecular configuration. It then decays with time, as expected from the fact that randomly oriented configurations are entropically more favored.
At longer timescale, $g_{\rm ori}(r)$ converges to $g_{\rm ori}(r) \approx 0.5$ at $r \gg 1$ and is overall similar to that of Fig.~\ref{fig:static_Pb0}(c) for a spherically-bonded system.

This aging process can be directly quantified by the total molecule polarization, given by
\begin{equation}
    P_{\rm tot} (t)= \frac{1}{N_{\rm d}} \sum_{i \in \mathcal{D}} P_i(t) \quad.
\end{equation}
Figure~\ref{fig:P_tot}(c) shows the time evolution of $P_{\rm tot}$. The system with $P_{\rm b}=0$ has random molecular orientations, producing $P_{\rm tot}\approx 0.5$ during the entire simulation.
On the contrary, the system with $P_{\rm b}=0.9$ takes a higher value, $P_{\rm tot}\approx 0.95$ at $t=0$, and it decays slowly with time, demonstrating a strong aging effect. At a longer time, it then reaches $P_{\rm tot}\approx 0.5$.

\section{Dynamical properties}

In the previous section we have confirmed that the random bonding protocol with a spherical cut-off produces a nearly equilibrium configuration right after bonding, and the subsequent aging process lasts only a short period of time. Thus, in practice, this procedure allows us to study equilibrium dynamical properties on a longer time scale. In the present section we hence study the (equilibrium) dynamical properties of randomly bonded glass-forming liquids generated by the spherical cut-off protocol.

\subsection{Intermediate scattering functions}

\begin{figure}[b]
\includegraphics[width=0.48\columnwidth]{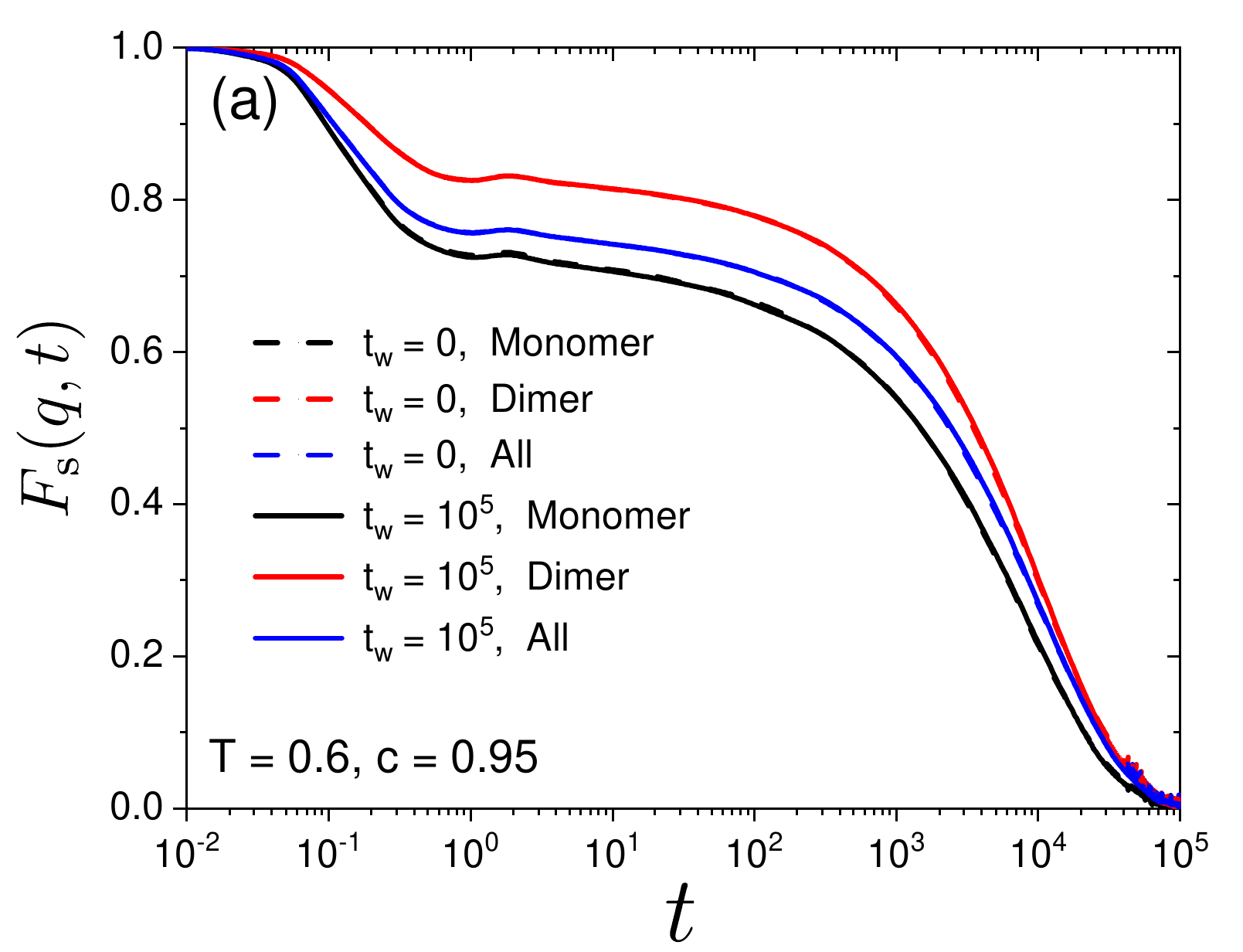}
\includegraphics[width=0.48\columnwidth]{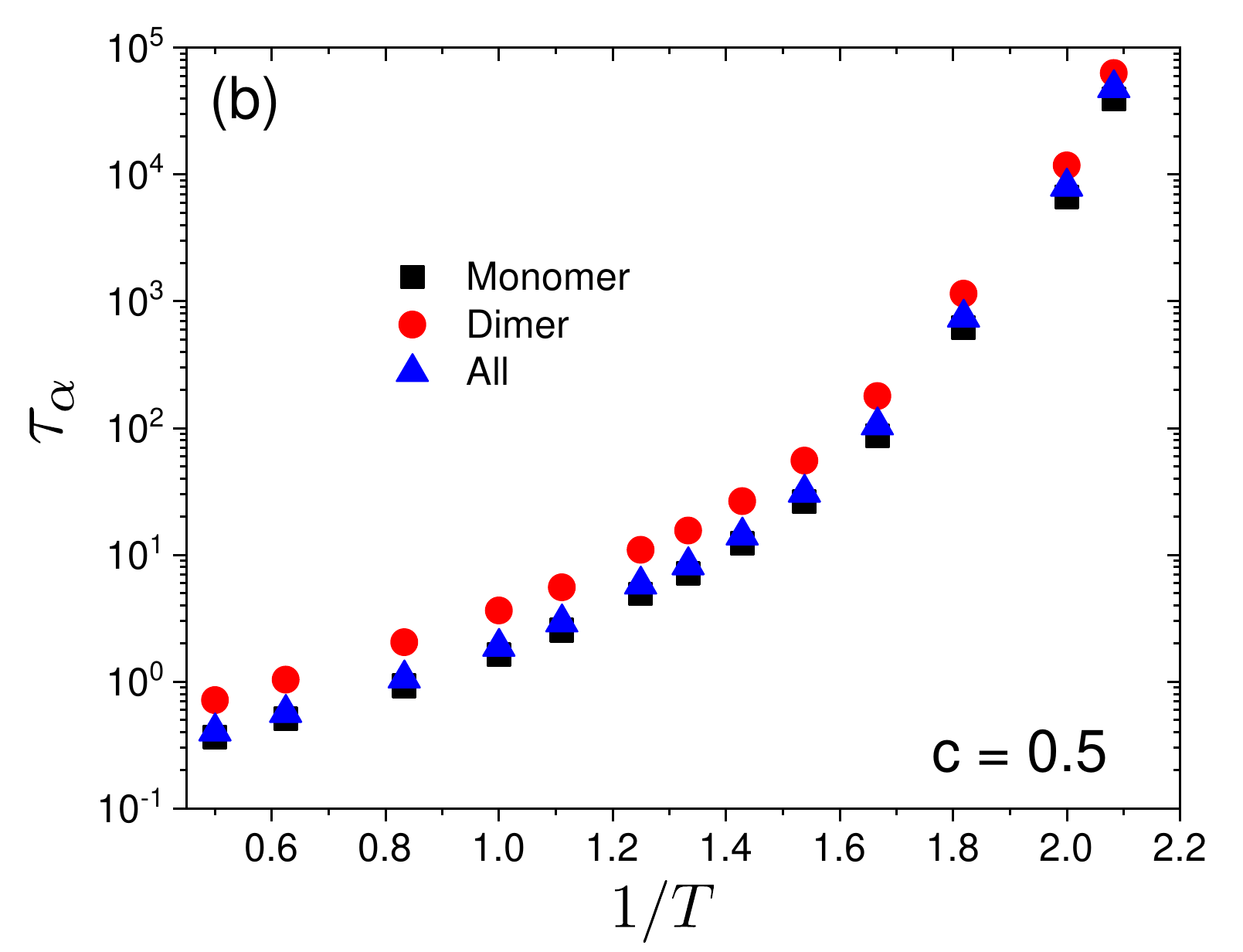}
\caption{
{\bf Dynamical correlations for dimers and monomers.}
(a): Self-intermediate scattering function $F_{\rm s}(q,t)$ for $T=0.6$ and $c=0.95$ for monomers, dimers, and all particles. Dashed and solid curves indicate $F_{\rm s}(q,t)$ computed from trajectories with the waiting time $t_{\rm w}=0$ and $t_{\rm w}=10^5$, respectively.
(b): Relaxation time $\tau_{\alpha}$ versus the inverse of temperature $1/T$ for $c=0.5$. 
}
\label{fig:absence_aging}
\end{figure}

To characterize the dynamic properties of the system, we compute the self part of the intermediate scattering functions for monomers, the center of mass of dimers, and all particles, given by
\begin{eqnarray}
F_{\rm s}^{\rm Mono}(q,t) &=& \overline{ \left\langle \frac{1}{N_{\rm m}} \sum_{j \in \mathcal{M}} e^{-i {\bf q}\cdot ({\bf r}_j(t)-{\bf r}_j(0))} \right\rangle_{\tilde{r}^{N_{\rm d}}} }\quad , \\
F_{\rm s}^{\rm Di}(q,t) &=& \overline{ \left\langle \frac{1}{N_{\rm d}} \sum_{j \in \mathcal{D}} e^{-i {\bf q}\cdot ({\bf R}_j(t)-{\bf R}_j(0))} \right\rangle_{\tilde{r}^{N_{\rm d}}} }\quad , \\
F_{\rm s}^{\rm All}(q,t) &=& \overline{ \left\langle \frac{1}{N} \sum_{j=1}^N e^{-i {\bf q}\cdot ({\bf r}_j(t)-{\bf r}_j(0))} \right\rangle_{\tilde{r}^{N_{\rm d}}} } \quad,
\end{eqnarray}
respectively.
We have averaged over 5–20 different realizations to calculate these time correlation functions.
The wave-vector $q$ is chosen to be $q=7.25$, the location of the main peak in the static structure factor~\cite{kob1995testing}.

Figure~\ref{fig:absence_aging}(a) shows these intermediate scattering functions at $T=0.6$ and $c=0.95$.
$F_{\rm s}^{\rm Di}(q,t)$ displays a higher value of the plateau than the one found in $F_{\rm s}^{\rm Mono}(q,t)$ or $F_{\rm s}^{\rm All}(q,t)$, which is reasonable since the effective cage size of the dimers is smaller than that of the monomers. More important is the observation that these functions relax on essentially the same timescale, which is evidence that the monomers and dimers have a very similar relaxation dynamics in terms of translational motions.
We define the relaxation time $\tau_{\alpha}$ as the time at which the intermediate scattering function decays to $1/e$ and
present the $\tau_{\alpha}$ versus $1/T$ plot for $c=0.5$ in Fig.~\ref{fig:absence_aging}(b).
The relaxation time for the center of mass of dimers exceeds $\tau_\alpha$ for the monomers by a factor around 2, independent of $T$, which shows that the dynamics of the two types of particles stays coupled in the whole accessed $T-$range, and we have checked that this is also the case for the other values of $c$.
Hence one can conclude that the three definitions of the intermediate scattering functions provide essentially the same information in terms of structural relaxation. Therefore we focus in the following on $F_{\rm s}^{\rm All}(q,t)$ and drop the superscript unless otherwise specified.

In Fig.~\ref{fig:absence_aging}(a), we also include data for a different waiting time $t_{\rm w}$, i.e., the time interval between the initial configuration ($t=0$), and the time when we start measuring the correlation ($t=t_{\rm w}$).
The graph demonstrates that there is no detectable waiting-time dependence in the time scale of structural relaxation, as expected from the structural analysis presented in the previous section.
Thus, the random bonding (using the spherical cut-off protocol) allows us to probe equilibrium dynamics relevant to structural relaxation by simulations right after bonding, akin to random pinning.

\begin{figure}[b]
\includegraphics[width=0.48\columnwidth]{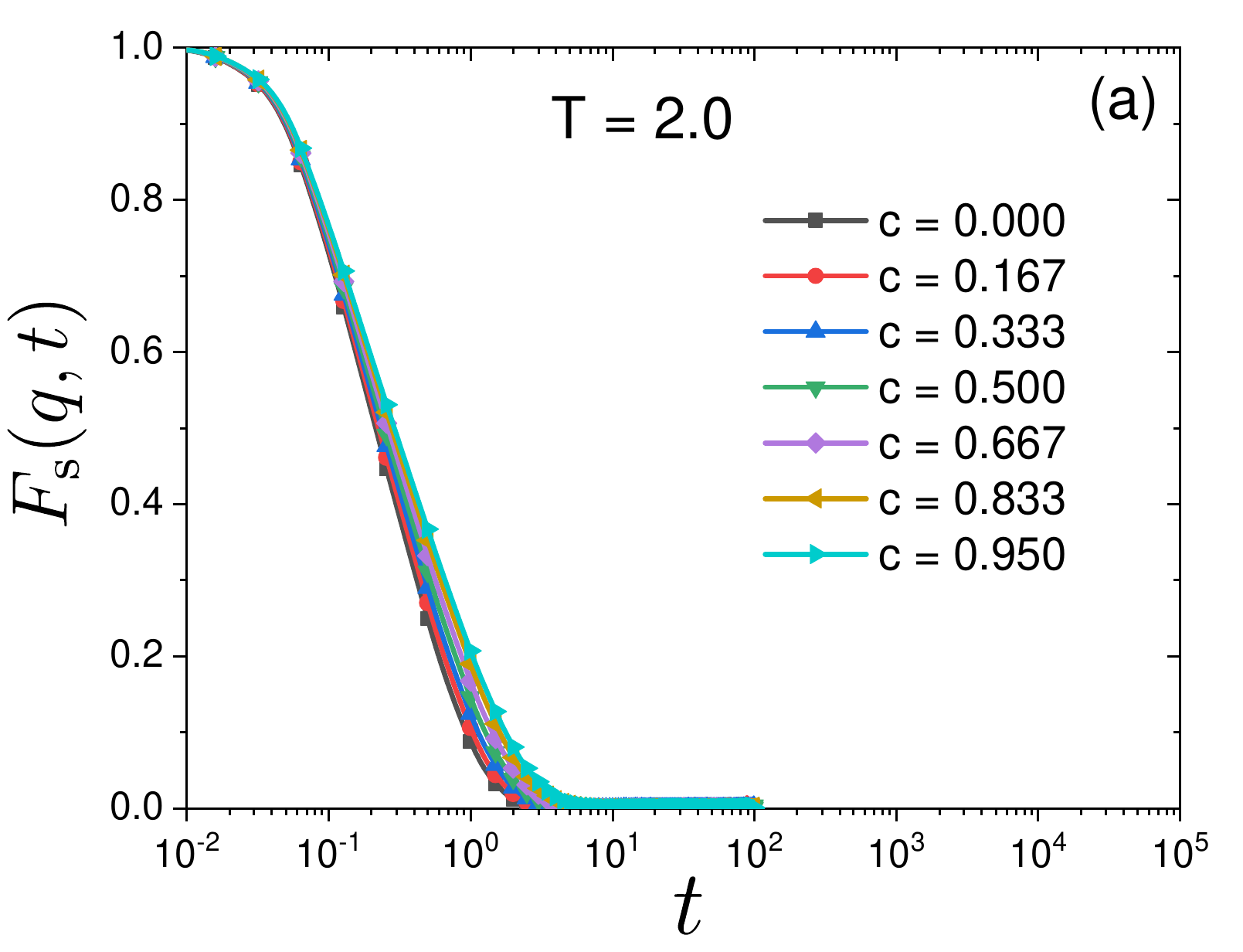}
\includegraphics[width=0.48\columnwidth]{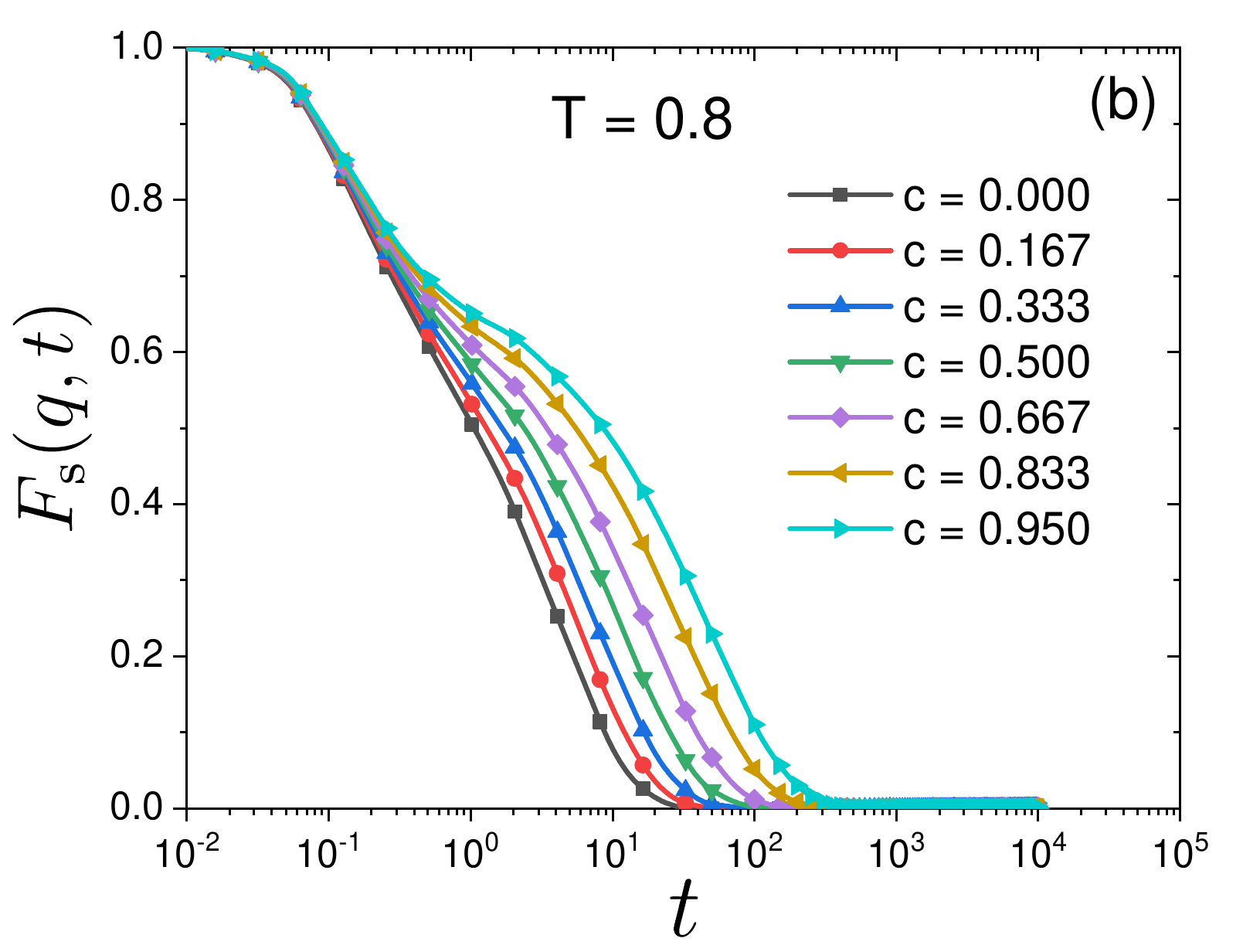}
\includegraphics[width=0.48\columnwidth]{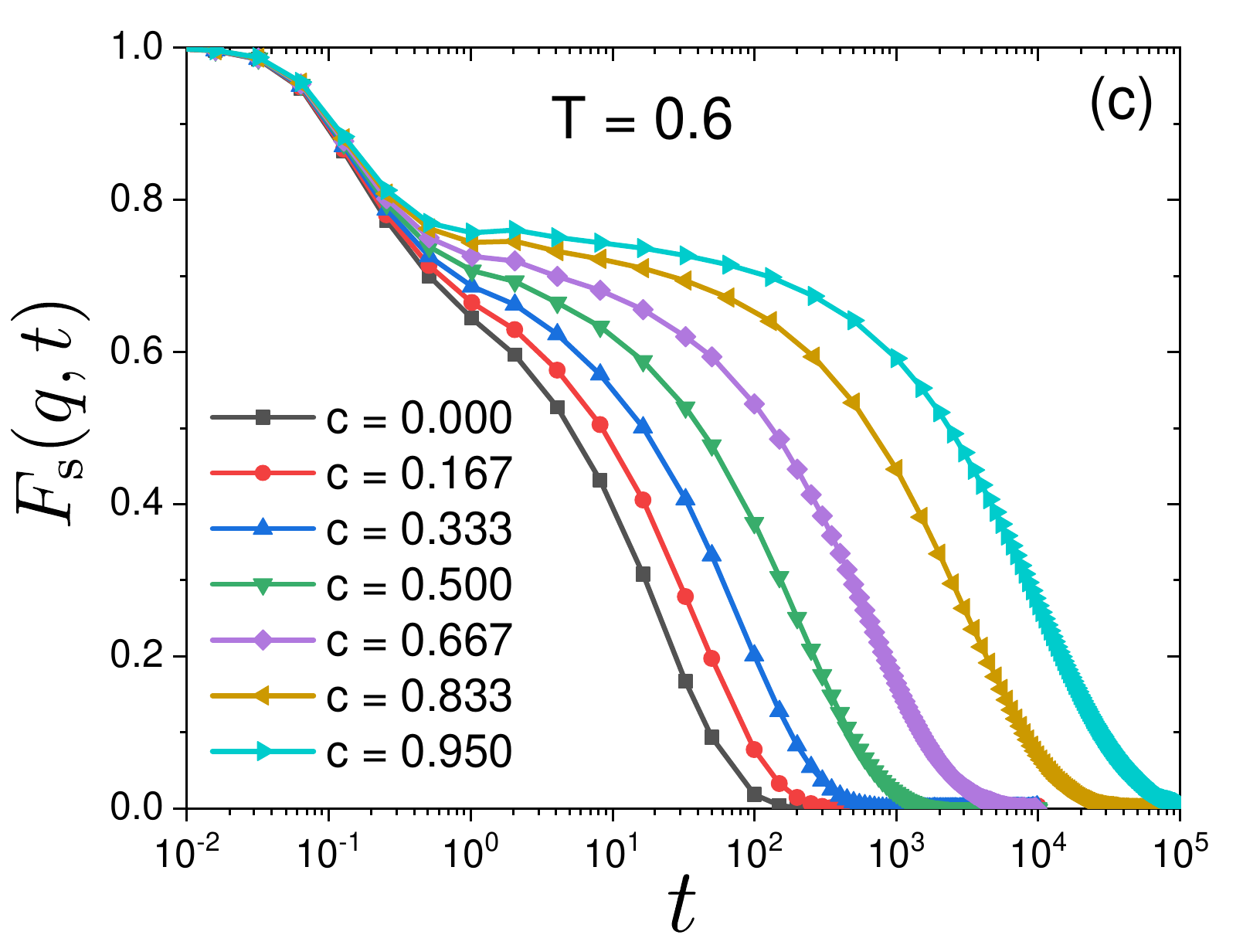}
\includegraphics[width=0.48\columnwidth]{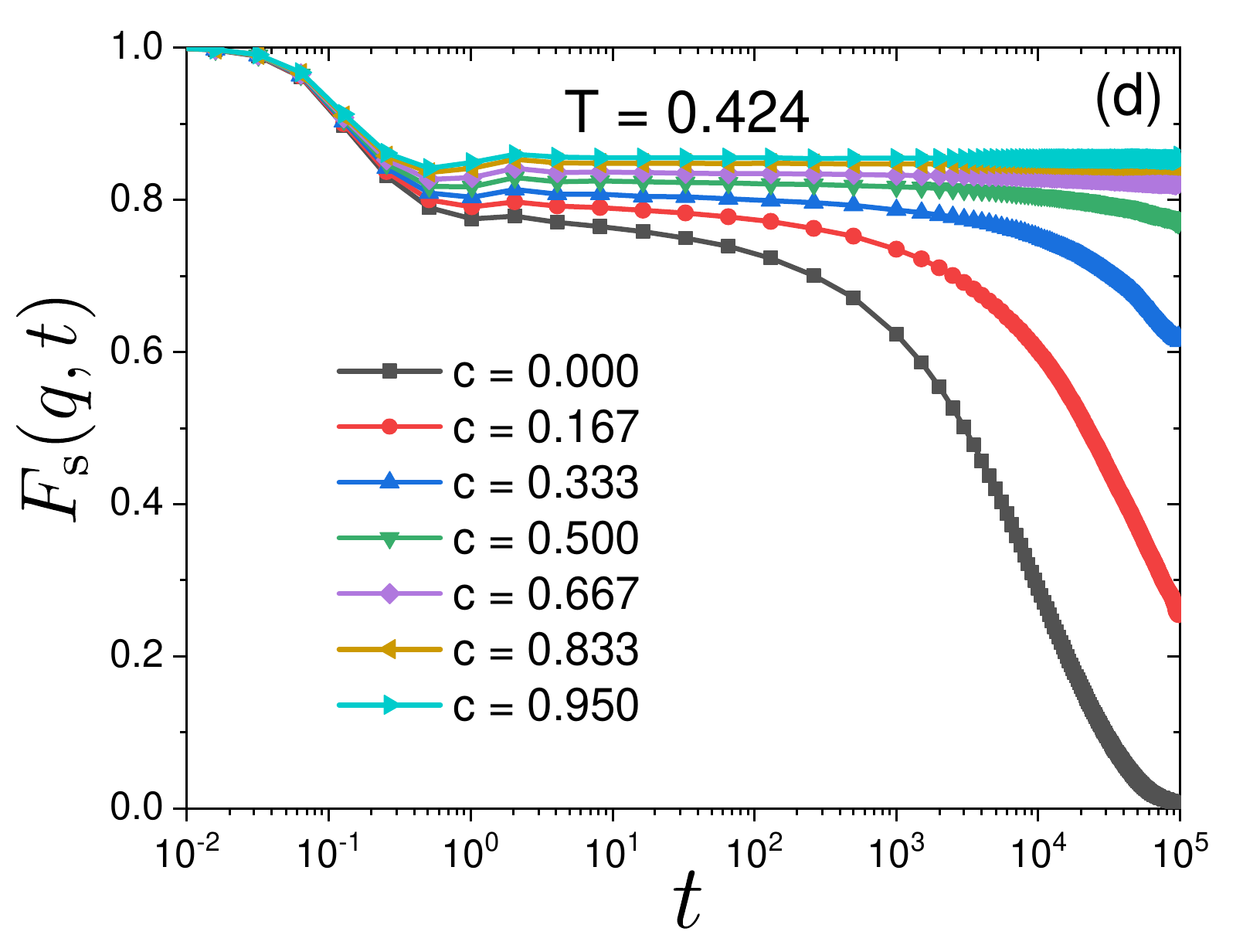}
\caption{{\bf Intermediate scattering function for various temperatures and dimer concentrations.}
$F_{\rm s}(q,t)$ for all particles for several values of $c$ for $T=2.0$ (a), $T=0.8$ (b), $T=0.6$ (c), and $T=0.424$ (d).
}
\label{fig:different_T}
\end{figure}

The temperature and $c-$dependence of $F_{\rm s}(q,t)$ is presented in Fig.~\ref{fig:different_T}. 
We find that the influence of bonding is very small if temperature is high, $T=2.0$, panel (a). 
However, once $T$ is decreased, panels (b)-(d), the bonding affects the dynamics very strongly, akin to the behavior of randomly pinned systems~\cite{jack2013dynamical,chakrabarty2016understanding}.
For example, at the lowest temperature $T=0.424$, one can see that for $c> 0.5$ the dynamics is completely frozen on the timescale of our simulation, demonstrating that the random bonding allows to access an extremely slow glassy dynamics in almost equilibrium. 
By using the data for $\tau_\alpha$ (see Fig.~\ref{fig:tau_scaling}(a) below), we estimate that the bonded system at $T=0.424$ and $c=0.95$ (that are prepared by making bonds from the original $T=0.424$ configurations) has an equilibrium relaxation time $\tau_\alpha \approx 10^{12}$, thus about a factor of $10^7$ larger than the largest $\tau_\alpha$ accessed in our simulations~\cite{Ozawa2023}. This demonstrates that the random bonding protocol indeed provides us with a huge gain in terms of computational time for the preparation of the initial equilibrium state.

\begin{figure}[t]
\includegraphics[width=0.5\columnwidth]{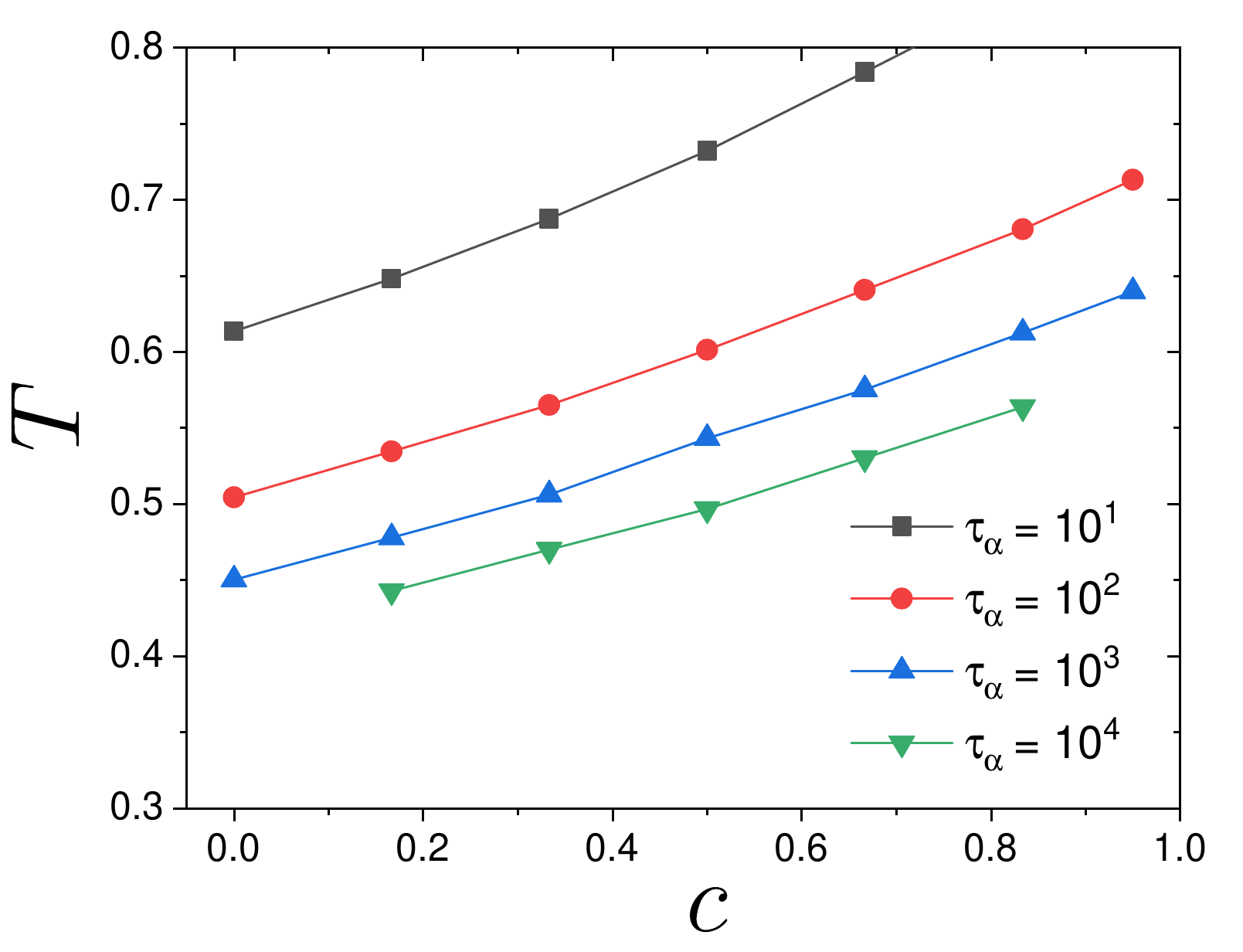}
\caption{{\bf Relaxation time as a function of temperature $T$ and dimer concentration $c$.} 
We report iso-$\tau_{\alpha}$ curves in the $T-c$ plane, obtained from the dynamical correlations shown in Fig.~\ref{fig:different_T}.
}
\label{fig:T_vs_c}
\end{figure}

\begin{figure}[t]
\includegraphics[width=0.48\columnwidth]{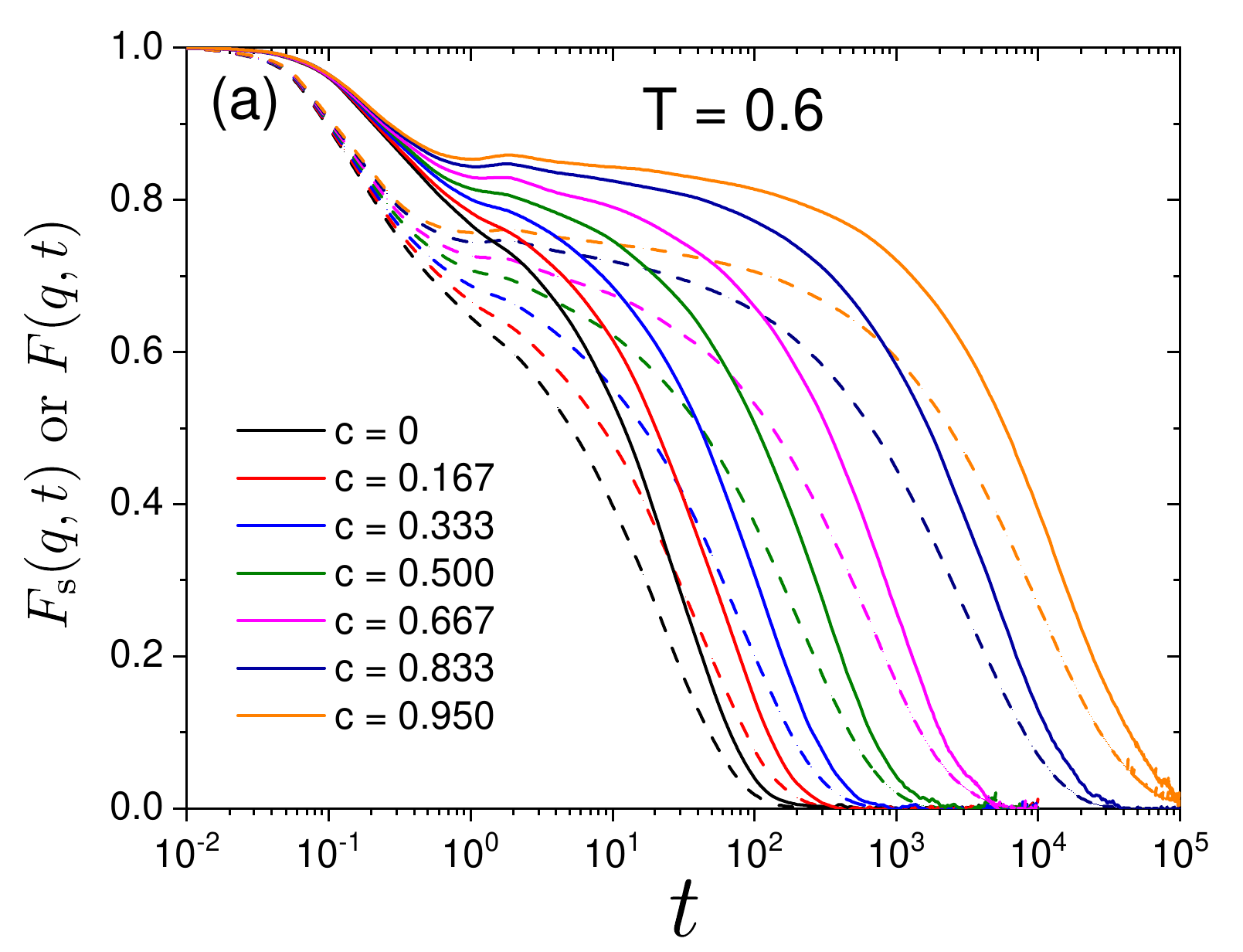}
\includegraphics[width=0.48\columnwidth]{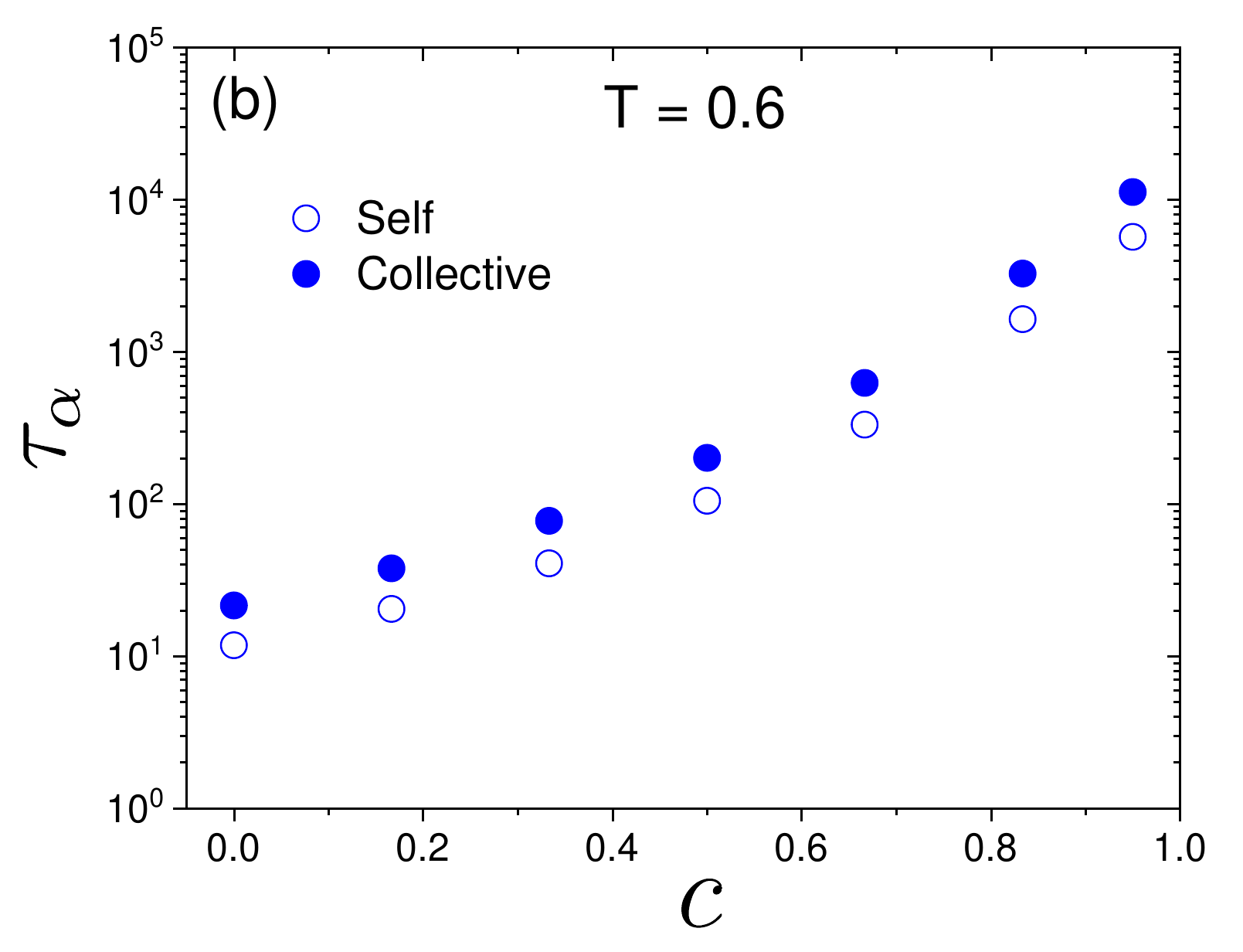}
\caption{
{\bf Comparison of the self and collective dynamics.}
(a):~Self (dashed curves) and collective (solid curves) intermediate scattering functions at $T=0.6$ for different $c$.
(b):~Relaxation time $\tau_{\alpha}$ versus $c$, computed from the data in (a).
}
\label{fig:self_vs_collective}
\end{figure}

In order to see the influence of $T$ and $c$ together we present in Fig.~\ref{fig:T_vs_c} the iso-$\tau_{\alpha}$ curves in the $T-c$ plane. These curves increase with increasing $c$, which is again similar to the results found for randomly pinned systems~\cite{kob2013probing,ozawa2015equilibrium}. The shape of the curves depends only mildly on $\tau_\alpha$, hinting at a simple functional relation between $\tau_\alpha$, $T$ and $c$. This point will be investigated in more detail below.

Finally, we present a comparison between the self and collective parts of the intermediate scattering functions. 
Previous studies have reported that in randomly pinned fluids the collective part shows apparent freezing while the self part did not, rendering the analysis of the system dynamics difficult~\cite{charbonneau2013decorrelation,ozawa2015equilibrium,chakrabarty2016understanding}. 
In Fig.~\ref{fig:self_vs_collective}(a), we show the self and collective parts for the randomly-bonded glass formers at $T=0.6$ for different values of $c$.
One recognizes that, in contrast to the pinned systems, the collective part also relaxes to zero and that the relaxation time is slightly larger than the one for the self part, at least for the wave-vector considered. 
Interestingly, however, the ratio between the two timescales is about a factor two irrespectively of $c$, as shown in Fig.~\ref{fig:self_vs_collective}(b). This suggests that self and collective correlators do not decouple and therefore we can conclude that the self part gives reliable dynamical information about the system.

\subsection{Dynamical scaling}

\begin{figure}[t]
\includegraphics[width=0.49\columnwidth]{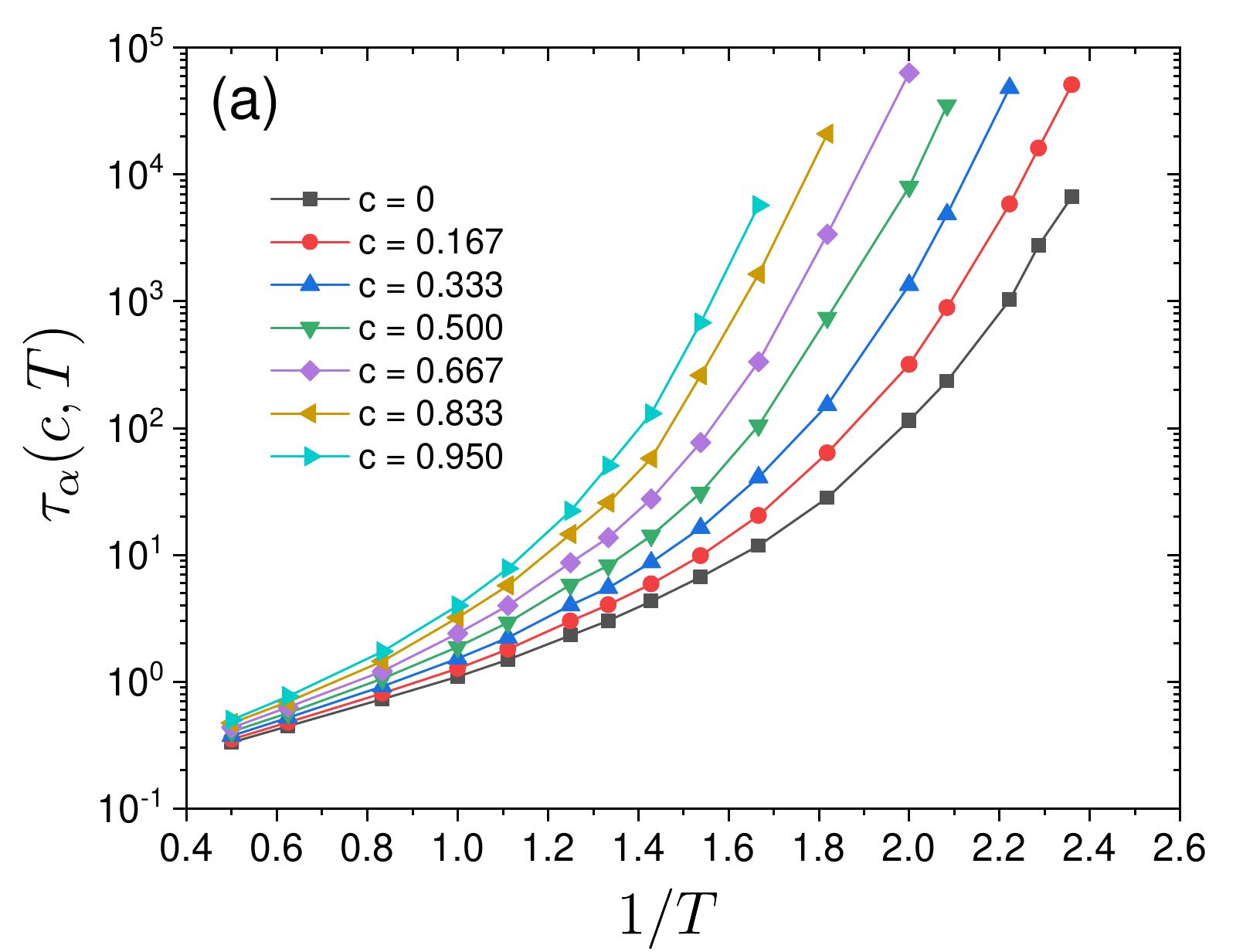}
\includegraphics[width=0.49\columnwidth]{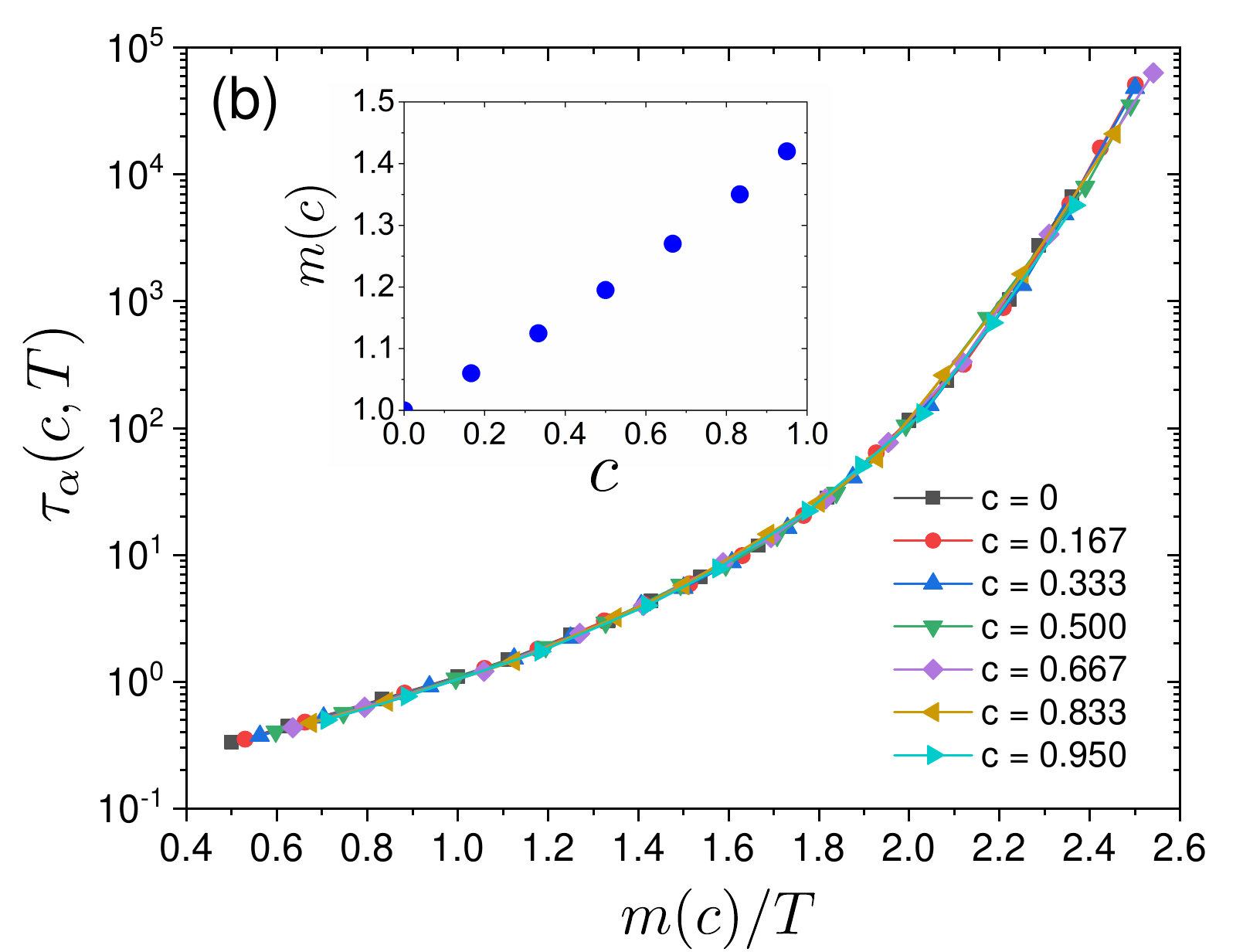}
\caption{{\bf Dynamical scaling of the relaxation time with dimer concentration.}
(a): Relaxation time $\tau_\alpha(c, T)$ obtained from the self-intermediate scattering function for all particles.
(b): The same data using a normalized abscissa, $m(c)/T$. The inset shows $m(c)$ versus $c$.
}
\label{fig:tau_scaling}
\end{figure}

In the following we use a scaling analysis to examine how $\tau_\alpha$ depends on $T$ and $c$.
In Fig.~\ref{fig:tau_scaling}(a), we present an Arrhenius plot for the structural relaxation time $\tau_\alpha(c,T)$ for different $c$. 
For the $c=0$ system, which we will call the ``original'' system, 
$\tau_\alpha$ follows the well-known non-Arrhenius temperature dependence~\cite{kob1995testing}: 
$\tau_\alpha(c=0,T)=\tau_0 \exp\left[\frac{E(c=0, T)}{T}\right]$, where $E(c=0, T)$ is a $T$-dependent activation energy accounting for the non-Arrhenius behavior.
With increasing $c$, $\tau_\alpha(c,T)$ increases as expected, which implies that the activation energy $E(c,T)$, defined by $\tau_\alpha(c, T)=\tau_0 \exp\left[\frac{E(c, T)}{T}\right]$, grows due to the addition of bonds.
Interestingly, we find that the temperature dependence of $\tau_\alpha(c, T)$ can be rescaled by an unknown function $m(c)$, namely, $\tau_\alpha(c, T)=\tau_\alpha(c=0, T/m(c))$, as shown in Fig.~\ref{fig:tau_scaling}(b). 
The inset shows $m(c)$, which has been determined manually for each $c$ such that $\tau_\alpha(c, T)$ superimposes with $\tau_\alpha(c=0, T)$. One sees that to a first approximation $m(c)$ is linear, but a slight upward bending can be noticed.
The existence of a master curve is so far empirical, and it holds at least for the $T$-and $c$-range probed by our simulations. Yet the scaling suggests that all relevant temperature scales, such as the mode-coupling crossover $T_{\rm mct}$~\cite{gotze2009complex} and the Kauzmann transition temperature $T_{\rm K}$~\cite{cammarota2023kauzmann} (if it exists) for the original system ($c=0$), are just scaled by $m(c)$.
This implies that the fragility of randomly-bonded glass-forming liquids does not change inherently across different $c$, in contrast to randomly pinned systems~\cite{kim2011slow,chakrabarty2015dynamics}. 
This scaling for the randomly bonded systems also implies that $E(c,T)=m(c)E(c=0, T/m(c))$. 
On the other hand, in the random pinning case, it was argued that $E(c,T)=q(c)E(c=0, T)$, where $q(c)$ is an increasing function of $c$ with $q(c=0)=1$~\cite{chakrabarty2016understanding}, which explains the fact that fragility decreases with increasing $c$.
Moreover, Ref.~\cite{chakrabarty2016understanding} argued that if $E(c=0, T)$ has a singularity at a finite $T_{\rm K}$, e.g., $E(c=0, T)/T \sim 1/(T-T_{\rm K})$, pinned systems at $c>0$ inherit the same singular temperature $T_{\rm K}$ irrespective of $c$.
Comparing these results with the $(T,c)$-dependence of our bonded system, one can conclude that the dynamics of randomly bonded glass-forming liquids is qualitatively quite different from the one of randomly pinned systems also in terms of the temperature dependence of structural relaxation.

\begin{figure}[t]
\includegraphics[width=0.49\columnwidth]{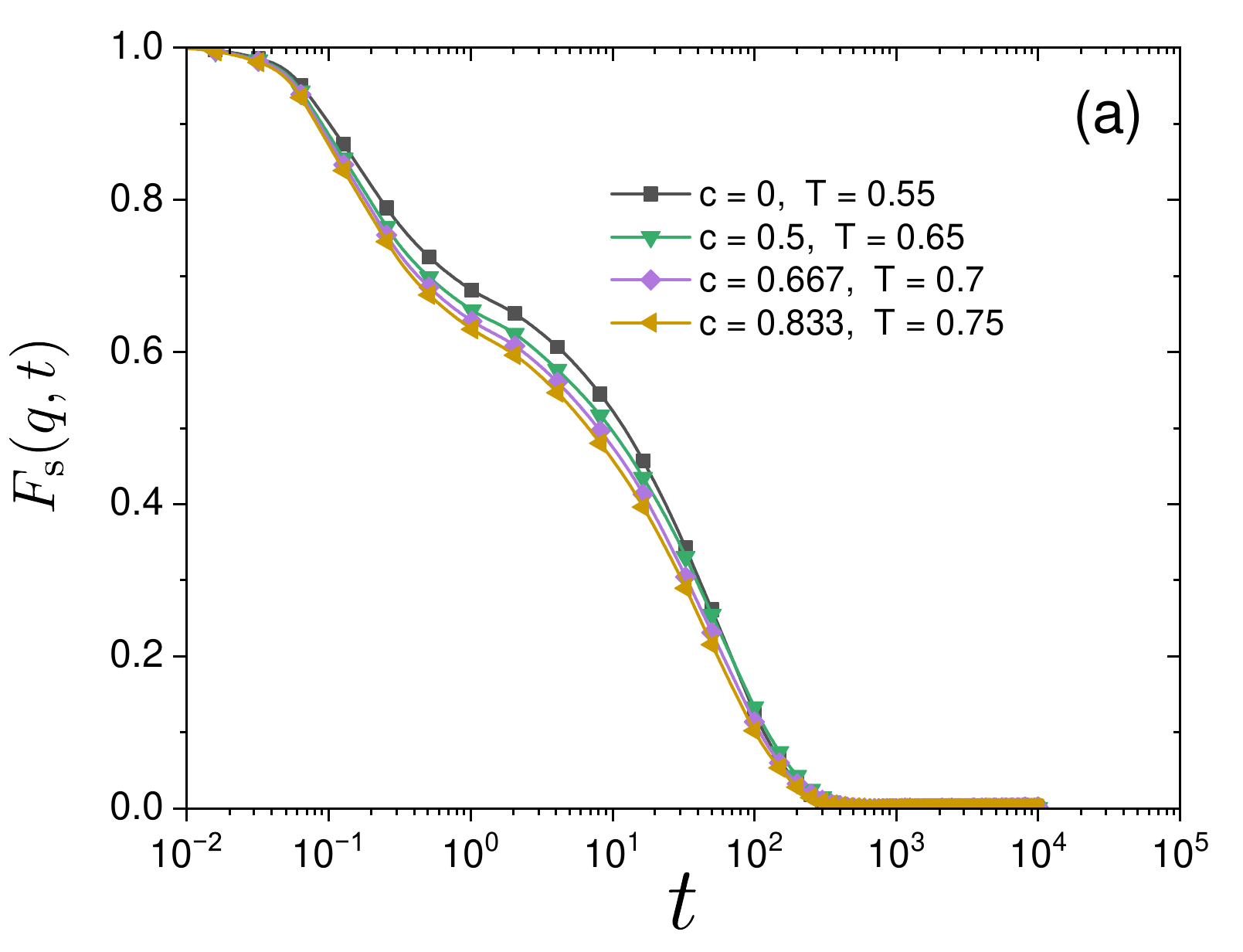}
\includegraphics[width=0.49\columnwidth]{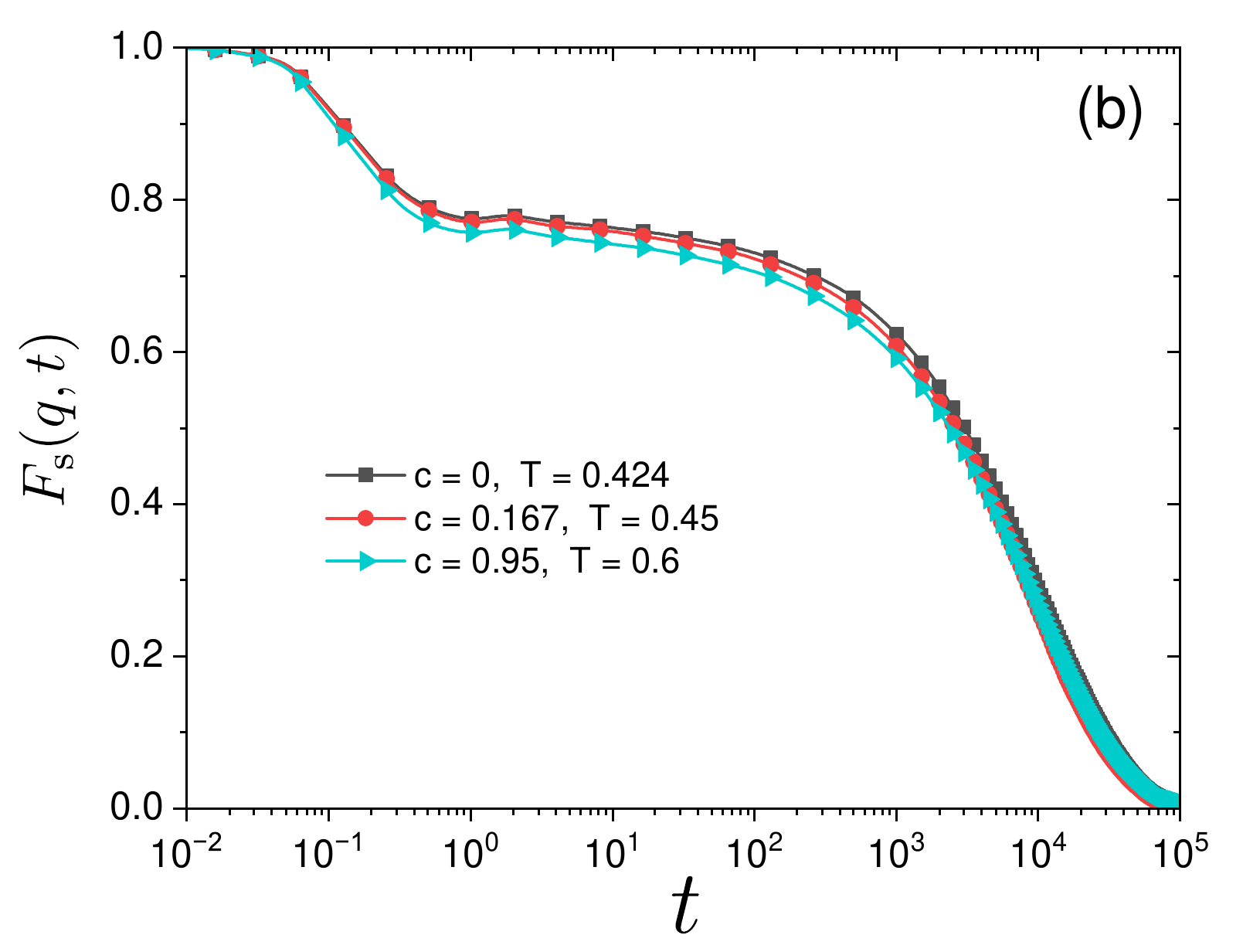}
\caption{
{\bf Comparison of dynamical correlations at corresponding state points under dynamical scaling.}
$F_{\rm s}(q,t)$ for different state points with similar $T/m(c)$. (a): $T/m(c) \approx 0.55$. (b): $T/m(c) \approx 0.42$.
}
\label{fig:superimpose}
\end{figure}

The dynamical scaling shown in Fig.~\ref{fig:tau_scaling}(b) reminds us of simple systems where structure and dynamics are invariant to a good approximation along isomorphs in the phase diagram~\cite{gnan2009pressure,schroder2014simplicity}. In these systems, the dynamics at different state points can be rescaled by a uniform scaling of space and time.
In contrast, the randomly bonded systems introduce strong constraints in the system (in the form of bonds) that alter dynamical relaxation processes significantly as $c$ is increased. The origin of the observed empirical scaling must then be different from the isomorph invariance.
Figure~\ref{fig:superimpose} shows $F_{\rm s}(q,t)$ having similar $T/m(c)$. Although the relaxation time $\tau_\alpha$ is similar, there is a trend that the correlators with higher $c$ show a lower plateau compared to those with smaller $c$. This suggests that the observed scaling collapse  cannot be understood as a simple uniform space-time rescaling. Further investigations are needed to understand the origin of the empirical dynamical scaling.

\section{Dynamical heterogeneity}

The dynamics of glassy liquids is accompanied by strong dynamical heterogeneities, the intensity of which grows with decreasing temperature~\cite{hurley1995kinetic,kob1997dynamical,yamamoto1998dynamics,karmakar2009growing}. Since our bonding procedure allows to generate configurations in the deeply supercooled regime, we can thus access these heterogeneities in the randomly bonded glass-forming liquids in (nearly) thermal equilibrium. 
Since bonded dimer systems involve rotational motion as an additional relaxation channel, we consider dynamical heterogeneities not only for the positional degrees of freedom but also the rotational ones.

\subsection{Positional degrees of freedom}

\begin{figure}[htbp]
\includegraphics[width=0.48\columnwidth]{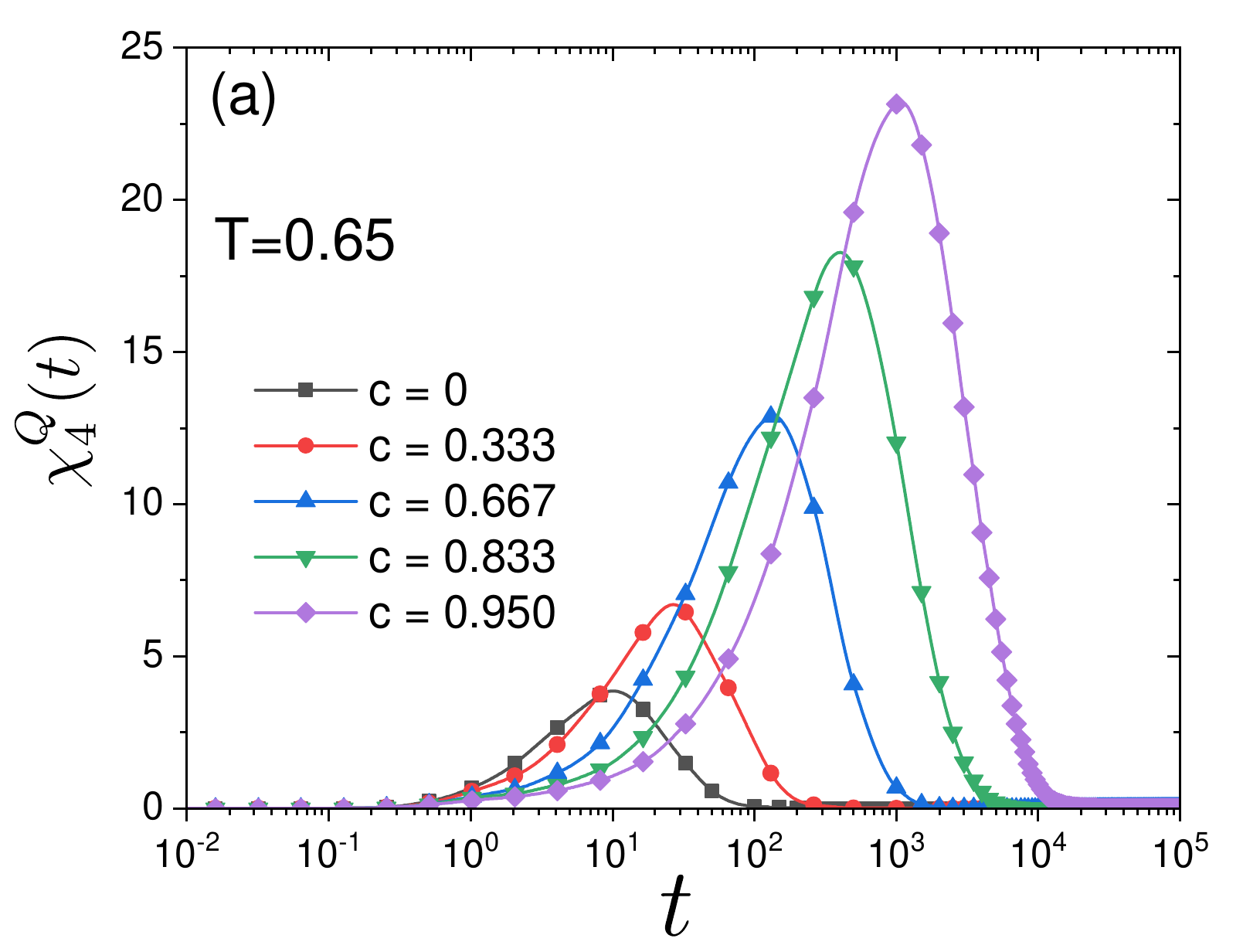}
\includegraphics[width=0.48\columnwidth]{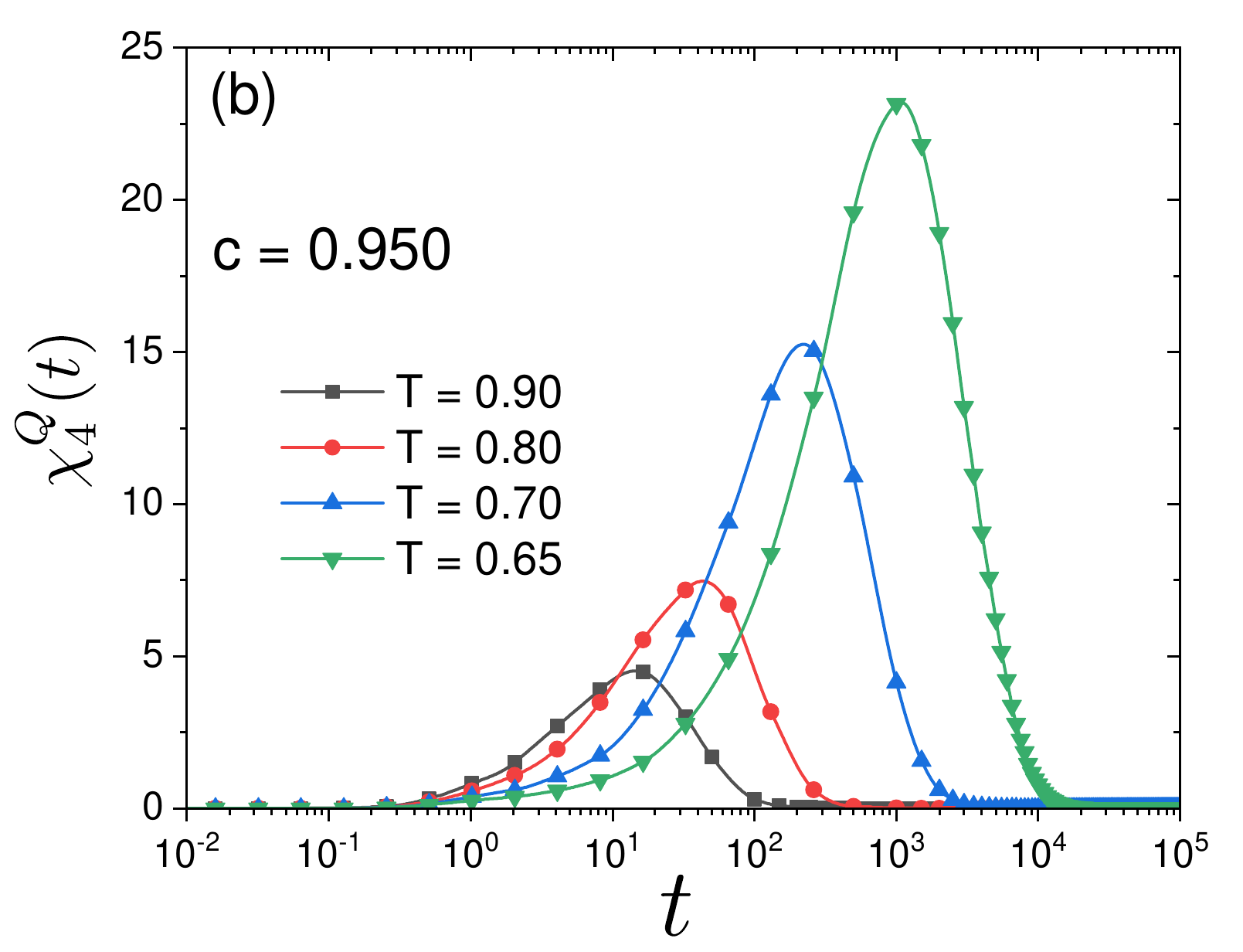}
\caption{
{\bf Dynamical heterogeneity of positional degrees of freedom.}
(a): Four-point correlation function $\chi_4^Q(t)$ associated with positional degrees of freedom computed from the overlap function $\widehat Q$ at a constant temperature ($T=0.65$), panel (a), and at different $T$ for constant dimer concentration $c=0.950$, panel (b).}
\label{fig:chi4_Fskt}
\end{figure}

First, we compute the standard four-point correlation function $\chi_4^Q(t)$ associated with positional degrees of freedom, which is given by
\begin{equation}
    \chi_4^Q (t) = N \overline{\left( \left\langle \widehat Q^2(t) \right\rangle_{\tilde{r}^{N_{\rm d}}} - \left\langle \widehat Q(t) \right\rangle^2_{\tilde{r}^{N_{\rm d}}} \right)},
    \label{eq:chi4_Fskt1}
\end{equation}
where $\widehat Q(t)= \frac{1}{N} \sum_{i=1}^N \theta(a- |{\bf r}_i(t)-{\bf r}_i(0)|)$ is an overlap function taking into account all particles and $\theta(x)$ is the Heaviside step function~\cite{donati2002theory}. We set the distance $a$ to the often used value $a=0.3$. 
We note that $\chi_4^Q(t)$ defined in Eq.~(\ref{eq:chi4_Fskt1}) does not contain contributions from sample-to-sample fluctuations associated with different realizations of bonds~\cite{kob2014nonlinear}.
Figure~\ref{fig:chi4_Fskt} shows $\chi_4^Q(t)$ for different values of  $c$ at a constant temperature, panel (a), and with decreasing $T$ at a constant $c$, panel (b).
We find systematic growth with increasing glassiness in both cases, which is in contrast to randomly pinned particle systems. 
It has been reported that the four-point correlation function of randomly pinned particle systems does not grow systematically or decreases approaching glass transition, while the relaxation time increases significantly~\cite{kim2011slow,jack2013dynamical,kob2014nonlinear,li2015decoupling}. This difference in the dynamical behavior is directly related to the fact that in pinned systems the size of the dynamical heterogeneities is hindered by the presence of the pinned particles, while in the present system the heterogeneities can grow unhindered.

\begin{figure}[htbp]
\includegraphics[width=0.48\columnwidth]{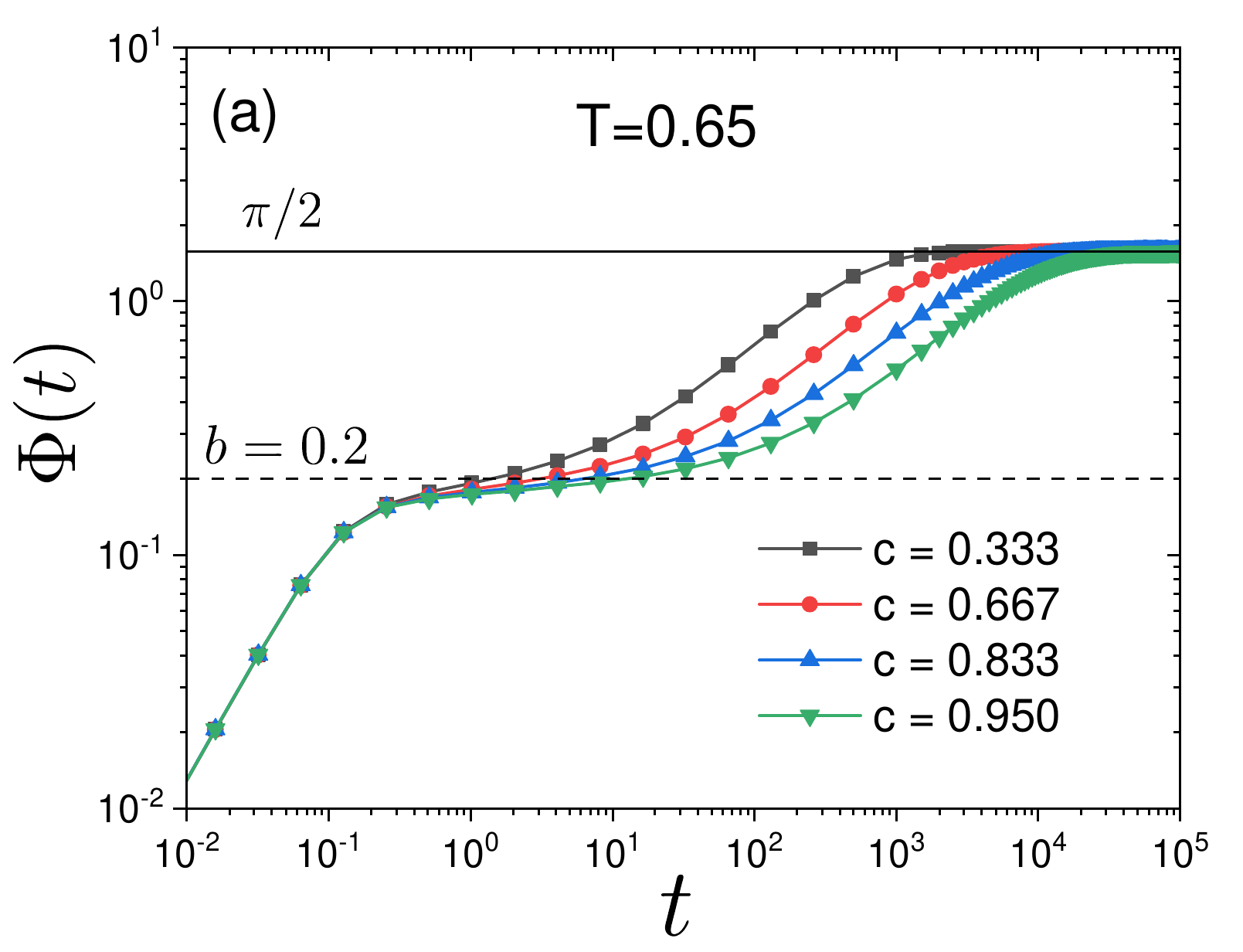}
\includegraphics[width=0.48\columnwidth]{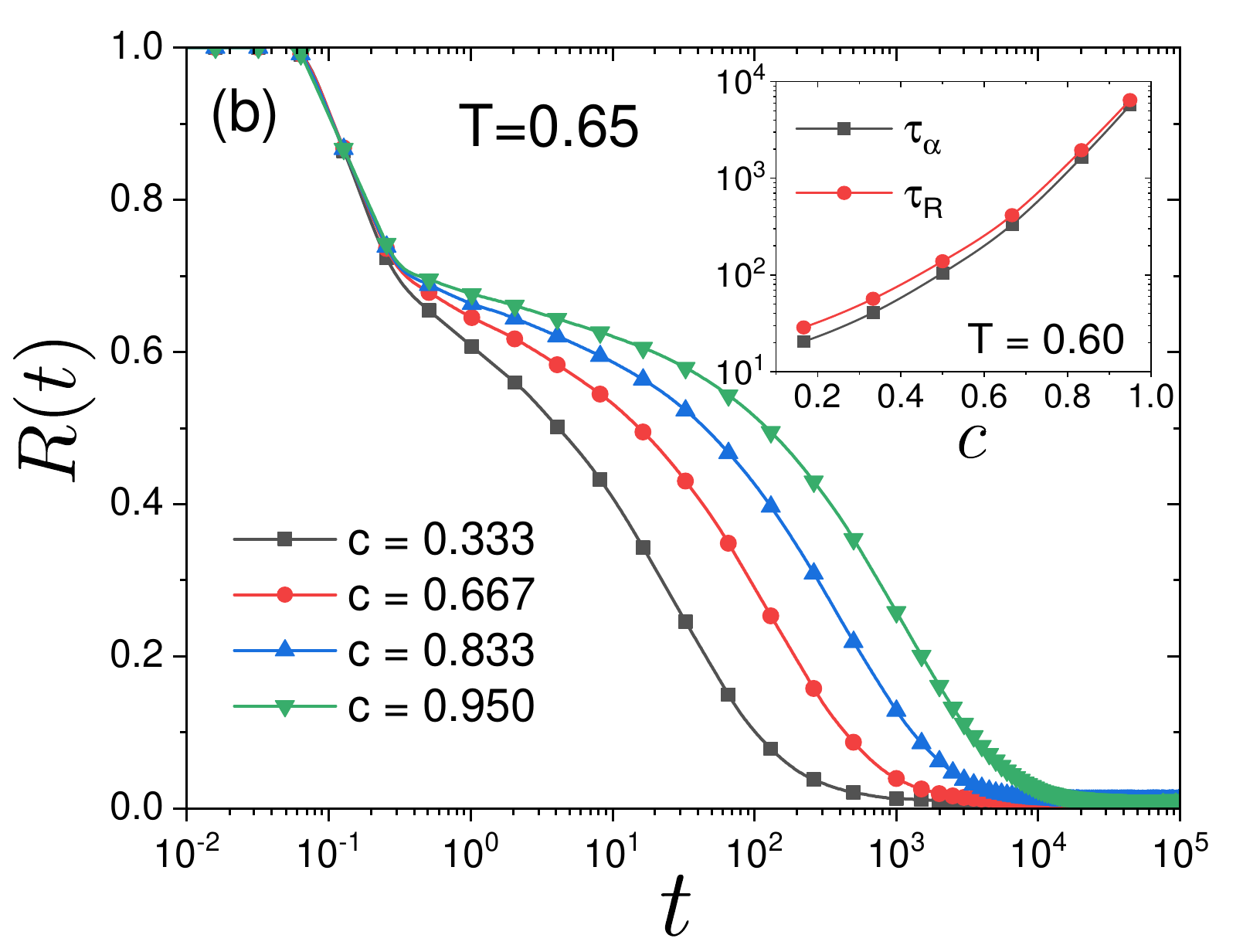}
\includegraphics[width=0.48\columnwidth]{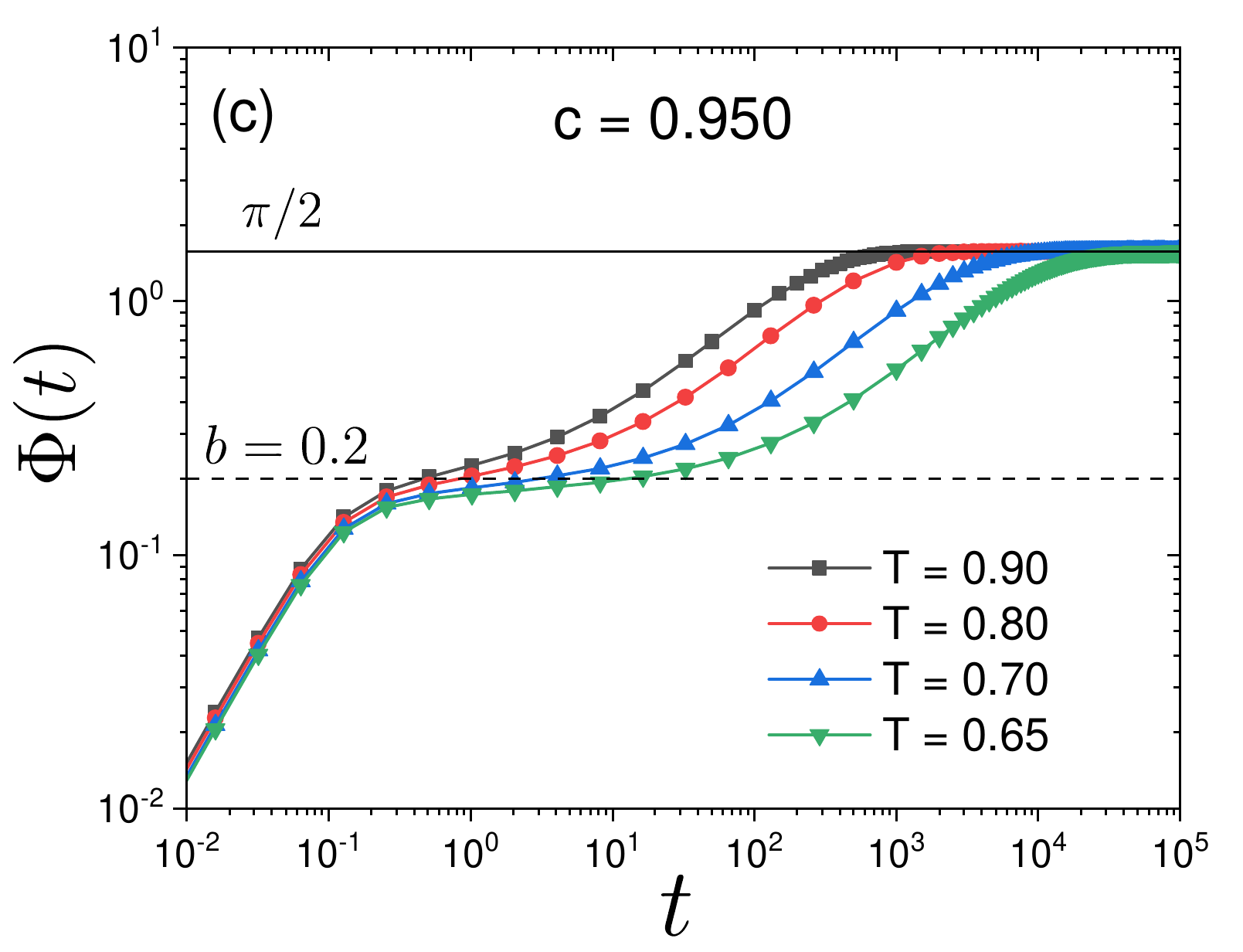}
\includegraphics[width=0.48\columnwidth]{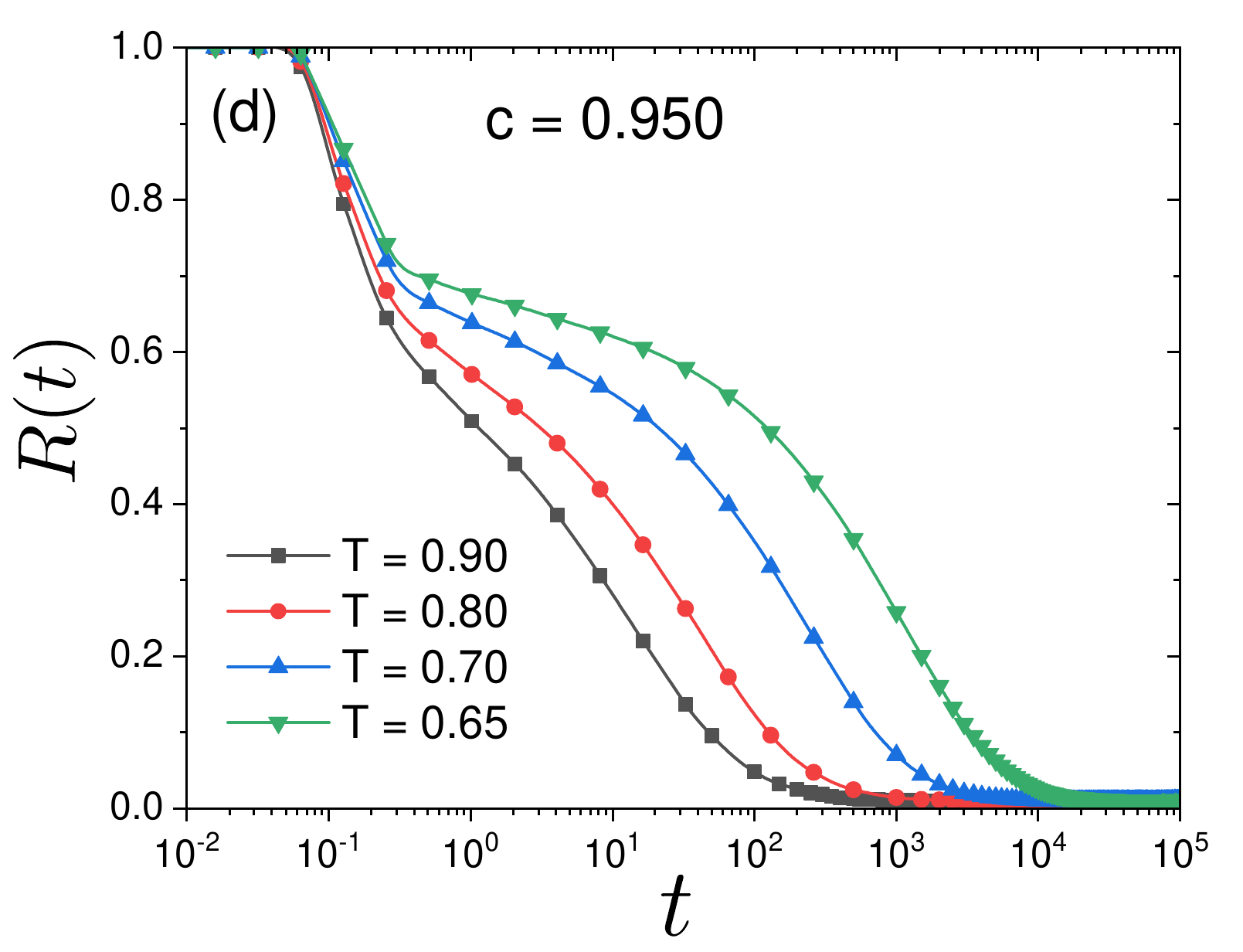}
\caption{
{\bf Dynamical correlations of the orientational degrees of freedom.}
(a, c): Mean rotational angle $\Phi(t)$ for (a) a constant temperature ($T=0.65$) while varying $c$ and (c) a constant concentration ($c=0.95$) varying $T$. The horizontal dashed line is at $\Phi(t)=0.2$, the threshold used to define the overlap function $R(t)$.
(b, d): The corresponding rotational overlap function $R(t)$ for (b) constant $T$ and (d) constant $c$. The inset in (b) compares the relaxation times $\tau_R$ measured by $R(\tau_R)=0.3$ and $\tau_\alpha$ measured by $F_s(q,\tau_\alpha)=1/e$ (computed from all particles) for $T=0.60$.
}
\label{fig:rotation_angle}
\end{figure}

\subsection{Rotational degrees of freedom}

We next consider the rotational degrees of freedom for the dimers.
We first compute the average dynamics by the mean rotational angle~\cite{shiraishi2023johari}, which is given by 
\begin{equation}
    \Phi(t) = \overline{ \left\langle \frac{1}{ N_{\rm d}} \sum_{\substack{i \in \mathcal{D}}} \phi_{i}(t) \right\rangle_{\tilde{r}^{N_{\rm d}}} } \quad,
\end{equation}
where $\phi_i(t)=\arccos\left({\bf n}_i(t) \cdot {\bf n}_i(0)\right)$. Figure~\ref{fig:rotation_angle}(a) shows $\Phi(t)$ for different $c$ at constant temperature. One finds that the correlator has a two-step relaxation with a plateau on an intermediate timescale, akin to the mean-squared displacement.
At sufficiently long times the correlator approaches the asymptotic value $\pi/2$, which is expected when ${\bf n}_i(t=0)$ and ${\bf n}_i(t \to \infty)$ are uncorrelated.
In short, $\Phi(t)$ can separate vibrational motion and structural rearrangement in terms of rotational relaxation. Besides, the separation becomes more distinct when $c$ is increased, which is also in qualitative agreement with the mean-squared displacement.

We can now define a mean overlap function associated with the rotational motion via
\begin{equation}
    R(t) = \overline{ \left\langle \frac{1}{ N_{\rm d}} \sum_{\substack{i \in \mathcal{D}}} \theta \left(b-\phi_{i}(t) \right) \right\rangle_{\tilde{r}^{N_{\rm d}}} } \quad ,
\end{equation}
where $b$ is a threshold separating vibrational motion and structural relaxation. In practice we have chosen the value $b=0.2$, see Fig.~\ref{fig:rotation_angle}(a).
Figure~\ref{fig:rotation_angle}(b) shows that $R(t)$ presents a two-step relaxation, similar to $F_s(q,t)$. 
This correlation function allows to compute a characteristic timescale for rotational relaxation, $\tau_R$, defined by $R(\tau_R)=0.3$, which is presented in the inset, together with $\tau_\alpha$ obtained by $F_s(q,\tau_\alpha)=1/e$.
Both $\tau_R$ and $\tau_\alpha$ track each other very well, particularly at larger $c$, suggesting that positional and rotational relaxations are strongly coupled in the deep glassy regime.
If instead of varying the concentration of the dimers one changes the temperature, one find qualitatively the same glassy slowing down phenomenology, see Figs.~\ref{fig:rotation_angle}(c, d).

\begin{figure}[htbp]
\includegraphics[width=0.48\columnwidth]{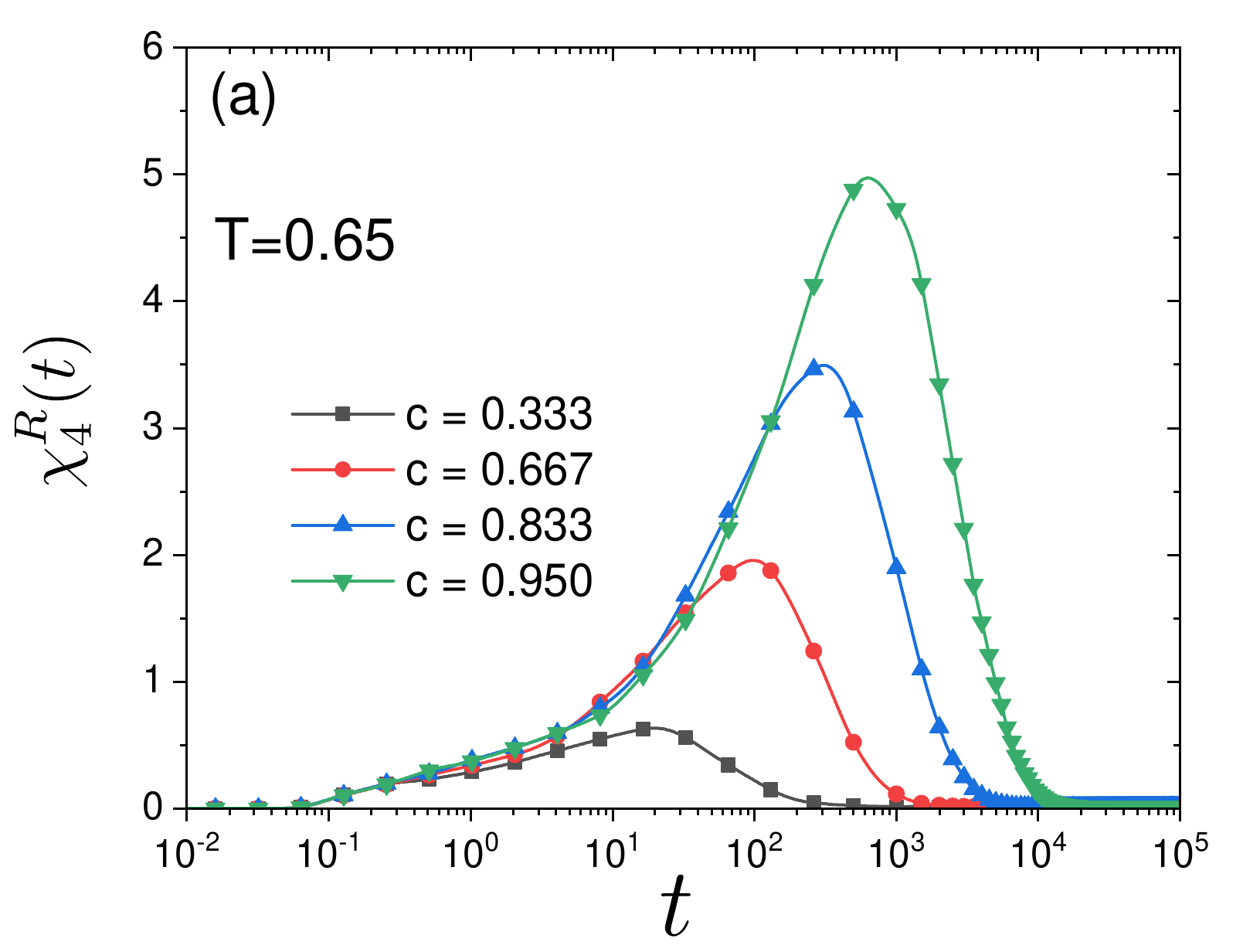}
\includegraphics[width=0.48\columnwidth]{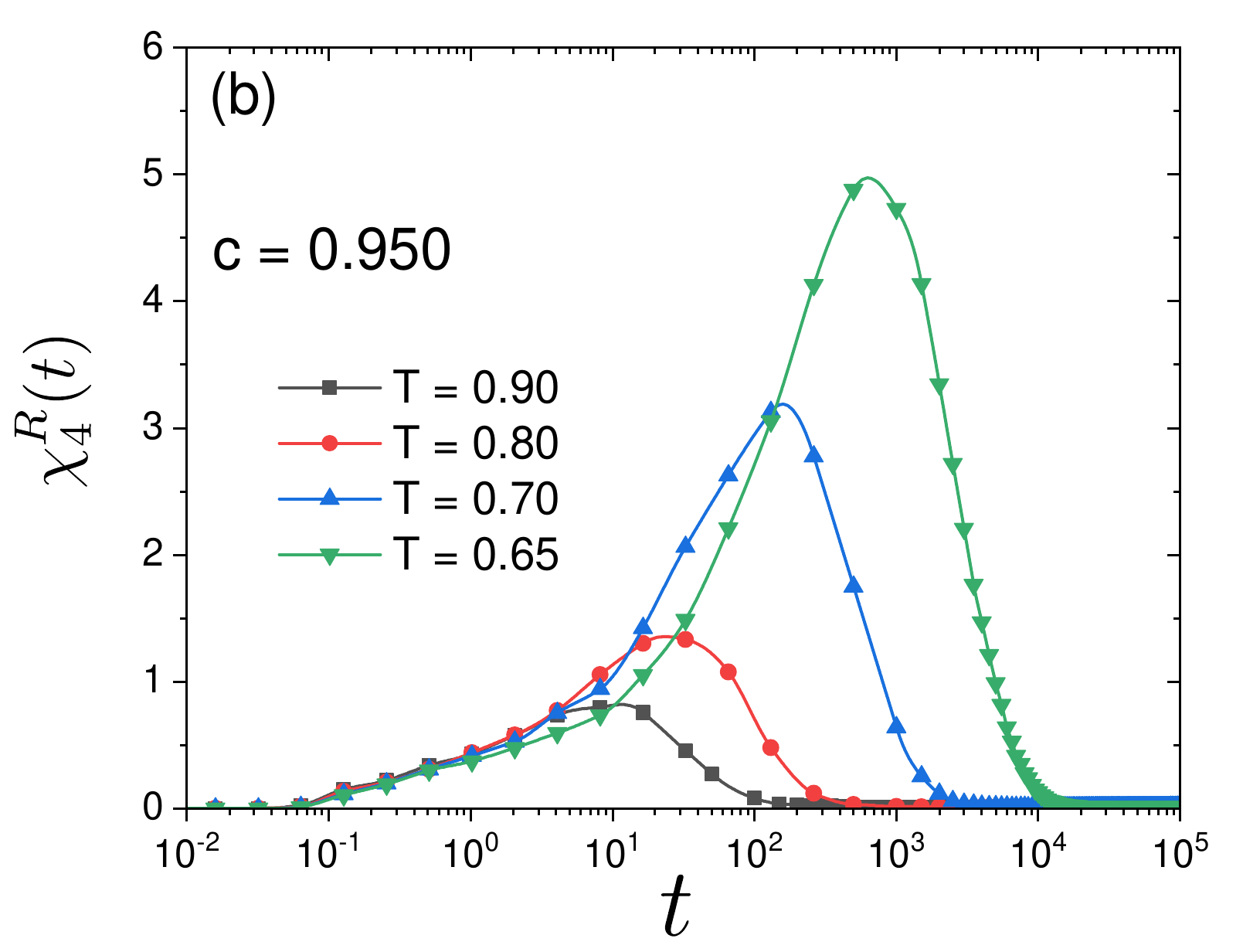}
\caption{{\bf Dynamical heterogeneity of the orientational degrees of freedom.} (a, b): Four-point correlation function $\chi_4^R(t)$ associated with the rotational degrees of freedom computed from the overlap function $\widehat R$ at a constant temperature ($T=0.65$) while varying $c$, panel (a), and a constant concentration ($c=0.95$), varying $T$, panel (b).}
\label{fig:rotation_angle_chi4}
\end{figure}

Finally, we define the corresponding four-point correlation function associated with the rotational degrees of freedom by
\begin{equation}
    \chi_4^R (t) = N_{\rm d} \overline{\left( \left\langle \widehat R^2(t) \right\rangle_{\tilde{r}^{N_{\rm d}}} - \left\langle \widehat R(t) \right\rangle^2_{\tilde{r}^{N_{\rm d}}} \right)},
    \label{eq:chi4_Fskt}
\end{equation}
where $\widehat R(t)= \frac{1}{N_{\rm d}} \sum_{\substack{i \in \mathcal{D}}} \theta \left(b-\phi_{i}(t) \right)$. 
Figure~\ref{fig:rotation_angle_chi4}(a) shows the time evolution of $\chi_4^R(t)$ varying $c$ at fixed $T$.
We find that $\chi_4^R(t)$ grows systematically with increasing glassiness, i.e., here concentration of dimers.
We find the same trend when $T$ is decreased while $c$ is fixed, see Fig.~\ref{fig:rotation_angle_chi4}(b). We thus conclude that randomly-bonded glass-forming liquids demonstrate growing dynamical heterogeneities approaching the glass transition in terms of both positional and rotational degrees of freedom. 
We note that dynamical heterogeneities in rotational motions have so far not be studied widely in computer simulations~\cite{kawasaki2019classification,kou2018translational},
while these are relevant for most molecular experiments~\cite{cicerone95dynhet}.

\section{Conclusion and Discussion}

We have studied randomly bonded glass-forming liquids where pairs of neighbor particles chosen from an equilibrium configuration are bonded permanently. We confirmed theoretically and numerically that random bonding with a neighbor cut-off is not in strict equilibrium right after bonding. However, if one generates the bonds using a spherical cut-off as in Ref.~\cite{Ozawa2023}, the deviation from equilibrium is very small, and the aging process stops soon after the timescale of vibrations. Therefore this random bonding method can be used to probe the (almost) equilibrium dynamics of stable bonded glass-forming liquids deep inside the energy landscape. 

Our detailed computer simulations demonstrated that 1) there is no decoupling between self and collective correlation functions, 2)  fragility does not change by increasing the concentration of bonds, and 3) dynamical heterogeneity keeps growing with approaching the glass transition. All these features are thus in contrast to the behavior found in the dynamics of randomly pinned systems. These discrepancies are (partly)  related to the preservation of the translational invariance in the random bonding process, emphasizing the importance of the details on how the quenched disorder is generated.

Most previous studies on low-temperature glassy dynamics have been performed in simple spherical particle systems, whereas most real molecular liquids experiments characterize orientational relaxation processes probed by dielectric measurements, making a conceptual gap between the simulation and experiment.
Instead, our randomly-bonded system with rotational relaxation allows us to study phenomena that are relevant for real experiments. As a future investigation, this approach permits thus to measure various rotational observables and, e.g., test the validity of the Stoke-Einstein-Debye relation in the deeply supercooled state~\cite{tarjus1995breakdown,kawasaki2019spurious}. On the more theoretical side, it would be interesting to compute a phase diagram of randomly bonded glass formers based on the framework developed in Ref.~\cite{hall2003microscopic} since the obtained results will be useful to connect the dynamics of molecular systems to the ones of gels. 
On the more applied side, it would be extremely interesting to revisit a series
of random bonding protocols that are routinely used to prepare amorphous solids such as epoxy resins~\cite{johari1994dynamics,corcione2006temperature}, vitrimers~\cite{kloxin2013covalent,denissen2016vitrimers}, colloidal or emulsion clusters~\cite{duguet2011design,peng2013colloidal,mcmullen2022self,mcmullen2018freely,yuan2016synthesis,goldbart1996randomly,corcione2006temperature,carbas2014effect,corezzi2002bond,mereu2015interplay}, in order to understand to what extend these protocols give rise to equilibrated glass samples whose properties remain stable over time.

In conclusion, we emphasize that the bonding approach presented in this work is not limited to the creation of dimers, since it can easily be extended to trimers, oligomers, etc, and this in contrast to methods that have been proposed earlier. This freedom will thus permit in the future to study the (nearly) equilibrium properties of glass-forming systems at thermodynamic state points which have so far been inaccessible to simulations.

\section*{Acknowledgements}

We thank Y.Iwashita for many useful discussions and for previous collaboration on this subject~\cite{Ozawa2023}. We also thank J.Brujic, S.Corezzi, D.Fioretto, F.Sciortino, and E.Zaccarelli for useful discussions and support.

\bibliography{bonding.bib}

\begin{thebibliography}{92}%
\makeatletter
\providecommand \@ifxundefined [1]{%
 \@ifx{#1\undefined}
}%
\providecommand \@ifnum [1]{%
 \ifnum #1\expandafter \@firstoftwo
 \else \expandafter \@secondoftwo
 \fi
}%
\providecommand \@ifx [1]{%
 \ifx #1\expandafter \@firstoftwo
 \else \expandafter \@secondoftwo
 \fi
}%
\providecommand \natexlab [1]{#1}%
\providecommand \enquote  [1]{``#1''}%
\providecommand \bibnamefont  [1]{#1}%
\providecommand \bibfnamefont [1]{#1}%
\providecommand \citenamefont [1]{#1}%
\providecommand \href@noop [0]{\@secondoftwo}%
\providecommand \href [0]{\begingroup \@sanitize@url \@href}%
\providecommand \@href[1]{\@@startlink{#1}\@@href}%
\providecommand \@@href[1]{\endgroup#1\@@endlink}%
\providecommand \@sanitize@url [0]{\catcode `\\12\catcode `\$12\catcode
  `\&12\catcode `\#12\catcode `\^12\catcode `\_12\catcode `\%12\relax}%
\providecommand \@@startlink[1]{}%
\providecommand \@@endlink[0]{}%
\providecommand \url  [0]{\begingroup\@sanitize@url \@url }%
\providecommand \@url [1]{\endgroup\@href {#1}{\urlprefix }}%
\providecommand \urlprefix  [0]{URL }%
\providecommand \Eprint [0]{\href }%
\providecommand \doibase [0]{http://dx.doi.org/}%
\providecommand \selectlanguage [0]{\@gobble}%
\providecommand \bibinfo  [0]{\@secondoftwo}%
\providecommand \bibfield  [0]{\@secondoftwo}%
\providecommand \translation [1]{[#1]}%
\providecommand \BibitemOpen [0]{}%
\providecommand \bibitemStop [0]{}%
\providecommand \bibitemNoStop [0]{.\EOS\space}%
\providecommand \EOS [0]{\spacefactor3000\relax}%
\providecommand \BibitemShut  [1]{\csname bibitem#1\endcsname}%
\let\auto@bib@innerbib\@empty
\bibitem [{\citenamefont {Ediger}\ \emph {et~al.}(1996)\citenamefont {Ediger},
  \citenamefont {Angell},\ and\ \citenamefont {Nagel}}]{ediger1996supercooled}%
  \BibitemOpen
  \bibfield  {author} {\bibinfo {author} {\bibfnamefont {M.~D.}\ \bibnamefont
  {Ediger}}, \bibinfo {author} {\bibfnamefont {C.~A.}\ \bibnamefont {Angell}},
  \ and\ \bibinfo {author} {\bibfnamefont {S.~R.}\ \bibnamefont {Nagel}},\
  }\href@noop {} {\bibfield  {journal} {\bibinfo  {journal} {J. Phys. Chem.}\
  }\textbf {\bibinfo {volume} {100}},\ \bibinfo {pages} {13200} (\bibinfo
  {year} {1996})}\BibitemShut {NoStop}%
\bibitem [{\citenamefont {Debenedetti}\ and\ \citenamefont
  {Stillinger}(2001)}]{debenedetti2001supercooled}%
  \BibitemOpen
  \bibfield  {author} {\bibinfo {author} {\bibfnamefont {P.~G.}\ \bibnamefont
  {Debenedetti}}\ and\ \bibinfo {author} {\bibfnamefont {F.~H.}\ \bibnamefont
  {Stillinger}},\ }\href@noop {} {\bibfield  {journal} {\bibinfo  {journal}
  {Nature}\ }\textbf {\bibinfo {volume} {410}},\ \bibinfo {pages} {259}
  (\bibinfo {year} {2001})}\BibitemShut {NoStop}%
\bibitem [{\citenamefont {Rodney}\ \emph {et~al.}(2011)\citenamefont {Rodney},
  \citenamefont {Tanguy},\ and\ \citenamefont
  {Vandembroucq}}]{rodney2011modeling}%
  \BibitemOpen
  \bibfield  {author} {\bibinfo {author} {\bibfnamefont {D.}~\bibnamefont
  {Rodney}}, \bibinfo {author} {\bibfnamefont {A.}~\bibnamefont {Tanguy}}, \
  and\ \bibinfo {author} {\bibfnamefont {D.}~\bibnamefont {Vandembroucq}},\
  }\href@noop {} {\bibfield  {journal} {\bibinfo  {journal} {Model. Simul.
  Mater. Sci. Eng.}\ }\textbf {\bibinfo {volume} {19}},\ \bibinfo {pages}
  {083001} (\bibinfo {year} {2011})}\BibitemShut {NoStop}%
\bibitem [{\citenamefont {Swallen}\ \emph {et~al.}(2007)\citenamefont
  {Swallen}, \citenamefont {Kearns}, \citenamefont {Mapes}, \citenamefont
  {Kim}, \citenamefont {McMahon}, \citenamefont {Ediger}, \citenamefont {Wu},
  \citenamefont {Yu},\ and\ \citenamefont {Satija}}]{swallen2007organic}%
  \BibitemOpen
  \bibfield  {author} {\bibinfo {author} {\bibfnamefont {S.~F.}\ \bibnamefont
  {Swallen}}, \bibinfo {author} {\bibfnamefont {K.~L.}\ \bibnamefont {Kearns}},
  \bibinfo {author} {\bibfnamefont {M.~K.}\ \bibnamefont {Mapes}}, \bibinfo
  {author} {\bibfnamefont {Y.~S.}\ \bibnamefont {Kim}}, \bibinfo {author}
  {\bibfnamefont {R.~J.}\ \bibnamefont {McMahon}}, \bibinfo {author}
  {\bibfnamefont {M.~D.}\ \bibnamefont {Ediger}}, \bibinfo {author}
  {\bibfnamefont {T.}~\bibnamefont {Wu}}, \bibinfo {author} {\bibfnamefont
  {L.}~\bibnamefont {Yu}}, \ and\ \bibinfo {author} {\bibfnamefont
  {S.}~\bibnamefont {Satija}},\ }\href@noop {} {\bibfield  {journal} {\bibinfo
  {journal} {Science}\ }\textbf {\bibinfo {volume} {315}},\ \bibinfo {pages}
  {353} (\bibinfo {year} {2007})}\BibitemShut {NoStop}%
\bibitem [{\citenamefont {Queen}\ \emph {et~al.}(2013)\citenamefont {Queen},
  \citenamefont {Liu}, \citenamefont {Karel}, \citenamefont {Metcalf},\ and\
  \citenamefont {Hellman}}]{queen2013excess}%
  \BibitemOpen
  \bibfield  {author} {\bibinfo {author} {\bibfnamefont {D.}~\bibnamefont
  {Queen}}, \bibinfo {author} {\bibfnamefont {X.}~\bibnamefont {Liu}}, \bibinfo
  {author} {\bibfnamefont {J.}~\bibnamefont {Karel}}, \bibinfo {author}
  {\bibfnamefont {T.}~\bibnamefont {Metcalf}}, \ and\ \bibinfo {author}
  {\bibfnamefont {F.}~\bibnamefont {Hellman}},\ }\href@noop {} {\bibfield
  {journal} {\bibinfo  {journal} {Phys. Rev. Lett.}\ }\textbf {\bibinfo
  {volume} {110}},\ \bibinfo {pages} {135901} (\bibinfo {year}
  {2013})}\BibitemShut {NoStop}%
\bibitem [{\citenamefont {Yu}\ \emph {et~al.}(2013)\citenamefont {Yu},
  \citenamefont {Luo},\ and\ \citenamefont {Samwer}}]{yu2013ultrastable}%
  \BibitemOpen
  \bibfield  {author} {\bibinfo {author} {\bibfnamefont {H.-B.}\ \bibnamefont
  {Yu}}, \bibinfo {author} {\bibfnamefont {Y.}~\bibnamefont {Luo}}, \ and\
  \bibinfo {author} {\bibfnamefont {K.}~\bibnamefont {Samwer}},\ }\href@noop {}
  {\bibfield  {journal} {\bibinfo  {journal} {Adv. Mater.}\ }\textbf {\bibinfo
  {volume} {25}},\ \bibinfo {pages} {5904} (\bibinfo {year}
  {2013})}\BibitemShut {NoStop}%
\bibitem [{\citenamefont {Yoon}\ and\ \citenamefont
  {McKenna}(2018)}]{yoon2018testing}%
  \BibitemOpen
  \bibfield  {author} {\bibinfo {author} {\bibfnamefont {H.}~\bibnamefont
  {Yoon}}\ and\ \bibinfo {author} {\bibfnamefont {G.~B.}\ \bibnamefont
  {McKenna}},\ }\href@noop {} {\bibfield  {journal} {\bibinfo  {journal} {Sci.
  Adv.}\ }\textbf {\bibinfo {volume} {4}},\ \bibinfo {pages} {eaau5423}
  (\bibinfo {year} {2018})}\BibitemShut {NoStop}%
\bibitem [{\citenamefont {Raegen}\ \emph {et~al.}(2020)\citenamefont {Raegen},
  \citenamefont {Yin}, \citenamefont {Zhou},\ and\ \citenamefont
  {Forrest}}]{raegen2020ultrastable}%
  \BibitemOpen
  \bibfield  {author} {\bibinfo {author} {\bibfnamefont {A.~N.}\ \bibnamefont
  {Raegen}}, \bibinfo {author} {\bibfnamefont {J.}~\bibnamefont {Yin}},
  \bibinfo {author} {\bibfnamefont {Q.}~\bibnamefont {Zhou}}, \ and\ \bibinfo
  {author} {\bibfnamefont {J.~A.}\ \bibnamefont {Forrest}},\ }\href@noop {}
  {\bibfield  {journal} {\bibinfo  {journal} {Nat. Mater.}\ }\textbf {\bibinfo
  {volume} {19}},\ \bibinfo {pages} {1110} (\bibinfo {year}
  {2020})}\BibitemShut {NoStop}%
\bibitem [{\citenamefont {Ediger}(2017)}]{ediger2017perspective}%
  \BibitemOpen
  \bibfield  {author} {\bibinfo {author} {\bibfnamefont {M.~D.}\ \bibnamefont
  {Ediger}},\ }\href@noop {} {\bibfield  {journal} {\bibinfo  {journal} {J.
  Chem. Phys.}\ }\textbf {\bibinfo {volume} {147}},\ \bibinfo {pages} {210901}
  (\bibinfo {year} {2017})}\BibitemShut {NoStop}%
\bibitem [{\citenamefont {Rodriguez-Tinoco}\ \emph {et~al.}(2022)\citenamefont
  {Rodriguez-Tinoco}, \citenamefont {Gonzalez-Silveira}, \citenamefont
  {Ramos},\ and\ \citenamefont {Rodriguez-Viejo}}]{rodriguez2022ultrastable}%
  \BibitemOpen
  \bibfield  {author} {\bibinfo {author} {\bibfnamefont {C.}~\bibnamefont
  {Rodriguez-Tinoco}}, \bibinfo {author} {\bibfnamefont {M.}~\bibnamefont
  {Gonzalez-Silveira}}, \bibinfo {author} {\bibfnamefont {M.~A.}\ \bibnamefont
  {Ramos}}, \ and\ \bibinfo {author} {\bibfnamefont {J.}~\bibnamefont
  {Rodriguez-Viejo}},\ }\href@noop {} {\bibfield  {journal} {\bibinfo
  {journal} {La Rivista del Nuovo Cimento}\ }\textbf {\bibinfo {volume} {45}},\
  \bibinfo {pages} {325} (\bibinfo {year} {2022})}\BibitemShut {NoStop}%
\bibitem [{\citenamefont {Marinari}\ and\ \citenamefont
  {Parisi}(1992)}]{marinari1992simulated}%
  \BibitemOpen
  \bibfield  {author} {\bibinfo {author} {\bibfnamefont {E.}~\bibnamefont
  {Marinari}}\ and\ \bibinfo {author} {\bibfnamefont {G.}~\bibnamefont
  {Parisi}},\ }\href@noop {} {\bibfield  {journal} {\bibinfo  {journal} {EPL}\
  }\textbf {\bibinfo {volume} {19}},\ \bibinfo {pages} {451} (\bibinfo {year}
  {1992})}\BibitemShut {NoStop}%
\bibitem [{\citenamefont {Hukushima}\ and\ \citenamefont
  {Nemoto}(1996)}]{hukushima1996exchange}%
  \BibitemOpen
  \bibfield  {author} {\bibinfo {author} {\bibfnamefont {K.}~\bibnamefont
  {Hukushima}}\ and\ \bibinfo {author} {\bibfnamefont {K.}~\bibnamefont
  {Nemoto}},\ }\href@noop {} {\bibfield  {journal} {\bibinfo  {journal} {J.
  Phys. Soc. Japan}\ }\textbf {\bibinfo {volume} {65}},\ \bibinfo {pages}
  {1604} (\bibinfo {year} {1996})}\BibitemShut {NoStop}%
\bibitem [{\citenamefont {Yamamoto}\ and\ \citenamefont
  {Kob}(2000)}]{yamamoto2000replica}%
  \BibitemOpen
  \bibfield  {author} {\bibinfo {author} {\bibfnamefont {R.}~\bibnamefont
  {Yamamoto}}\ and\ \bibinfo {author} {\bibfnamefont {W.}~\bibnamefont {Kob}},\
  }\href@noop {} {\bibfield  {journal} {\bibinfo  {journal} {Phys. Rev. E}\
  }\textbf {\bibinfo {volume} {61}},\ \bibinfo {pages} {5473} (\bibinfo {year}
  {2000})}\BibitemShut {NoStop}%
\bibitem [{\citenamefont {Santen}\ and\ \citenamefont
  {Krauth}(2000)}]{santen2000absence}%
  \BibitemOpen
  \bibfield  {author} {\bibinfo {author} {\bibfnamefont {L.}~\bibnamefont
  {Santen}}\ and\ \bibinfo {author} {\bibfnamefont {W.}~\bibnamefont
  {Krauth}},\ }\href@noop {} {\bibfield  {journal} {\bibinfo  {journal}
  {Nature}\ }\textbf {\bibinfo {volume} {405}},\ \bibinfo {pages} {550}
  (\bibinfo {year} {2000})}\BibitemShut {NoStop}%
\bibitem [{\citenamefont {Grigera}\ and\ \citenamefont
  {Parisi}(2001)}]{grigera2001fast}%
  \BibitemOpen
  \bibfield  {author} {\bibinfo {author} {\bibfnamefont {T.~S.}\ \bibnamefont
  {Grigera}}\ and\ \bibinfo {author} {\bibfnamefont {G.}~\bibnamefont
  {Parisi}},\ }\href@noop {} {\bibfield  {journal} {\bibinfo  {journal} {Phys.
  Rev. E}\ }\textbf {\bibinfo {volume} {63}},\ \bibinfo {pages} {045102}
  (\bibinfo {year} {2001})}\BibitemShut {NoStop}%
\bibitem [{\citenamefont {Guti{\'e}rrez}\ \emph {et~al.}(2015)\citenamefont
  {Guti{\'e}rrez}, \citenamefont {Karmakar}, \citenamefont {Pollack},\ and\
  \citenamefont {Procaccia}}]{gutierrez2015static}%
  \BibitemOpen
  \bibfield  {author} {\bibinfo {author} {\bibfnamefont {R.}~\bibnamefont
  {Guti{\'e}rrez}}, \bibinfo {author} {\bibfnamefont {S.}~\bibnamefont
  {Karmakar}}, \bibinfo {author} {\bibfnamefont {Y.~G.}\ \bibnamefont
  {Pollack}}, \ and\ \bibinfo {author} {\bibfnamefont {I.}~\bibnamefont
  {Procaccia}},\ }\href@noop {} {\bibfield  {journal} {\bibinfo  {journal}
  {EPL}\ }\textbf {\bibinfo {volume} {111}},\ \bibinfo {pages} {56009}
  (\bibinfo {year} {2015})}\BibitemShut {NoStop}%
\bibitem [{\citenamefont {Ninarello}\ \emph {et~al.}(2017)\citenamefont
  {Ninarello}, \citenamefont {Berthier},\ and\ \citenamefont
  {Coslovich}}]{ninarello2017models}%
  \BibitemOpen
  \bibfield  {author} {\bibinfo {author} {\bibfnamefont {A.}~\bibnamefont
  {Ninarello}}, \bibinfo {author} {\bibfnamefont {L.}~\bibnamefont {Berthier}},
  \ and\ \bibinfo {author} {\bibfnamefont {D.}~\bibnamefont {Coslovich}},\
  }\href@noop {} {\bibfield  {journal} {\bibinfo  {journal} {Phys. Rev. X}\
  }\textbf {\bibinfo {volume} {7}},\ \bibinfo {pages} {021039} (\bibinfo {year}
  {2017})}\BibitemShut {NoStop}%
\bibitem [{\citenamefont {Kim}(2003)}]{kim2003effects}%
  \BibitemOpen
  \bibfield  {author} {\bibinfo {author} {\bibfnamefont {K.}~\bibnamefont
  {Kim}},\ }\href@noop {} {\bibfield  {journal} {\bibinfo  {journal} {EPL}\
  }\textbf {\bibinfo {volume} {61}},\ \bibinfo {pages} {790} (\bibinfo {year}
  {2003})}\BibitemShut {NoStop}%
\bibitem [{\citenamefont {Cammarota}\ and\ \citenamefont
  {Biroli}(2012)}]{cammarota2012ideal}%
  \BibitemOpen
  \bibfield  {author} {\bibinfo {author} {\bibfnamefont {C.}~\bibnamefont
  {Cammarota}}\ and\ \bibinfo {author} {\bibfnamefont {G.}~\bibnamefont
  {Biroli}},\ }\href@noop {} {\bibfield  {journal} {\bibinfo  {journal} {PNAS}\
  }\textbf {\bibinfo {volume} {109}},\ \bibinfo {pages} {8850} (\bibinfo {year}
  {2012})}\BibitemShut {NoStop}%
\bibitem [{\citenamefont {No{\'e}}\ \emph {et~al.}(2019)\citenamefont
  {No{\'e}}, \citenamefont {Olsson}, \citenamefont {K{\"o}hler},\ and\
  \citenamefont {Wu}}]{noe2019boltzmann}%
  \BibitemOpen
  \bibfield  {author} {\bibinfo {author} {\bibfnamefont {F.}~\bibnamefont
  {No{\'e}}}, \bibinfo {author} {\bibfnamefont {S.}~\bibnamefont {Olsson}},
  \bibinfo {author} {\bibfnamefont {J.}~\bibnamefont {K{\"o}hler}}, \ and\
  \bibinfo {author} {\bibfnamefont {H.}~\bibnamefont {Wu}},\ }\href@noop {}
  {\bibfield  {journal} {\bibinfo  {journal} {Science}\ }\textbf {\bibinfo
  {volume} {365}},\ \bibinfo {pages} {eaaw1147} (\bibinfo {year}
  {2019})}\BibitemShut {NoStop}%
\bibitem [{\citenamefont {Wu}\ \emph {et~al.}(2019)\citenamefont {Wu},
  \citenamefont {Wang},\ and\ \citenamefont {Zhang}}]{wu2019solving}%
  \BibitemOpen
  \bibfield  {author} {\bibinfo {author} {\bibfnamefont {D.}~\bibnamefont
  {Wu}}, \bibinfo {author} {\bibfnamefont {L.}~\bibnamefont {Wang}}, \ and\
  \bibinfo {author} {\bibfnamefont {P.}~\bibnamefont {Zhang}},\ }\href@noop {}
  {\bibfield  {journal} {\bibinfo  {journal} {Phys. Rev. Lett.}\ }\textbf
  {\bibinfo {volume} {122}},\ \bibinfo {pages} {080602} (\bibinfo {year}
  {2019})}\BibitemShut {NoStop}%
\bibitem [{\citenamefont {McNaughton}\ \emph {et~al.}(2020)\citenamefont
  {McNaughton}, \citenamefont {Milo{\v{s}}evi{\'c}}, \citenamefont {Perali},\
  and\ \citenamefont {Pilati}}]{mcnaughton2020boosting}%
  \BibitemOpen
  \bibfield  {author} {\bibinfo {author} {\bibfnamefont {B.}~\bibnamefont
  {McNaughton}}, \bibinfo {author} {\bibfnamefont {M.}~\bibnamefont
  {Milo{\v{s}}evi{\'c}}}, \bibinfo {author} {\bibfnamefont {A.}~\bibnamefont
  {Perali}}, \ and\ \bibinfo {author} {\bibfnamefont {S.}~\bibnamefont
  {Pilati}},\ }\href@noop {} {\bibfield  {journal} {\bibinfo  {journal} {Phys.
  Rev. E}\ }\textbf {\bibinfo {volume} {101}},\ \bibinfo {pages} {053312}
  (\bibinfo {year} {2020})}\BibitemShut {NoStop}%
\bibitem [{\citenamefont {Wu}\ \emph {et~al.}(2021)\citenamefont {Wu},
  \citenamefont {Rossi},\ and\ \citenamefont {Carleo}}]{wu2021unbiased}%
  \BibitemOpen
  \bibfield  {author} {\bibinfo {author} {\bibfnamefont {D.}~\bibnamefont
  {Wu}}, \bibinfo {author} {\bibfnamefont {R.}~\bibnamefont {Rossi}}, \ and\
  \bibinfo {author} {\bibfnamefont {G.}~\bibnamefont {Carleo}},\ }\href@noop {}
  {\bibfield  {journal} {\bibinfo  {journal} {Phys. Rev. Res.}\ }\textbf
  {\bibinfo {volume} {3}},\ \bibinfo {pages} {L042024} (\bibinfo {year}
  {2021})}\BibitemShut {NoStop}%
\bibitem [{\citenamefont {Hibat-Allah}\ \emph {et~al.}(2021)\citenamefont
  {Hibat-Allah}, \citenamefont {Inack}, \citenamefont {Wiersema}, \citenamefont
  {Melko},\ and\ \citenamefont {Carrasquilla}}]{hibat2021variational}%
  \BibitemOpen
  \bibfield  {author} {\bibinfo {author} {\bibfnamefont {M.}~\bibnamefont
  {Hibat-Allah}}, \bibinfo {author} {\bibfnamefont {E.~M.}\ \bibnamefont
  {Inack}}, \bibinfo {author} {\bibfnamefont {R.}~\bibnamefont {Wiersema}},
  \bibinfo {author} {\bibfnamefont {R.~G.}\ \bibnamefont {Melko}}, \ and\
  \bibinfo {author} {\bibfnamefont {J.}~\bibnamefont {Carrasquilla}},\
  }\href@noop {} {\bibfield  {journal} {\bibinfo  {journal} {Nat. Mach.
  Intell}\ }\textbf {\bibinfo {volume} {3}},\ \bibinfo {pages} {952} (\bibinfo
  {year} {2021})}\BibitemShut {NoStop}%
\bibitem [{\citenamefont {Gabri{\'e}}\ \emph {et~al.}(2022)\citenamefont
  {Gabri{\'e}}, \citenamefont {Rotskoff},\ and\ \citenamefont
  {Vanden-Eijnden}}]{gabrie2022adaptive}%
  \BibitemOpen
  \bibfield  {author} {\bibinfo {author} {\bibfnamefont {M.}~\bibnamefont
  {Gabri{\'e}}}, \bibinfo {author} {\bibfnamefont {G.~M.}\ \bibnamefont
  {Rotskoff}}, \ and\ \bibinfo {author} {\bibfnamefont {E.}~\bibnamefont
  {Vanden-Eijnden}},\ }\href@noop {} {\bibfield  {journal} {\bibinfo  {journal}
  {PNAS}\ }\textbf {\bibinfo {volume} {119}},\ \bibinfo {pages} {e2109420119}
  (\bibinfo {year} {2022})}\BibitemShut {NoStop}%
\bibitem [{\citenamefont {Ciarella}\ \emph {et~al.}(2023)\citenamefont
  {Ciarella}, \citenamefont {Trinquier}, \citenamefont {Weigt},\ and\
  \citenamefont {Zamponi}}]{ciarella2023machine}%
  \BibitemOpen
  \bibfield  {author} {\bibinfo {author} {\bibfnamefont {S.}~\bibnamefont
  {Ciarella}}, \bibinfo {author} {\bibfnamefont {J.}~\bibnamefont {Trinquier}},
  \bibinfo {author} {\bibfnamefont {M.}~\bibnamefont {Weigt}}, \ and\ \bibinfo
  {author} {\bibfnamefont {F.}~\bibnamefont {Zamponi}},\ }\href@noop {}
  {\bibfield  {journal} {\bibinfo  {journal} {Mach. learn.: sci. technol.}\
  }\textbf {\bibinfo {volume} {4}},\ \bibinfo {pages} {010501} (\bibinfo {year}
  {2023})}\BibitemShut {NoStop}%
\bibitem [{\citenamefont {Brito}\ \emph {et~al.}(2018)\citenamefont {Brito},
  \citenamefont {Lerner},\ and\ \citenamefont {Wyart}}]{brito2018theory}%
  \BibitemOpen
  \bibfield  {author} {\bibinfo {author} {\bibfnamefont {C.}~\bibnamefont
  {Brito}}, \bibinfo {author} {\bibfnamefont {E.}~\bibnamefont {Lerner}}, \
  and\ \bibinfo {author} {\bibfnamefont {M.}~\bibnamefont {Wyart}},\
  }\href@noop {} {\bibfield  {journal} {\bibinfo  {journal} {Phys. Rev. X}\
  }\textbf {\bibinfo {volume} {8}},\ \bibinfo {pages} {031050} (\bibinfo {year}
  {2018})}\BibitemShut {NoStop}%
\bibitem [{\citenamefont {Hagh}\ \emph {et~al.}(2022)\citenamefont {Hagh},
  \citenamefont {Nagel}, \citenamefont {Liu}, \citenamefont {Manning},\ and\
  \citenamefont {Corwin}}]{hagh2022transient}%
  \BibitemOpen
  \bibfield  {author} {\bibinfo {author} {\bibfnamefont {V.~F.}\ \bibnamefont
  {Hagh}}, \bibinfo {author} {\bibfnamefont {S.~R.}\ \bibnamefont {Nagel}},
  \bibinfo {author} {\bibfnamefont {A.~J.}\ \bibnamefont {Liu}}, \bibinfo
  {author} {\bibfnamefont {M.~L.}\ \bibnamefont {Manning}}, \ and\ \bibinfo
  {author} {\bibfnamefont {E.~I.}\ \bibnamefont {Corwin}},\ }\href@noop {}
  {\bibfield  {journal} {\bibinfo  {journal} {PNAS}\ }\textbf {\bibinfo
  {volume} {119}},\ \bibinfo {pages} {e2117622119} (\bibinfo {year}
  {2022})}\BibitemShut {NoStop}%
\bibitem [{\citenamefont {Kob}\ and\ \citenamefont
  {Berthier}(2013)}]{kob2013probing}%
  \BibitemOpen
  \bibfield  {author} {\bibinfo {author} {\bibfnamefont {W.}~\bibnamefont
  {Kob}}\ and\ \bibinfo {author} {\bibfnamefont {L.}~\bibnamefont {Berthier}},\
  }\href@noop {} {\bibfield  {journal} {\bibinfo  {journal} {Phys. Rev. Lett.}\
  }\textbf {\bibinfo {volume} {110}},\ \bibinfo {pages} {245702} (\bibinfo
  {year} {2013})}\BibitemShut {NoStop}%
\bibitem [{\citenamefont {Ozawa}\ \emph {et~al.}(2015)\citenamefont {Ozawa},
  \citenamefont {Kob}, \citenamefont {Ikeda},\ and\ \citenamefont
  {Miyazaki}}]{ozawa2015equilibrium}%
  \BibitemOpen
  \bibfield  {author} {\bibinfo {author} {\bibfnamefont {M.}~\bibnamefont
  {Ozawa}}, \bibinfo {author} {\bibfnamefont {W.}~\bibnamefont {Kob}}, \bibinfo
  {author} {\bibfnamefont {A.}~\bibnamefont {Ikeda}}, \ and\ \bibinfo {author}
  {\bibfnamefont {K.}~\bibnamefont {Miyazaki}},\ }\href@noop {} {\bibfield
  {journal} {\bibinfo  {journal} {PNAS}\ }\textbf {\bibinfo {volume} {112}},\
  \bibinfo {pages} {6914} (\bibinfo {year} {2015})}\BibitemShut {NoStop}%
\bibitem [{\citenamefont {Ozawa}\ \emph
  {et~al.}(2018{\natexlab{a}})\citenamefont {Ozawa}, \citenamefont {Ikeda},
  \citenamefont {Miyazaki},\ and\ \citenamefont {Kob}}]{ozawa2018ideal}%
  \BibitemOpen
  \bibfield  {author} {\bibinfo {author} {\bibfnamefont {M.}~\bibnamefont
  {Ozawa}}, \bibinfo {author} {\bibfnamefont {A.}~\bibnamefont {Ikeda}},
  \bibinfo {author} {\bibfnamefont {K.}~\bibnamefont {Miyazaki}}, \ and\
  \bibinfo {author} {\bibfnamefont {W.}~\bibnamefont {Kob}},\ }\href@noop {}
  {\bibfield  {journal} {\bibinfo  {journal} {Phys. Rev. Lett.}\ }\textbf
  {\bibinfo {volume} {121}},\ \bibinfo {pages} {205501} (\bibinfo {year}
  {2018}{\natexlab{a}})}\BibitemShut {NoStop}%
\bibitem [{\citenamefont {Biroli}\ and\ \citenamefont
  {Bouchaud}(2023)}]{biroli2023rfot}%
  \BibitemOpen
  \bibfield  {author} {\bibinfo {author} {\bibfnamefont {G.}~\bibnamefont
  {Biroli}}\ and\ \bibinfo {author} {\bibfnamefont {J.-P.}\ \bibnamefont
  {Bouchaud}},\ }\href@noop {} {\bibfield  {journal} {\bibinfo  {journal} {C.
  R. Phys.}\ }\textbf {\bibinfo {volume} {24}},\ \bibinfo {pages} {1} (\bibinfo
  {year} {2023})}\BibitemShut {NoStop}%
\bibitem [{\citenamefont {Gokhale}\ \emph {et~al.}(2014)\citenamefont
  {Gokhale}, \citenamefont {Nagamanasa}, \citenamefont {Ganapathy},\ and\
  \citenamefont {Sood}}]{gokhale2014growing}%
  \BibitemOpen
  \bibfield  {author} {\bibinfo {author} {\bibfnamefont {S.}~\bibnamefont
  {Gokhale}}, \bibinfo {author} {\bibfnamefont {K.~H.}\ \bibnamefont
  {Nagamanasa}}, \bibinfo {author} {\bibfnamefont {R.}~\bibnamefont
  {Ganapathy}}, \ and\ \bibinfo {author} {\bibfnamefont {A.}~\bibnamefont
  {Sood}},\ }\href@noop {} {\bibfield  {journal} {\bibinfo  {journal} {Nat.
  Commun.}\ }\textbf {\bibinfo {volume} {5}},\ \bibinfo {pages} {1} (\bibinfo
  {year} {2014})}\BibitemShut {NoStop}%
\bibitem [{\citenamefont {Williams}\ \emph {et~al.}(2018)\citenamefont
  {Williams}, \citenamefont {Turci}, \citenamefont {Hallett}, \citenamefont
  {Crowther}, \citenamefont {Cammarota}, \citenamefont {Biroli},\ and\
  \citenamefont {Royall}}]{williams2018experimental}%
  \BibitemOpen
  \bibfield  {author} {\bibinfo {author} {\bibfnamefont {I.}~\bibnamefont
  {Williams}}, \bibinfo {author} {\bibfnamefont {F.}~\bibnamefont {Turci}},
  \bibinfo {author} {\bibfnamefont {J.~E.}\ \bibnamefont {Hallett}}, \bibinfo
  {author} {\bibfnamefont {P.}~\bibnamefont {Crowther}}, \bibinfo {author}
  {\bibfnamefont {C.}~\bibnamefont {Cammarota}}, \bibinfo {author}
  {\bibfnamefont {G.}~\bibnamefont {Biroli}}, \ and\ \bibinfo {author}
  {\bibfnamefont {C.~P.}\ \bibnamefont {Royall}},\ }\href@noop {} {\bibfield
  {journal} {\bibinfo  {journal} {J. Condens. Matter Phys.}\ }\textbf {\bibinfo
  {volume} {30}},\ \bibinfo {pages} {094003} (\bibinfo {year}
  {2018})}\BibitemShut {NoStop}%
\bibitem [{\citenamefont {Kikumoto}\ \emph {et~al.}(2020)\citenamefont
  {Kikumoto}, \citenamefont {Torii}, \citenamefont {Fukao}, \citenamefont
  {Royall}, \citenamefont {Yao}, \citenamefont {Saruyama},\ and\ \citenamefont
  {Tatsumi}}]{kikumoto2020towards}%
  \BibitemOpen
  \bibfield  {author} {\bibinfo {author} {\bibfnamefont {G.}~\bibnamefont
  {Kikumoto}}, \bibinfo {author} {\bibfnamefont {N.}~\bibnamefont {Torii}},
  \bibinfo {author} {\bibfnamefont {K.}~\bibnamefont {Fukao}}, \bibinfo
  {author} {\bibfnamefont {C.~P.}\ \bibnamefont {Royall}}, \bibinfo {author}
  {\bibfnamefont {H.}~\bibnamefont {Yao}}, \bibinfo {author} {\bibfnamefont
  {Y.}~\bibnamefont {Saruyama}}, \ and\ \bibinfo {author} {\bibfnamefont
  {S.}~\bibnamefont {Tatsumi}},\ }\href@noop {} {\bibfield  {journal} {\bibinfo
   {journal} {arXiv:2003.06089}\ } (\bibinfo {year} {2020})}\BibitemShut
  {NoStop}%
\bibitem [{\citenamefont {Das}\ \emph {et~al.}(2023)\citenamefont {Das},
  \citenamefont {Bhowmik}, \citenamefont {Puthirath}, \citenamefont
  {Narayanan},\ and\ \citenamefont {Karmakar}}]{das2023soft}%
  \BibitemOpen
  \bibfield  {author} {\bibinfo {author} {\bibfnamefont {R.}~\bibnamefont
  {Das}}, \bibinfo {author} {\bibfnamefont {B.~P.}\ \bibnamefont {Bhowmik}},
  \bibinfo {author} {\bibfnamefont {A.~B.}\ \bibnamefont {Puthirath}}, \bibinfo
  {author} {\bibfnamefont {T.~N.}\ \bibnamefont {Narayanan}}, \ and\ \bibinfo
  {author} {\bibfnamefont {S.}~\bibnamefont {Karmakar}},\ }\href@noop {}
  {\bibfield  {journal} {\bibinfo  {journal} {PNAS nexus}\ }\textbf {\bibinfo
  {volume} {2}},\ \bibinfo {pages} {pgad277} (\bibinfo {year}
  {2023})}\BibitemShut {NoStop}%
\bibitem [{\citenamefont {Charbonneau}\ and\ \citenamefont
  {Tarjus}(2013)}]{charbonneau2013decorrelation}%
  \BibitemOpen
  \bibfield  {author} {\bibinfo {author} {\bibfnamefont {P.}~\bibnamefont
  {Charbonneau}}\ and\ \bibinfo {author} {\bibfnamefont {G.}~\bibnamefont
  {Tarjus}},\ }\href@noop {} {\bibfield  {journal} {\bibinfo  {journal} {Phys.
  Rev. E}\ }\textbf {\bibinfo {volume} {87}},\ \bibinfo {pages} {042305}
  (\bibinfo {year} {2013})}\BibitemShut {NoStop}%
\bibitem [{\citenamefont {Chakrabarty}\ \emph {et~al.}(2016)\citenamefont
  {Chakrabarty}, \citenamefont {Das}, \citenamefont {Karmakar},\ and\
  \citenamefont {Dasgupta}}]{chakrabarty2016understanding}%
  \BibitemOpen
  \bibfield  {author} {\bibinfo {author} {\bibfnamefont {S.}~\bibnamefont
  {Chakrabarty}}, \bibinfo {author} {\bibfnamefont {R.}~\bibnamefont {Das}},
  \bibinfo {author} {\bibfnamefont {S.}~\bibnamefont {Karmakar}}, \ and\
  \bibinfo {author} {\bibfnamefont {C.}~\bibnamefont {Dasgupta}},\ }\href@noop
  {} {\bibfield  {journal} {\bibinfo  {journal} {J. Chem. Phys.}\ }\textbf
  {\bibinfo {volume} {145}},\ \bibinfo {pages} {034507} (\bibinfo {year}
  {2016})}\BibitemShut {NoStop}%
\bibitem [{\citenamefont {Kim}\ \emph {et~al.}(2011)\citenamefont {Kim},
  \citenamefont {Miyazaki},\ and\ \citenamefont {Saito}}]{kim2011slow}%
  \BibitemOpen
  \bibfield  {author} {\bibinfo {author} {\bibfnamefont {K.}~\bibnamefont
  {Kim}}, \bibinfo {author} {\bibfnamefont {K.}~\bibnamefont {Miyazaki}}, \
  and\ \bibinfo {author} {\bibfnamefont {S.}~\bibnamefont {Saito}},\
  }\href@noop {} {\bibfield  {journal} {\bibinfo  {journal} {J. Condens. Matter
  Phys.}\ }\textbf {\bibinfo {volume} {23}},\ \bibinfo {pages} {234123}
  (\bibinfo {year} {2011})}\BibitemShut {NoStop}%
\bibitem [{\citenamefont {Chakrabarty}\ \emph {et~al.}(2015)\citenamefont
  {Chakrabarty}, \citenamefont {Karmakar},\ and\ \citenamefont
  {Dasgupta}}]{chakrabarty2015dynamics}%
  \BibitemOpen
  \bibfield  {author} {\bibinfo {author} {\bibfnamefont {S.}~\bibnamefont
  {Chakrabarty}}, \bibinfo {author} {\bibfnamefont {S.}~\bibnamefont
  {Karmakar}}, \ and\ \bibinfo {author} {\bibfnamefont {C.}~\bibnamefont
  {Dasgupta}},\ }\href@noop {} {\bibfield  {journal} {\bibinfo  {journal} {Sci.
  Rep.}\ }\textbf {\bibinfo {volume} {5}},\ \bibinfo {pages} {1} (\bibinfo
  {year} {2015})}\BibitemShut {NoStop}%
\bibitem [{\citenamefont {Jack}\ and\ \citenamefont
  {Fullerton}(2013)}]{jack2013dynamical}%
  \BibitemOpen
  \bibfield  {author} {\bibinfo {author} {\bibfnamefont {R.~L.}\ \bibnamefont
  {Jack}}\ and\ \bibinfo {author} {\bibfnamefont {C.~J.}\ \bibnamefont
  {Fullerton}},\ }\href@noop {} {\bibfield  {journal} {\bibinfo  {journal}
  {Phys. Rev. E}\ }\textbf {\bibinfo {volume} {88}},\ \bibinfo {pages} {042304}
  (\bibinfo {year} {2013})}\BibitemShut {NoStop}%
\bibitem [{\citenamefont {Kob}\ and\ \citenamefont
  {Coslovich}(2014)}]{kob2014nonlinear}%
  \BibitemOpen
  \bibfield  {author} {\bibinfo {author} {\bibfnamefont {W.}~\bibnamefont
  {Kob}}\ and\ \bibinfo {author} {\bibfnamefont {D.}~\bibnamefont
  {Coslovich}},\ }\href@noop {} {\bibfield  {journal} {\bibinfo  {journal}
  {Phys. Rev. E}\ }\textbf {\bibinfo {volume} {90}},\ \bibinfo {pages} {052305}
  (\bibinfo {year} {2014})}\BibitemShut {NoStop}%
\bibitem [{\citenamefont {Li}\ \emph {et~al.}(2015)\citenamefont {Li},
  \citenamefont {Zhu},\ and\ \citenamefont {Sun}}]{li2015decoupling}%
  \BibitemOpen
  \bibfield  {author} {\bibinfo {author} {\bibfnamefont {Y.-W.}\ \bibnamefont
  {Li}}, \bibinfo {author} {\bibfnamefont {Y.-L.}\ \bibnamefont {Zhu}}, \ and\
  \bibinfo {author} {\bibfnamefont {Z.-Y.}\ \bibnamefont {Sun}},\ }\href@noop
  {} {\bibfield  {journal} {\bibinfo  {journal} {J. Chem. Phys.}\ }\textbf
  {\bibinfo {volume} {142}} (\bibinfo {year} {2015})}\BibitemShut {NoStop}%
\bibitem [{\citenamefont {Ozawa}\ \emph {et~al.}(2023)\citenamefont {Ozawa},
  \citenamefont {Iwashita}, \citenamefont {Kob},\ and\ \citenamefont
  {Zamponi}}]{Ozawa2023}%
  \BibitemOpen
  \bibfield  {author} {\bibinfo {author} {\bibfnamefont {M.}~\bibnamefont
  {Ozawa}}, \bibinfo {author} {\bibfnamefont {Y.}~\bibnamefont {Iwashita}},
  \bibinfo {author} {\bibfnamefont {W.}~\bibnamefont {Kob}}, \ and\ \bibinfo
  {author} {\bibfnamefont {F.}~\bibnamefont {Zamponi}},\ }\href@noop {}
  {\bibfield  {journal} {\bibinfo  {journal} {Nat. Commun.}\ }\textbf {\bibinfo
  {volume} {14}} (\bibinfo {year} {2023})}\BibitemShut {NoStop}%
\bibitem [{\citenamefont {Johari}(1994)}]{johari1994dynamics}%
  \BibitemOpen
  \bibfield  {author} {\bibinfo {author} {\bibfnamefont {G.}~\bibnamefont
  {Johari}},\ }in\ \href@noop {} {\emph {\bibinfo {booktitle} {Disorder Effects
  on Relaxational Processes: Glasses, Polymers, Proteins}}}\ (\bibinfo
  {publisher} {Springer},\ \bibinfo {year} {1994})\ pp.\ \bibinfo {pages}
  {627--657}\BibitemShut {NoStop}%
\bibitem [{\citenamefont {Corcione}\ \emph {et~al.}(2006)\citenamefont
  {Corcione}, \citenamefont {Greco},\ and\ \citenamefont
  {Maffezzoli}}]{corcione2006temperature}%
  \BibitemOpen
  \bibfield  {author} {\bibinfo {author} {\bibfnamefont {C.~E.}\ \bibnamefont
  {Corcione}}, \bibinfo {author} {\bibfnamefont {A.}~\bibnamefont {Greco}}, \
  and\ \bibinfo {author} {\bibfnamefont {A.}~\bibnamefont {Maffezzoli}},\
  }\href@noop {} {\bibfield  {journal} {\bibinfo  {journal} {Polym. Eng. Sci.}\
  }\textbf {\bibinfo {volume} {46}},\ \bibinfo {pages} {493} (\bibinfo {year}
  {2006})}\BibitemShut {NoStop}%
\bibitem [{\citenamefont {Kloxin}\ and\ \citenamefont
  {Bowman}(2013)}]{kloxin2013covalent}%
  \BibitemOpen
  \bibfield  {author} {\bibinfo {author} {\bibfnamefont {C.~J.}\ \bibnamefont
  {Kloxin}}\ and\ \bibinfo {author} {\bibfnamefont {C.~N.}\ \bibnamefont
  {Bowman}},\ }\href@noop {} {\bibfield  {journal} {\bibinfo  {journal} {Chem.
  Soc. Rev.}\ }\textbf {\bibinfo {volume} {42}},\ \bibinfo {pages} {7161}
  (\bibinfo {year} {2013})}\BibitemShut {NoStop}%
\bibitem [{\citenamefont {Denissen}\ \emph {et~al.}(2016)\citenamefont
  {Denissen}, \citenamefont {Winne},\ and\ \citenamefont
  {Du~Prez}}]{denissen2016vitrimers}%
  \BibitemOpen
  \bibfield  {author} {\bibinfo {author} {\bibfnamefont {W.}~\bibnamefont
  {Denissen}}, \bibinfo {author} {\bibfnamefont {J.~M.}\ \bibnamefont {Winne}},
  \ and\ \bibinfo {author} {\bibfnamefont {F.~E.}\ \bibnamefont {Du~Prez}},\
  }\href@noop {} {\bibfield  {journal} {\bibinfo  {journal} {Chem. Sci.}\
  }\textbf {\bibinfo {volume} {7}},\ \bibinfo {pages} {30} (\bibinfo {year}
  {2016})}\BibitemShut {NoStop}%
\bibitem [{\citenamefont {Duguet}\ \emph {et~al.}(2011)\citenamefont {Duguet},
  \citenamefont {D{\'e}sert}, \citenamefont {Perro},\ and\ \citenamefont
  {Ravaine}}]{duguet2011design}%
  \BibitemOpen
  \bibfield  {author} {\bibinfo {author} {\bibfnamefont {E.}~\bibnamefont
  {Duguet}}, \bibinfo {author} {\bibfnamefont {A.}~\bibnamefont {D{\'e}sert}},
  \bibinfo {author} {\bibfnamefont {A.}~\bibnamefont {Perro}}, \ and\ \bibinfo
  {author} {\bibfnamefont {S.}~\bibnamefont {Ravaine}},\ }\href@noop {}
  {\bibfield  {journal} {\bibinfo  {journal} {Chem. Soc. Rev.}\ }\textbf
  {\bibinfo {volume} {40}},\ \bibinfo {pages} {941} (\bibinfo {year}
  {2011})}\BibitemShut {NoStop}%
\bibitem [{\citenamefont {Peng}\ \emph {et~al.}(2013)\citenamefont {Peng},
  \citenamefont {Smallenburg}, \citenamefont {Imhof}, \citenamefont
  {Dijkstra},\ and\ \citenamefont {van Blaaderen}}]{peng2013colloidal}%
  \BibitemOpen
  \bibfield  {author} {\bibinfo {author} {\bibfnamefont {B.}~\bibnamefont
  {Peng}}, \bibinfo {author} {\bibfnamefont {F.}~\bibnamefont {Smallenburg}},
  \bibinfo {author} {\bibfnamefont {A.}~\bibnamefont {Imhof}}, \bibinfo
  {author} {\bibfnamefont {M.}~\bibnamefont {Dijkstra}}, \ and\ \bibinfo
  {author} {\bibfnamefont {A.}~\bibnamefont {van Blaaderen}},\ }\href@noop {}
  {\bibfield  {journal} {\bibinfo  {journal} {Angew. Chem.}\ }\textbf {\bibinfo
  {volume} {125}},\ \bibinfo {pages} {6841} (\bibinfo {year}
  {2013})}\BibitemShut {NoStop}%
\bibitem [{\citenamefont {McMullen}\ \emph {et~al.}(2022)\citenamefont
  {McMullen}, \citenamefont {Mu{\~n}oz~Basagoiti}, \citenamefont {Zeravcic},\
  and\ \citenamefont {Brujic}}]{mcmullen2022self}%
  \BibitemOpen
  \bibfield  {author} {\bibinfo {author} {\bibfnamefont {A.}~\bibnamefont
  {McMullen}}, \bibinfo {author} {\bibfnamefont {M.}~\bibnamefont
  {Mu{\~n}oz~Basagoiti}}, \bibinfo {author} {\bibfnamefont {Z.}~\bibnamefont
  {Zeravcic}}, \ and\ \bibinfo {author} {\bibfnamefont {J.}~\bibnamefont
  {Brujic}},\ }\href@noop {} {\bibfield  {journal} {\bibinfo  {journal}
  {Nature}\ }\textbf {\bibinfo {volume} {610}},\ \bibinfo {pages} {502}
  (\bibinfo {year} {2022})}\BibitemShut {NoStop}%
\bibitem [{\citenamefont {McMullen}\ \emph {et~al.}(2018)\citenamefont
  {McMullen}, \citenamefont {Holmes-Cerfon}, \citenamefont {Sciortino},
  \citenamefont {Grosberg},\ and\ \citenamefont {Brujic}}]{mcmullen2018freely}%
  \BibitemOpen
  \bibfield  {author} {\bibinfo {author} {\bibfnamefont {A.}~\bibnamefont
  {McMullen}}, \bibinfo {author} {\bibfnamefont {M.}~\bibnamefont
  {Holmes-Cerfon}}, \bibinfo {author} {\bibfnamefont {F.}~\bibnamefont
  {Sciortino}}, \bibinfo {author} {\bibfnamefont {A.~Y.}\ \bibnamefont
  {Grosberg}}, \ and\ \bibinfo {author} {\bibfnamefont {J.}~\bibnamefont
  {Brujic}},\ }\href@noop {} {\bibfield  {journal} {\bibinfo  {journal} {Phys.
  Rev. Lett.}\ }\textbf {\bibinfo {volume} {121}},\ \bibinfo {pages} {138002}
  (\bibinfo {year} {2018})}\BibitemShut {NoStop}%
\bibitem [{\citenamefont {Yuan}\ \emph {et~al.}(2016)\citenamefont {Yuan},
  \citenamefont {Gu}, \citenamefont {Zhao}, \citenamefont {Yao}, \citenamefont
  {Guan},\ and\ \citenamefont {Zhang}}]{yuan2016synthesis}%
  \BibitemOpen
  \bibfield  {author} {\bibinfo {author} {\bibfnamefont {Q.}~\bibnamefont
  {Yuan}}, \bibinfo {author} {\bibfnamefont {J.}~\bibnamefont {Gu}}, \bibinfo
  {author} {\bibfnamefont {Y.-n.}\ \bibnamefont {Zhao}}, \bibinfo {author}
  {\bibfnamefont {L.}~\bibnamefont {Yao}}, \bibinfo {author} {\bibfnamefont
  {Y.}~\bibnamefont {Guan}}, \ and\ \bibinfo {author} {\bibfnamefont
  {Y.}~\bibnamefont {Zhang}},\ }\href@noop {} {\bibfield  {journal} {\bibinfo
  {journal} {ACS Macro Lett.}\ }\textbf {\bibinfo {volume} {5}},\ \bibinfo
  {pages} {565} (\bibinfo {year} {2016})}\BibitemShut {NoStop}%
\bibitem [{\citenamefont {Goldbart}\ \emph {et~al.}(1996)\citenamefont
  {Goldbart}, \citenamefont {Castillo},\ and\ \citenamefont
  {Zippelius}}]{goldbart1996randomly}%
  \BibitemOpen
  \bibfield  {author} {\bibinfo {author} {\bibfnamefont {P.~M.}\ \bibnamefont
  {Goldbart}}, \bibinfo {author} {\bibfnamefont {H.~E.}\ \bibnamefont
  {Castillo}}, \ and\ \bibinfo {author} {\bibfnamefont {A.}~\bibnamefont
  {Zippelius}},\ }\href@noop {} {\bibfield  {journal} {\bibinfo  {journal}
  {Advances in Physics}\ }\textbf {\bibinfo {volume} {45}},\ \bibinfo {pages}
  {393} (\bibinfo {year} {1996})}\BibitemShut {NoStop}%
\bibitem [{\citenamefont {Carbas}\ \emph {et~al.}(2014)\citenamefont {Carbas},
  \citenamefont {Marques}, \citenamefont {Da~Silva},\ and\ \citenamefont
  {Lopes}}]{carbas2014effect}%
  \BibitemOpen
  \bibfield  {author} {\bibinfo {author} {\bibfnamefont {R.}~\bibnamefont
  {Carbas}}, \bibinfo {author} {\bibfnamefont {E.}~\bibnamefont {Marques}},
  \bibinfo {author} {\bibfnamefont {L.}~\bibnamefont {Da~Silva}}, \ and\
  \bibinfo {author} {\bibfnamefont {A.}~\bibnamefont {Lopes}},\ }\href@noop {}
  {\bibfield  {journal} {\bibinfo  {journal} {J. Adhes.}\ }\textbf {\bibinfo
  {volume} {90}},\ \bibinfo {pages} {104} (\bibinfo {year} {2014})}\BibitemShut
  {NoStop}%
\bibitem [{\citenamefont {Corezzi}\ \emph {et~al.}(2002)\citenamefont
  {Corezzi}, \citenamefont {Fioretto},\ and\ \citenamefont
  {Rolla}}]{corezzi2002bond}%
  \BibitemOpen
  \bibfield  {author} {\bibinfo {author} {\bibfnamefont {S.}~\bibnamefont
  {Corezzi}}, \bibinfo {author} {\bibfnamefont {D.}~\bibnamefont {Fioretto}}, \
  and\ \bibinfo {author} {\bibfnamefont {P.}~\bibnamefont {Rolla}},\
  }\href@noop {} {\bibfield  {journal} {\bibinfo  {journal} {Nature}\ }\textbf
  {\bibinfo {volume} {420}},\ \bibinfo {pages} {653} (\bibinfo {year}
  {2002})}\BibitemShut {NoStop}%
\bibitem [{\citenamefont {Mereu}\ \emph {et~al.}(2015)\citenamefont {Mereu},
  \citenamefont {Liotta}, \citenamefont {Comez},\ and\ \citenamefont
  {Corezzi}}]{mereu2015interplay}%
  \BibitemOpen
  \bibfield  {author} {\bibinfo {author} {\bibfnamefont {I.}~\bibnamefont
  {Mereu}}, \bibinfo {author} {\bibfnamefont {A.}~\bibnamefont {Liotta}},
  \bibinfo {author} {\bibfnamefont {L.}~\bibnamefont {Comez}}, \ and\ \bibinfo
  {author} {\bibfnamefont {S.}~\bibnamefont {Corezzi}},\ }\href@noop {}
  {\bibfield  {journal} {\bibinfo  {journal} {The Journal of Chemical Physics}\
  }\textbf {\bibinfo {volume} {142}} (\bibinfo {year} {2015})}\BibitemShut
  {NoStop}%
\bibitem [{\citenamefont {Krakoviack}(2010)}]{krakoviack2010statistical}%
  \BibitemOpen
  \bibfield  {author} {\bibinfo {author} {\bibfnamefont {V.}~\bibnamefont
  {Krakoviack}},\ }\href@noop {} {\bibfield  {journal} {\bibinfo  {journal}
  {Phys. Rev. E}\ }\textbf {\bibinfo {volume} {82}},\ \bibinfo {pages} {061501}
  (\bibinfo {year} {2010})}\BibitemShut {NoStop}%
\bibitem [{\citenamefont {Scheidler}\ \emph {et~al.}(2004)\citenamefont
  {Scheidler}, \citenamefont {Kob},\ and\ \citenamefont
  {Binder}}]{scheidler2004relaxation}%
  \BibitemOpen
  \bibfield  {author} {\bibinfo {author} {\bibfnamefont {P.}~\bibnamefont
  {Scheidler}}, \bibinfo {author} {\bibfnamefont {W.}~\bibnamefont {Kob}}, \
  and\ \bibinfo {author} {\bibfnamefont {K.}~\bibnamefont {Binder}},\
  }\href@noop {} {\bibfield  {journal} {\bibinfo  {journal} {J. Phys. Chem.}\
  }\textbf {\bibinfo {volume} {108}},\ \bibinfo {pages} {6673} (\bibinfo {year}
  {2004})}\BibitemShut {NoStop}%
\bibitem [{\citenamefont {Ikeda}\ \emph {et~al.}(2017)\citenamefont {Ikeda},
  \citenamefont {Zamponi},\ and\ \citenamefont {Ikeda}}]{ikeda2017mean}%
  \BibitemOpen
  \bibfield  {author} {\bibinfo {author} {\bibfnamefont {H.}~\bibnamefont
  {Ikeda}}, \bibinfo {author} {\bibfnamefont {F.}~\bibnamefont {Zamponi}}, \
  and\ \bibinfo {author} {\bibfnamefont {A.}~\bibnamefont {Ikeda}},\
  }\href@noop {} {\bibfield  {journal} {\bibinfo  {journal} {J. Chem. Phys.}\
  }\textbf {\bibinfo {volume} {147}},\ \bibinfo {pages} {234506} (\bibinfo
  {year} {2017})}\BibitemShut {NoStop}%
\bibitem [{\citenamefont {Szamel}(2019)}]{szamel2019theory}%
  \BibitemOpen
  \bibfield  {author} {\bibinfo {author} {\bibfnamefont {G.}~\bibnamefont
  {Szamel}},\ }\href@noop {} {\bibfield  {journal} {\bibinfo  {journal}
  {JSTAT}\ }\textbf {\bibinfo {volume} {2019}},\ \bibinfo {pages} {104016}
  (\bibinfo {year} {2019})}\BibitemShut {NoStop}%
\bibitem [{\citenamefont {Kapteijns}\ \emph {et~al.}(2019)\citenamefont
  {Kapteijns}, \citenamefont {Ji}, \citenamefont {Brito}, \citenamefont
  {Wyart},\ and\ \citenamefont {Lerner}}]{kapteijns2019fast}%
  \BibitemOpen
  \bibfield  {author} {\bibinfo {author} {\bibfnamefont {G.}~\bibnamefont
  {Kapteijns}}, \bibinfo {author} {\bibfnamefont {W.}~\bibnamefont {Ji}},
  \bibinfo {author} {\bibfnamefont {C.}~\bibnamefont {Brito}}, \bibinfo
  {author} {\bibfnamefont {M.}~\bibnamefont {Wyart}}, \ and\ \bibinfo {author}
  {\bibfnamefont {E.}~\bibnamefont {Lerner}},\ }\href@noop {} {\bibfield
  {journal} {\bibinfo  {journal} {Phys. Rev. E}\ }\textbf {\bibinfo {volume}
  {99}},\ \bibinfo {pages} {012106} (\bibinfo {year} {2019})}\BibitemShut
  {NoStop}%
\bibitem [{\citenamefont {Marchand}\ \emph {et~al.}(2022)\citenamefont
  {Marchand}, \citenamefont {Ozawa}, \citenamefont {Biroli},\ and\
  \citenamefont {Mallat}}]{marchand2022wavelet}%
  \BibitemOpen
  \bibfield  {author} {\bibinfo {author} {\bibfnamefont {T.}~\bibnamefont
  {Marchand}}, \bibinfo {author} {\bibfnamefont {M.}~\bibnamefont {Ozawa}},
  \bibinfo {author} {\bibfnamefont {G.}~\bibnamefont {Biroli}}, \ and\ \bibinfo
  {author} {\bibfnamefont {S.}~\bibnamefont {Mallat}},\ }\href@noop {}
  {\bibfield  {journal} {\bibinfo  {journal} {arXiv:2207.04941}\ } (\bibinfo
  {year} {2022})}\BibitemShut {NoStop}%
\bibitem [{\citenamefont {Zdeborov{\'a}}\ and\ \citenamefont
  {Krzakala}(2016)}]{zdeborova2016statistical}%
  \BibitemOpen
  \bibfield  {author} {\bibinfo {author} {\bibfnamefont {L.}~\bibnamefont
  {Zdeborov{\'a}}}\ and\ \bibinfo {author} {\bibfnamefont {F.}~\bibnamefont
  {Krzakala}},\ }\href@noop {} {\bibfield  {journal} {\bibinfo  {journal} {Adv.
  Phys.}\ }\textbf {\bibinfo {volume} {65}},\ \bibinfo {pages} {453} (\bibinfo
  {year} {2016})}\BibitemShut {NoStop}%
\bibitem [{\citenamefont {Cammarota}\ and\ \citenamefont
  {Biroli}(2013)}]{cammarota2013random}%
  \BibitemOpen
  \bibfield  {author} {\bibinfo {author} {\bibfnamefont {C.}~\bibnamefont
  {Cammarota}}\ and\ \bibinfo {author} {\bibfnamefont {G.}~\bibnamefont
  {Biroli}},\ }\href@noop {} {\bibfield  {journal} {\bibinfo  {journal} {J.
  Chem. Phys.}\ }\textbf {\bibinfo {volume} {138}} (\bibinfo {year}
  {2013})}\BibitemShut {NoStop}%
\bibitem [{\citenamefont {Kob}\ and\ \citenamefont
  {Andersen}(1995)}]{kob1995testing}%
  \BibitemOpen
  \bibfield  {author} {\bibinfo {author} {\bibfnamefont {W.}~\bibnamefont
  {Kob}}\ and\ \bibinfo {author} {\bibfnamefont {H.~C.}\ \bibnamefont
  {Andersen}},\ }\href@noop {} {\bibfield  {journal} {\bibinfo  {journal}
  {Phys. Rev. E}\ }\textbf {\bibinfo {volume} {51}},\ \bibinfo {pages} {4626}
  (\bibinfo {year} {1995})}\BibitemShut {NoStop}%
\bibitem [{\citenamefont {Hiroike}(1960)}]{hiroike1960new}%
  \BibitemOpen
  \bibfield  {author} {\bibinfo {author} {\bibfnamefont {K.}~\bibnamefont
  {Hiroike}},\ }\href@noop {} {\bibfield  {journal} {\bibinfo  {journal} {Prog.
  Theor. Phys.}\ }\textbf {\bibinfo {volume} {24}},\ \bibinfo {pages} {317}
  (\bibinfo {year} {1960})}\BibitemShut {NoStop}%
\bibitem [{\citenamefont {Morita}\ and\ \citenamefont
  {Hiroike}(1961)}]{morita1961new}%
  \BibitemOpen
  \bibfield  {author} {\bibinfo {author} {\bibfnamefont {T.}~\bibnamefont
  {Morita}}\ and\ \bibinfo {author} {\bibfnamefont {K.}~\bibnamefont
  {Hiroike}},\ }\href@noop {} {\bibfield  {journal} {\bibinfo  {journal} {Prog.
  Theor. Phys.}\ }\textbf {\bibinfo {volume} {25}},\ \bibinfo {pages} {537}
  (\bibinfo {year} {1961})}\BibitemShut {NoStop}%
\bibitem [{\citenamefont {Ozawa}\ \emph
  {et~al.}(2018{\natexlab{b}})\citenamefont {Ozawa}, \citenamefont {Parisi},\
  and\ \citenamefont {Berthier}}]{ozawa2018configurational}%
  \BibitemOpen
  \bibfield  {author} {\bibinfo {author} {\bibfnamefont {M.}~\bibnamefont
  {Ozawa}}, \bibinfo {author} {\bibfnamefont {G.}~\bibnamefont {Parisi}}, \
  and\ \bibinfo {author} {\bibfnamefont {L.}~\bibnamefont {Berthier}},\
  }\href@noop {} {\bibfield  {journal} {\bibinfo  {journal} {J. Chem. Phys.}\
  }\textbf {\bibinfo {volume} {149}},\ \bibinfo {pages} {154501} (\bibinfo
  {year} {2018}{\natexlab{b}})}\BibitemShut {NoStop}%
\bibitem [{\citenamefont {Biroli}\ \emph {et~al.}(2008)\citenamefont {Biroli},
  \citenamefont {Bouchaud}, \citenamefont {Cavagna}, \citenamefont {Grigera},\
  and\ \citenamefont {Verrocchio}}]{biroli2008thermodynamic}%
  \BibitemOpen
  \bibfield  {author} {\bibinfo {author} {\bibfnamefont {G.}~\bibnamefont
  {Biroli}}, \bibinfo {author} {\bibfnamefont {J.-P.}\ \bibnamefont
  {Bouchaud}}, \bibinfo {author} {\bibfnamefont {A.}~\bibnamefont {Cavagna}},
  \bibinfo {author} {\bibfnamefont {T.~S.}\ \bibnamefont {Grigera}}, \ and\
  \bibinfo {author} {\bibfnamefont {P.}~\bibnamefont {Verrocchio}},\
  }\href@noop {} {\bibfield  {journal} {\bibinfo  {journal} {Nat. Phys.}\
  }\textbf {\bibinfo {volume} {4}},\ \bibinfo {pages} {771} (\bibinfo {year}
  {2008})}\BibitemShut {NoStop}%
\bibitem [{\citenamefont {Hocky}\ \emph {et~al.}(2012)\citenamefont {Hocky},
  \citenamefont {Markland},\ and\ \citenamefont {Reichman}}]{hocky2012growing}%
  \BibitemOpen
  \bibfield  {author} {\bibinfo {author} {\bibfnamefont {G.~M.}\ \bibnamefont
  {Hocky}}, \bibinfo {author} {\bibfnamefont {T.~E.}\ \bibnamefont {Markland}},
  \ and\ \bibinfo {author} {\bibfnamefont {D.~R.}\ \bibnamefont {Reichman}},\
  }\href@noop {} {\bibfield  {journal} {\bibinfo  {journal} {Phys. Rev. Lett.}\
  }\textbf {\bibinfo {volume} {108}},\ \bibinfo {pages} {225506} (\bibinfo
  {year} {2012})}\BibitemShut {NoStop}%
\bibitem [{\citenamefont {Hocky}\ \emph {et~al.}(2014)\citenamefont {Hocky},
  \citenamefont {Berthier}, \citenamefont {Kob},\ and\ \citenamefont
  {Reichman}}]{hocky2014crossovers}%
  \BibitemOpen
  \bibfield  {author} {\bibinfo {author} {\bibfnamefont {G.~M.}\ \bibnamefont
  {Hocky}}, \bibinfo {author} {\bibfnamefont {L.}~\bibnamefont {Berthier}},
  \bibinfo {author} {\bibfnamefont {W.}~\bibnamefont {Kob}}, \ and\ \bibinfo
  {author} {\bibfnamefont {D.~R.}\ \bibnamefont {Reichman}},\ }\href@noop {}
  {\bibfield  {journal} {\bibinfo  {journal} {Phys. Rev. E}\ }\textbf {\bibinfo
  {volume} {89}},\ \bibinfo {pages} {052311} (\bibinfo {year}
  {2014})}\BibitemShut {NoStop}%
\bibitem [{\citenamefont {Yaida}\ \emph {et~al.}(2016)\citenamefont {Yaida},
  \citenamefont {Berthier}, \citenamefont {Charbonneau},\ and\ \citenamefont
  {Tarjus}}]{yaida2016point}%
  \BibitemOpen
  \bibfield  {author} {\bibinfo {author} {\bibfnamefont {S.}~\bibnamefont
  {Yaida}}, \bibinfo {author} {\bibfnamefont {L.}~\bibnamefont {Berthier}},
  \bibinfo {author} {\bibfnamefont {P.}~\bibnamefont {Charbonneau}}, \ and\
  \bibinfo {author} {\bibfnamefont {G.}~\bibnamefont {Tarjus}},\ }\href@noop {}
  {\bibfield  {journal} {\bibinfo  {journal} {Phys. Rev. E}\ }\textbf {\bibinfo
  {volume} {94}},\ \bibinfo {pages} {032605} (\bibinfo {year}
  {2016})}\BibitemShut {NoStop}%
\bibitem [{\citenamefont {Nandi}\ \emph {et~al.}(2021)\citenamefont {Nandi},
  \citenamefont {Kob},\ and\ \citenamefont
  {Maitra~Bhattacharyya}}]{nandi2021connecting}%
  \BibitemOpen
  \bibfield  {author} {\bibinfo {author} {\bibfnamefont {U.~K.}\ \bibnamefont
  {Nandi}}, \bibinfo {author} {\bibfnamefont {W.}~\bibnamefont {Kob}}, \ and\
  \bibinfo {author} {\bibfnamefont {S.}~\bibnamefont {Maitra~Bhattacharyya}},\
  }\href@noop {} {\bibfield  {journal} {\bibinfo  {journal} {J. Chem. Phys.}\
  }\textbf {\bibinfo {volume} {154}},\ \bibinfo {pages} {094506} (\bibinfo
  {year} {2021})}\BibitemShut {NoStop}%
\bibitem [{\citenamefont {Nandi}\ \emph {et~al.}(2022)\citenamefont {Nandi},
  \citenamefont {Patel}, \citenamefont {Moid}, \citenamefont {Nandi},
  \citenamefont {Sengupta}, \citenamefont {Karmakar}, \citenamefont {Maiti},
  \citenamefont {Dasgupta},\ and\ \citenamefont
  {Maitra~Bhattacharyya}}]{nandi2022thermodynamics}%
  \BibitemOpen
  \bibfield  {author} {\bibinfo {author} {\bibfnamefont {U.~K.}\ \bibnamefont
  {Nandi}}, \bibinfo {author} {\bibfnamefont {P.}~\bibnamefont {Patel}},
  \bibinfo {author} {\bibfnamefont {M.}~\bibnamefont {Moid}}, \bibinfo {author}
  {\bibfnamefont {M.~K.}\ \bibnamefont {Nandi}}, \bibinfo {author}
  {\bibfnamefont {S.}~\bibnamefont {Sengupta}}, \bibinfo {author}
  {\bibfnamefont {S.}~\bibnamefont {Karmakar}}, \bibinfo {author}
  {\bibfnamefont {P.~K.}\ \bibnamefont {Maiti}}, \bibinfo {author}
  {\bibfnamefont {C.}~\bibnamefont {Dasgupta}}, \ and\ \bibinfo {author}
  {\bibfnamefont {S.}~\bibnamefont {Maitra~Bhattacharyya}},\ }\href@noop {}
  {\bibfield  {journal} {\bibinfo  {journal} {J. Chem. Phys.}\ }\textbf
  {\bibinfo {volume} {156}} (\bibinfo {year} {2022})}\BibitemShut {NoStop}%
\bibitem [{\citenamefont {Andersen}(1983)}]{andersen1983rattle}%
  \BibitemOpen
  \bibfield  {author} {\bibinfo {author} {\bibfnamefont {H.~C.}\ \bibnamefont
  {Andersen}},\ }\href@noop {} {\bibfield  {journal} {\bibinfo  {journal} {J.
  Comput. Phys.}\ }\textbf {\bibinfo {volume} {52}},\ \bibinfo {pages} {24}
  (\bibinfo {year} {1983})}\BibitemShut {NoStop}%
\bibitem [{\citenamefont {G{\"o}tze}(2009)}]{gotze2009complex}%
  \BibitemOpen
  \bibfield  {author} {\bibinfo {author} {\bibfnamefont {W.}~\bibnamefont
  {G{\"o}tze}},\ }\href@noop {} {\emph {\bibinfo {title} {Complex dynamics of
  glass-forming liquids: A mode-coupling theory}}},\ Vol.\ \bibinfo {volume}
  {143}\ (\bibinfo  {publisher} {Oxford University Press, USA},\ \bibinfo
  {year} {2009})\BibitemShut {NoStop}%
\bibitem [{\citenamefont {Cammarota}\ \emph {et~al.}(2023)\citenamefont
  {Cammarota}, \citenamefont {Ozawa},\ and\ \citenamefont
  {Tarjus}}]{cammarota2023kauzmann}%
  \BibitemOpen
  \bibfield  {author} {\bibinfo {author} {\bibfnamefont {C.}~\bibnamefont
  {Cammarota}}, \bibinfo {author} {\bibfnamefont {M.}~\bibnamefont {Ozawa}}, \
  and\ \bibinfo {author} {\bibfnamefont {G.}~\bibnamefont {Tarjus}},\ }in\
  \href@noop {} {\emph {\bibinfo {booktitle} {Spin Glass Theory and Far Beyond:
  Replica Symmetry Breaking After 40 Years}}}\ (\bibinfo  {publisher} {World
  Scientific},\ \bibinfo {year} {2023})\ pp.\ \bibinfo {pages}
  {203--218}\BibitemShut {NoStop}%
\bibitem [{\citenamefont {Gnan}\ \emph {et~al.}(2009)\citenamefont {Gnan},
  \citenamefont {Schr{\o}der}, \citenamefont {Pedersen}, \citenamefont
  {Bailey},\ and\ \citenamefont {Dyre}}]{gnan2009pressure}%
  \BibitemOpen
  \bibfield  {author} {\bibinfo {author} {\bibfnamefont {N.}~\bibnamefont
  {Gnan}}, \bibinfo {author} {\bibfnamefont {T.~B.}\ \bibnamefont
  {Schr{\o}der}}, \bibinfo {author} {\bibfnamefont {U.~R.}\ \bibnamefont
  {Pedersen}}, \bibinfo {author} {\bibfnamefont {N.~P.}\ \bibnamefont
  {Bailey}}, \ and\ \bibinfo {author} {\bibfnamefont {J.~C.}\ \bibnamefont
  {Dyre}},\ }\href@noop {} {\bibfield  {journal} {\bibinfo  {journal} {J. Chem.
  Phys.}\ }\textbf {\bibinfo {volume} {131}},\ \bibinfo {pages} {234504}
  (\bibinfo {year} {2009})}\BibitemShut {NoStop}%
\bibitem [{\citenamefont {Schr{\o}der}\ and\ \citenamefont
  {Dyre}(2014)}]{schroder2014simplicity}%
  \BibitemOpen
  \bibfield  {author} {\bibinfo {author} {\bibfnamefont {T.~B.}\ \bibnamefont
  {Schr{\o}der}}\ and\ \bibinfo {author} {\bibfnamefont {J.~C.}\ \bibnamefont
  {Dyre}},\ }\href@noop {} {\bibfield  {journal} {\bibinfo  {journal} {J. Chem.
  Phys.}\ }\textbf {\bibinfo {volume} {141}},\ \bibinfo {pages} {204502}
  (\bibinfo {year} {2014})}\BibitemShut {NoStop}%
\bibitem [{\citenamefont {Hurley}\ and\ \citenamefont
  {Harrowell}(1995)}]{hurley1995kinetic}%
  \BibitemOpen
  \bibfield  {author} {\bibinfo {author} {\bibfnamefont {M.}~\bibnamefont
  {Hurley}}\ and\ \bibinfo {author} {\bibfnamefont {P.}~\bibnamefont
  {Harrowell}},\ }\href@noop {} {\bibfield  {journal} {\bibinfo  {journal}
  {Phys. Rev. E}\ }\textbf {\bibinfo {volume} {52}},\ \bibinfo {pages} {1694}
  (\bibinfo {year} {1995})}\BibitemShut {NoStop}%
\bibitem [{\citenamefont {Kob}\ \emph {et~al.}(1997)\citenamefont {Kob},
  \citenamefont {Donati}, \citenamefont {Plimpton}, \citenamefont {Poole},\
  and\ \citenamefont {Glotzer}}]{kob1997dynamical}%
  \BibitemOpen
  \bibfield  {author} {\bibinfo {author} {\bibfnamefont {W.}~\bibnamefont
  {Kob}}, \bibinfo {author} {\bibfnamefont {C.}~\bibnamefont {Donati}},
  \bibinfo {author} {\bibfnamefont {S.~J.}\ \bibnamefont {Plimpton}}, \bibinfo
  {author} {\bibfnamefont {P.~H.}\ \bibnamefont {Poole}}, \ and\ \bibinfo
  {author} {\bibfnamefont {S.~C.}\ \bibnamefont {Glotzer}},\ }\href@noop {}
  {\bibfield  {journal} {\bibinfo  {journal} {Phys. Rev. Lett.}\ }\textbf
  {\bibinfo {volume} {79}},\ \bibinfo {pages} {2827} (\bibinfo {year}
  {1997})}\BibitemShut {NoStop}%
\bibitem [{\citenamefont {Yamamoto}\ and\ \citenamefont
  {Onuki}(1998)}]{yamamoto1998dynamics}%
  \BibitemOpen
  \bibfield  {author} {\bibinfo {author} {\bibfnamefont {R.}~\bibnamefont
  {Yamamoto}}\ and\ \bibinfo {author} {\bibfnamefont {A.}~\bibnamefont
  {Onuki}},\ }\href@noop {} {\bibfield  {journal} {\bibinfo  {journal} {Phys.
  Rev. E}\ }\textbf {\bibinfo {volume} {58}},\ \bibinfo {pages} {3515}
  (\bibinfo {year} {1998})}\BibitemShut {NoStop}%
\bibitem [{\citenamefont {Karmakar}\ \emph {et~al.}(2009)\citenamefont
  {Karmakar}, \citenamefont {Dasgupta},\ and\ \citenamefont
  {Sastry}}]{karmakar2009growing}%
  \BibitemOpen
  \bibfield  {author} {\bibinfo {author} {\bibfnamefont {S.}~\bibnamefont
  {Karmakar}}, \bibinfo {author} {\bibfnamefont {C.}~\bibnamefont {Dasgupta}},
  \ and\ \bibinfo {author} {\bibfnamefont {S.}~\bibnamefont {Sastry}},\
  }\href@noop {} {\bibfield  {journal} {\bibinfo  {journal} {PNAS}\ }\textbf
  {\bibinfo {volume} {106}},\ \bibinfo {pages} {3675} (\bibinfo {year}
  {2009})}\BibitemShut {NoStop}%
\bibitem [{\citenamefont {Donati}\ \emph {et~al.}(2002)\citenamefont {Donati},
  \citenamefont {Franz}, \citenamefont {Glotzer},\ and\ \citenamefont
  {Parisi}}]{donati2002theory}%
  \BibitemOpen
  \bibfield  {author} {\bibinfo {author} {\bibfnamefont {C.}~\bibnamefont
  {Donati}}, \bibinfo {author} {\bibfnamefont {S.}~\bibnamefont {Franz}},
  \bibinfo {author} {\bibfnamefont {S.~C.}\ \bibnamefont {Glotzer}}, \ and\
  \bibinfo {author} {\bibfnamefont {G.}~\bibnamefont {Parisi}},\ }\href@noop {}
  {\bibfield  {journal} {\bibinfo  {journal} {J. Non-Cryst.}\ }\textbf
  {\bibinfo {volume} {307}},\ \bibinfo {pages} {215} (\bibinfo {year}
  {2002})}\BibitemShut {NoStop}%
\bibitem [{\citenamefont {Shiraishi}\ \emph {et~al.}(2023)\citenamefont
  {Shiraishi}, \citenamefont {Mizuno},\ and\ \citenamefont
  {Ikeda}}]{shiraishi2023johari}%
  \BibitemOpen
  \bibfield  {author} {\bibinfo {author} {\bibfnamefont {K.}~\bibnamefont
  {Shiraishi}}, \bibinfo {author} {\bibfnamefont {H.}~\bibnamefont {Mizuno}}, \
  and\ \bibinfo {author} {\bibfnamefont {A.}~\bibnamefont {Ikeda}},\
  }\href@noop {} {\bibfield  {journal} {\bibinfo  {journal} {PNAS}\ }\textbf
  {\bibinfo {volume} {120}},\ \bibinfo {pages} {e2215153120} (\bibinfo {year}
  {2023})}\BibitemShut {NoStop}%
\bibitem [{\citenamefont {Kawasaki}\ and\ \citenamefont
  {Kim}(2019{\natexlab{a}})}]{kawasaki2019classification}%
  \BibitemOpen
  \bibfield  {author} {\bibinfo {author} {\bibfnamefont {T.}~\bibnamefont
  {Kawasaki}}\ and\ \bibinfo {author} {\bibfnamefont {K.}~\bibnamefont {Kim}},\
  }\href@noop {} {\bibfield  {journal} {\bibinfo  {journal} {JSTAT}\ }\textbf
  {\bibinfo {volume} {2019}},\ \bibinfo {pages} {084004} (\bibinfo {year}
  {2019}{\natexlab{a}})}\BibitemShut {NoStop}%
\bibitem [{\citenamefont {Kou}\ \emph {et~al.}(2018)\citenamefont {Kou},
  \citenamefont {Cao}, \citenamefont {Li}, \citenamefont {Xia}, \citenamefont
  {Li}, \citenamefont {Dong}, \citenamefont {Zhang}, \citenamefont {Zhang},
  \citenamefont {Kob},\ and\ \citenamefont {Wang}}]{kou2018translational}%
  \BibitemOpen
  \bibfield  {author} {\bibinfo {author} {\bibfnamefont {B.}~\bibnamefont
  {Kou}}, \bibinfo {author} {\bibfnamefont {Y.}~\bibnamefont {Cao}}, \bibinfo
  {author} {\bibfnamefont {J.}~\bibnamefont {Li}}, \bibinfo {author}
  {\bibfnamefont {C.}~\bibnamefont {Xia}}, \bibinfo {author} {\bibfnamefont
  {Z.}~\bibnamefont {Li}}, \bibinfo {author} {\bibfnamefont {H.}~\bibnamefont
  {Dong}}, \bibinfo {author} {\bibfnamefont {A.}~\bibnamefont {Zhang}},
  \bibinfo {author} {\bibfnamefont {J.}~\bibnamefont {Zhang}}, \bibinfo
  {author} {\bibfnamefont {W.}~\bibnamefont {Kob}}, \ and\ \bibinfo {author}
  {\bibfnamefont {Y.}~\bibnamefont {Wang}},\ }\href@noop {} {\bibfield
  {journal} {\bibinfo  {journal} {Phys. Rev. Lett.}\ }\textbf {\bibinfo
  {volume} {121}},\ \bibinfo {pages} {018002} (\bibinfo {year}
  {2018})}\BibitemShut {NoStop}%
\bibitem [{\citenamefont {Cicerone}\ and\ \citenamefont
  {Ediger}(1995)}]{cicerone95dynhet}%
  \BibitemOpen
  \bibfield  {author} {\bibinfo {author} {\bibfnamefont {M.~T.}\ \bibnamefont
  {Cicerone}}\ and\ \bibinfo {author} {\bibfnamefont {M.~D.}\ \bibnamefont
  {Ediger}},\ }\href@noop {} {\bibfield  {journal} {\bibinfo  {journal} {J.
  Chem. Phys.}\ }\textbf {\bibinfo {volume} {103}},\ \bibinfo {pages} {5684}
  (\bibinfo {year} {1995})}\BibitemShut {NoStop}%
\bibitem [{\citenamefont {Tarjus}\ and\ \citenamefont
  {Kivelson}(1995)}]{tarjus1995breakdown}%
  \BibitemOpen
  \bibfield  {author} {\bibinfo {author} {\bibfnamefont {G.}~\bibnamefont
  {Tarjus}}\ and\ \bibinfo {author} {\bibfnamefont {D.}~\bibnamefont
  {Kivelson}},\ }\href@noop {} {\bibfield  {journal} {\bibinfo  {journal} {J.
  Chem. Phys.}\ }\textbf {\bibinfo {volume} {103}},\ \bibinfo {pages} {3071}
  (\bibinfo {year} {1995})}\BibitemShut {NoStop}%
\bibitem [{\citenamefont {Kawasaki}\ and\ \citenamefont
  {Kim}(2019{\natexlab{b}})}]{kawasaki2019spurious}%
  \BibitemOpen
  \bibfield  {author} {\bibinfo {author} {\bibfnamefont {T.}~\bibnamefont
  {Kawasaki}}\ and\ \bibinfo {author} {\bibfnamefont {K.}~\bibnamefont {Kim}},\
  }\href@noop {} {\bibfield  {journal} {\bibinfo  {journal} {Sci. Rep.}\
  }\textbf {\bibinfo {volume} {9}},\ \bibinfo {pages} {8118} (\bibinfo {year}
  {2019}{\natexlab{b}})}\BibitemShut {NoStop}%
\bibitem [{\citenamefont {Hall}\ and\ \citenamefont
  {Wolynes}(2003)}]{hall2003microscopic}%
  \BibitemOpen
  \bibfield  {author} {\bibinfo {author} {\bibfnamefont {R.~W.}\ \bibnamefont
  {Hall}}\ and\ \bibinfo {author} {\bibfnamefont {P.~G.}\ \bibnamefont
  {Wolynes}},\ }\href@noop {} {\bibfield  {journal} {\bibinfo  {journal} {Phys.
  Rev. Lett.}\ }\textbf {\bibinfo {volume} {90}},\ \bibinfo {pages} {085505}
  (\bibinfo {year} {2003})}\BibitemShut {NoStop}%
\end{thebibliography}%

\end{document}